\documentclass[manuscript,a4paper,article]{aastex}

\newcommand{\Chandra}{{\it Chandra}}

\newcommand{\XMM}{{\it XMM-Newton}}
\newcommand{\Suzaku}{{\it Suzaku}}
\newcommand{\ROSAT}{{\it ROSAT}}
\newcommand{\ASCA}{{\it ASCA}}

\usepackage{epsfig}
\usepackage{amsmath}
\usepackage[breaklinks]{hyperref}
\usepackage[title]{appendix}
\usepackage{color}

\graphicspath{{figures/}}

\shorttitle{A study of merger history of the Galaxy Group HCG~62}
\shortauthors{Hu et al.}

\begin{document}

\title{A Study of the Merger History of the Galaxy Group HCG~62 Based on X-Ray Observations and SPH Simulations}

\author{\sc Dan Hu\altaffilmark{1}, Haiguang Xu\altaffilmark{1,2,3}, Xi Kang\altaffilmark{4},  Weitian Li\altaffilmark{1}, Zhenghao Zhu\altaffilmark{1}, Zhixian Ma\altaffilmark{5}, Chenxi Shan\altaffilmark{1}, Zhongli Zhang\altaffilmark{6}, Liyi Gu\altaffilmark{7}, Chengze Liu\altaffilmark{1}, Qian Zheng\altaffilmark{6}, and Xiang-ping Wu\altaffilmark{8}}

\altaffiltext{1}{School of Physics and Astronomy, Shanghai Jiao Tong University, 800 Dongchuan Road, Minhang, Shanghai 200240, China; email: hudan\_bazhaoyu@sjtu.edu.cn, hgxu@sjtu.edu.cn}
\altaffiltext{2}{Tsung-Dao Lee Institute, Shanghai Jiao Tong University, 800 Dongchuan Road, Minhang, Shanghai 200240, China}
\altaffiltext{3}{IFSA Collaborative Innovation Center, Shanghai Jiao Tong University, Minhang, Shanghai 200240, China}
\altaffiltext{4}{Purple Mountain Observatory, the partner group of MPI fuer Astronomie, 2 west beijing road, nanjing 210008, China}
\altaffiltext{5}{Department of Electronic Engineering, Shanghai Jiao Tong University, 800 Dongchuan Road, Minhang, Shanghai 200240, China}
\altaffiltext{6}{Shanghai Astronomical Observatory, Chinese Academy of Sciences, 80 Nandan Road, Shanghai 200030, China}
\altaffiltext{7}{SRON Netherlands Institute for Space Research, Sorbonnelaan 2, 3584 CA Utrecht, the Netherlands}
\altaffiltext{8}{National Astronomical Observatories, Chinese Academy of Sciences, 20A Datun Road, Beijing 100012, China}

\begin{abstract}
We choose the bright compact group HCG~62, which was found to exhibit both excess X-ray emission and high Fe abundance to the southwest of its core, as an example to study the impact of mergers on chemical enrichment in the intragroup medium. We first reanalyze the high-quality \Chandra\ and \XMM\ archive data to search for the evidence for additional SN II yields, which is expected as a direct result of the possible merger-induced starburst. We reveal that, similar to the Fe abundance, the Mg abundance also shows a high value in both the innermost region and the southwest substructure, forming a high-abundance plateau, meanwhile all the SN Ia and SN II yields show rather flat distributions in $>0.1r_{200}$ in favor of an early enrichment. Then we carry out a series of idealized numerical simulations to model the collision of two initially isolated galaxy groups by using the TreePM-SPH GADGET-3 code. We find that the observed X-ray emission and metal distributions, as well as the relative positions of the two bright central galaxies with reference to the X-ray peak, can be well reproduced in a major merger with a mass ratio of 3 when the merger-induced starburst is assumed. The `best-match' snapshot is pinpointed after the third pericentric passage when the southwest substructure is formed due to gas sloshing. By following the evolution of the simulated merging system, we conclude that the effects of such a major merger on chemical enrichment are mostly restricted within the core region when the final relaxed state is reached.
\end{abstract}


%#####################
\section{INTRODUCTION}
%#####################

Nearly all of the elements in the Universe heavier than helium are created through nucleosynthesis during explosive burning in supernova events, as well as through steady fusion of lighter elements into heavier ones in evolving stars \citep[e.g.,][]{woosley86, tsujimoto95}. In galaxy clusters and groups, however, a considerable amount of the synthesized metals do not accumulate in the interior of member galaxies, but have been mixed into the intracluster medium (see \citealt{sarazin88} for an earlier review), indicating either rapid transport and mix or gas enrichment at early evolution stages. In order to trace the chemical evolution of galaxy clusters and groups, studies of the spatial distributions of metal abundances and their ratios are essential, because the metal enrichment caused by Type Ia supernovae (SN Ia) is expected to differ significantly from that by Type II supernovae (SN II) both in time and space.

Results of early X-ray observations show that there exists an excessive amount of Fe in the intracluster medium. In order to explain the origin of this excess, \citet{vigroux77} proposed to introduce the concept of early enrichment, which can be caused by the first-generation objects during or prior to the assembly of galaxy clusters and groups, and by the follow-up supernovae events, especially during the peak period of star formation at redshifts of $2-3$ \citep[e.g.,][]{romeo06}. This scenario is strongly favored by recently found uniform distributions of Fe abundance and abundance ratios (e.g., Mg/Fe and Si/Fe) as a function of radius out to the virial radius in Abell 399 and Abell 401 \citep{fujita08}, the Perseus cluster \citep{werner13}, and the Virgo cluster \citep{simionescu15}. In many other sources \citep{sasaki14, urban17}, this phenomenon has also been confirmed out to about half of the virial radius. 

In the core regions of many galaxy clusters and groups, however, a remarkable inward Fe abundance increase (e.g., see \citealt{makishima01}; \citealt{werner08}, for reviews; Recent results include, e.g., \citealt{bohringer04}; \citealt{tamura04}; \citealt{LM08,RP07,humphrey12}; \citealt{sasaki14,lagana15,mernier15,mernier17}) has been reported. Since this phenomenon turns to be observed in cool core systems \citep[e.g.,][]{vikhlinin05,pratt07}, it is speculated that, in addition to early enrichment, later enrichment caused by SN Ia events \citep{werner08,RP09}, which produced a large amount of Fe, Ni and Si-group elements, may have played a role in the formation of the central Fe increase. If true, a period ranging from several Gyr to about 10~Gyr without major dynamic disturbance is required to ensure a remarkable central metal accumulations as estimated by, e.g., \citet{bohringer04}, \citet{wang05}, and \citet{matsushita07a}. 

As for the alpha elements O and/or Mg, early studies using the \XMM\ and \Suzaku\ data suggested that their abundances turn to exhibit rather low values in the core region and/or flat distributions in many cases (e.g., Abell 496, \citealt{tamura01,lovisari11}; M87, \citealt{bohringer01,finoguenov02,matsushita03,werner06a}; a sample of 19 galaxy clusters, \citealt{tamura04}; the Perseus cluster \citealt{sanders04}; Abell 85, \citealt{durret05}; 2A 0335+096, \citealt{werner06b}; Abell 1060, \citealt{sato07}; AWM 7, \citealt{sato08}; the Fornax cluster, \citealt{matsushita07b}). However, recent analyses of \XMM\ and \Suzaku\ data have revealed centrally enhanced O and/or Mg abundance in Sérsic 159-03 \citep{deplaa06}, in the Hydra A cluster \citep{simionescu09}, in the Virgo cluster \citep{million11}, in Abell 2029 \citep{lovisari11}, in the Centaurus cluster \citep{SF06,lovisari11}, in the NGC 507 group \citep{sato09}, and in a sample of 44 cool-core systems \citep{mernier17}. A possible explanation for this phenomenon is that the centrally peaked SN II products are enriched by past (and sometimes present) extra star formation \citep{SF06,million11,mernier17}, which could be triggered by a major merger\footnote{~AGN activity, sometimes triggered by the merger, may also be able to trigger star formation and adjust metal redistribution \citep[e.g.,][]{schiminovich94,charmandaris00,bicknell00,debreuck05,papadopoulos05}. This interesting topic, however, is beyond the scope of this paper and therefore is not included here.}.

During a merger metals may be mixed and redistributed via the processes of ram-pressure stripping \citep{cui10,mernier15}, galaxy-galaxy interactions \citep{kapferer05,SD08}, outflows \citep{veilleux05,kapferer09}, and gas sloshing \citep{MV07,simionescu10}, and in some cases these may trigger massive star formation \citep{durret05,sun07} that can form remarkable abundance substructures.

As suggested by \citet{LT78}, the tidal force interaction between galaxies in a merger often triggers intense bursts of star-formation activity via radial inflows and spatially extended gas turbulence, when at least one of the galaxies contains a substantial amount of gas (i.e., wet merger). On the other hand, almost all starburst galaxies, including Luminous InfraRed Galaxies (LIRG), Ultra-Luminous InfraRed Galaxies (ULIRG) located at low redshifts ($z < 1 - 2$), and strongest starbursts (ULIRGs and Hyper-LIRGs) located at higher redshifts are predominantly merging systems (see reviews of \citealt{KE12}, \citealt{MD14}, and \citealt{bournaud11}). During such starbursts, which commonly last a few $10^{7}-10^{8}$~years, stars can form at rates tens or even hundreds of times greater than those observed in normal galaxies \citep{drexler09}. During a typical merger-induced starburst, the bulk of SN II events occur less than a few tens of Myr after the trigger since their massive progenitors evolve very fast. Although the SN Ia explosions of the old stellar population keep occurring continuously, those related to the newly formed stellar population shall require a considerable time delay (up to several Gyr; \citealt{montuori10}). Thus, once the star formation is triggered by the merger, the gas will be enriched primarily by the subsequent SN II events, which may be characterized by substantial concentrations of O, Ne, and Mg, before an additional Fe enrichment occurs at the post-merger stage.

Recent observations have revealed that cluster mergers can act as an important mechanism not only to drive metal redistribution, but also to trigger starburst, which may accounts for additional enrichment. On one hand, by analyzing \Chandra\ and \XMM\ data jointly \citet{su171} demonstrated that in the Fornax cluster merger-induced gas sloshing is effective in lifting a significant amount of metals from the core. Similar phenomena have also been found in Abell 496 \citep{ghizzardi14} and the Centaurus cluster \citep{sanders16}, as well as predicted in numerical simulations \citep[e.g.,][]{AM06,roediger11}. On the other hand, in the optical band, \citet{sobral15} showed that the star-forming galaxies in the merging cluster CIZA J2242.8+5301 (`Sausage') exhibit significantly higher metallicities (nearly one solar) than those located in the outskirts or in the field environment by using the $\rm [N_{II}]/H_{\alpha}$ emission line ratio. \citet{shimakawa15} find the similar results that galaxies in the dense environment more likely have higher metallicity than those in the field.

Compared with the case of galaxy clusters, the effects of mergers on the chemical enrichment in the intragroup medium are less studied. In this work we examine these issues using the smoothed particle hydrodynamics (SPH) code TreePM-SPH GADGET-3 \citep{springel05} to carry out three-dimensional numerical simulations. We choose the X-ray bright compact group HCG~62 as our example, because (1) the optical kinematic study of \citet{spavone06} has shown that its S0 member NGC~4778 (one of the two interacting central dominating galaxies; the other is NGC~4776) may have been undergoing a recent merger, as hinted by both kinematical and morphological peculiarities in its central region (i.e., a decoupled stellar component revealed by nuclear counterrotation) and outer halo (i.e., an asymmetric rotation curve and a velocity dispersion profile with a clear rise toward NGC~4776; see also \citealt{johnson07}); and (2) in addition to a central Fe enhancement, previous \Chandra\ and \XMM\ observations \citep{gu07,gitti10,rafferty13} have identified a special region located at about 2\arcmin\ southwest of the X-ray peak, which simultaneously exhibits excess X-ray emission and high Fe abundance. The authors argued that the origin of this special region is possibly related to a recent merger. 
Although \citet{tokoi08} failed to confirm the existence of this metal substructure by using \Suzaku\ data (see \S5.2, where we show that it is difficult to detect such an abundance substructure with \Suzaku's limited spatial resolution), they indicated that some of the hard sources identified within $r < 3.3'$ maybe remnants of previous minor merger (i.e., black holes of the merged galaxies).

In order to search for the evidence for additional SN II enrichment and to help reconstruct the merger history of HCG~62 via simulations, we reanalyzed the high-quality \Chandra\ and \XMM\ archive data with a total exposure of about 334.4~ks, which provide X-ray measurements (e.g., X-ray surface brightness, SN Ia and SN II yields, etc.) with sufficient spatial resolution that are required in the simulations. We describe the observations in \S2 and present the results of the X-ray imaging spectroscopic study in \S3. In \S4 we describe the procedure, technical details, and results of the hydrodynamical simulations. Finally, we present discussion and summary in \S5 and \S6, respectively. 

Throughout this paper we quote errors at the 90$\%$ confidence level unless otherwise stated and adopt cosmological parameters $H_0$ = 70 km s$^{-1}$ Mpc$^{-1}$ and $\Omega_m$ = 1 $-$ $\Omega_{\Lambda}$ = 0.27 for a flat universe. 
At the redshift of HCG~62 ($z$ = 0.0137), these parameters yield an angular diameter distance of 57.7~Mpc (i.e., $1\arcmin$ corresponds to 16.8~kpc) and a luminosity distance of 59.3~Mpc. We use the solar abundance standards of \citet{GS98}, according to which the iron abundance relative to hydrogen is $3.16\times10^{-5}$ in number.

%##############################################
\section{X-RAY OBSERVATIONS AND DATA REDUCTION}
%##############################################
We reanalyzed the archive X-ray data to characterize the spatial distributions of X-ray surface brightness, gas temperature, metal abundances, gas density, and dark matter density with sufficient spatial resolutions, which will be used in \S4 to constrain our numerical simulations, since corresponding information is not complete in literature.
For more detailed information about the X-ray imaging and spectral properties of HCG~62
(e.g., X-ray cavities, two-dimensional maps of temperature and abundance, azimuthally averaged spectral analysis, etc.), please refer to \citet{morita06}, \citet{gu07}, \citet{tokoi08}, \citet{gitti10}, and \citet{rafferty13}.

%==========================
\subsection{\Chandra\ Data}
%==========================
We analyzed the data obtained with the S3 chip of the Advanced CCD Imaging Spectrometer (ACIS) on board the \Chandra\ X-ray observatory in three pointing observations, which were performed on 2000 January 25 (ObsID 921, 48.5~ks, FAINT mode), 2009 March 2 (ObsID 10462, 67.1~ks, VFAINT mode), and 2009 March 3 (ObsID 10874, 51.4~ks, VFAINT mode). For each observation we followed the standard \Chandra\ data reduction procedure to process the data by using CIAO v4.6 and CALDB v4.6.2. After executing corrections for gain, CTI (except for ObsID 921), and astrometry, removing events with \ASCA\ grades 1, 5, and 7, and removing bad pixels and columns by running the CIAO script \textit{chandra$\_$repro}, we examined the light curves extracted in $0.5-12.0$~keV from source-free regions near the CCD edges and excluded time intervals contaminated by occasional particle background flares, during which the count rate deviates from the mean value by $20\%$. Then we applied CIAO tools \textit{wavedetect} and \textit{celldetect} to identify and exclude all the point sources detected beyond the $3\sigma$ threshold in the ACIS images. These steps yielded a total of 165~ks clean exposure from the three observations (Table~\ref{tbl-1}).

%======================
\subsection{\XMM\ Data}
%======================
We also analyzed the data obtained with the European Photon Imaging Camera (EPIC) on board the \XMM\ observatory in two observations, which were performed on 2007 June (ObsID 0504780501, 129.4~ks and ObsID 0504780601, 38.0~ks). In the observations the EPIC MOS detectors were set in Full Frame Mode, and the EPIC PN detector was set in Extended Full Frame Mode, both with the MEDIUM filter. We followed the standard procedure \citep{snowden08} to carry out data reduction and calibration by using SAS v14.0.0. We used the \XMM\ Extended Source Analysis Software (XMM-ESAS) tasks \textit{emchain} and \textit{epchain} to generate calibrated MOS and PN event files from raw data, respectively. In the screening process we set FLAG = 0 and kept events with PATTERNs $0-12$ for MOS detectors and events with PATTERNs $0-4$ for PN detector. By examining light curves extracted from source free regions in $1.0-5.0$~keV and $10.0-14.0$~keV, we rejected time intervals affected by soft and hard band flares, in which detector count rate exceeds the $2\sigma$ limit above the quiescent mean value \citep[e.g.,][]{gu12}. We also removed point sources by applying the tasks \textit{cheese} and \textit{cheese-bands}, and the results were cross-checked by comparing them with those obtained with the \Chandra\ ACIS images (\S2.1). The resulting total clean exposures from the two observations are $\sim$ 137.6~ks for MOS1, $\sim$ 138.8~ks for MOS2, and $\sim$ 125.5~ks for PN (Table~\ref{tbl-1}).

%==================================
\subsection{Background Modeling}
%==================================
%-----------------------------------
\subsubsection{\Chandra\ Background}
%-----------------------------------
For each \Chandra\ observation, a set of spectra were extracted from the boundary regions located on the S3 CCD, where the thermal emission from the group is relatively weak, and were used to construct the background model that consists of both the Cosmic X-ray Background (CXB) and the instrumental background. We fitted the group emission remaining in the extracted spectra with an absorbed thermal APEC model by fixing $N_{\rm H}$ at the Galactic value of $3.00 \times 10^{20}$~$\rm cm^{-2}$ \citep{DL90} and modeled the parameters ($kT$ and $Z$) at the corresponding average values for the outermost regions, which are given in \citet{gu07} and \citet{rafferty13}.
We modeled the CXB component by using two unabsorbed APEC components ($kT = 0.08$~keV and $0.2$~keV, respectively, and $Z = 0$; \citealp{lumb02,gu12}) to approximate the Galactic soft emission, and using an absorbed power-law component with index $\Gamma=1.4$ \citep{mushotzky00,CR07} to describe the unresolved CXB. The \ROSAT\ All-Sky Survey (RASS) spectra extracted from the same boundary regions were also jointly fitted to help constrain the X-ray background models.
As for the instrumental background, we modeled it by utilizing the spectra extracted from the \Chandra\ stowed background datasets using the same CCD regions, the $9.5-12.0$~keV count rate of which is renormalized with respect to that of ACIS S3 boundary spectra for each observation.
In addition, we adopted a 3\% uncertainty on the normalization of the instrumental background to represent the systematic uncertainty in the modeling and propagate it into our final results \citep[see, e.g.,][]{vikhlinin05,sun09}. By using the best-fit background model obtained in fitting the ACIS S3 boundary spectra, we were able to construct the \Chandra\ background templates for each observation.

%-------------------------------
\subsubsection{\XMM\ Background}
%-------------------------------
We followed the official ESAS-Cookbook\footnote{~https://heasarc.gsfc.nasa.gov/docs/xmm/esas/cookbook/xmm-esas.html} and the approach of \citet{snowden08} to model the \XMM\ EPIC background, which is basically similar to that adopted for \Chandra\ ACIS data, except that the data of two MOS and one PN detectors were jointly fitted and the instrumental background, including the strong fluorescent lines, must be carefully modeled.

For each observation, a set of MOS and PN spectra were extracted from a region located at about $11\arcmin - 14\arcmin$ ($\sim 185-235$~kpc) away from the group center in $0.3-11.0$~keV and $0.4-11.0$~keV, respectively, and were used to construct the background model that consists of both the CXB and the instrumental background. These spectra, together with the $0.2-2.0$~keV RASS spectrum, were fitted jointly with the same model described in \S2.3.1 (group emission + CXB).
The quiescent particle background (QPB) makes the main contribution to the instrumental background and was modeled with the spectra obtained in the \XMM\ filter wheel closed (FWC) observations, which were extracted from the same regions as the corresponding background spectrum by using the \textit{mos\_back} and \textit{pn\_back} tools for the MOS and PN detectors, respectively. Since the QPB spectral regions affected by the strong instrumental lines were cut out and replaced by the interpolated power-law components (for the reasons of this approach, see \citealt{snowden08}), we added several Gaussian lines into the QPB model; for the MOS spectra two Gaussians at energies of $\sim 1.49$~keV (Al~K$\alpha$) and $\sim 1.75$~keV (Si~K$\alpha$) were added, while for the PN spectra six Gaussians at energies of $\sim 1.49$~keV, $\sim 7.11$~keV, $\sim 7.49$~keV, $\sim 8.05$~keV, $\sim 8.62$~keV, and $\sim 8.90$~keV were taken into account. The normalizations of these lines were left free in the fittings. We also evaluated the systematic errors of the \XMM\ background resulted from the instrumental lines by making use of Monte Carlo simulations to randomly vary the normalizations of these lines according to the uncertainties allowed in the fittings (see Appendix A).
In addition, an unfolded broken power-law component was employed to account for the possible residual soft-proton contamination.
The best-fit background model achieved in the fittings was used to create the corresponding background templates for each \XMM\ observation.

%############################################
\section{X-RAY IMAGING AND SPECTRAL ANALYSIS}
%############################################

%=======================================================
\subsection{\Chandra\ X-ray Surface Brightness Profiles} 
\begin{figure}
\centering
\graphicspath{{figures/}}
\includegraphics[scale=.25]{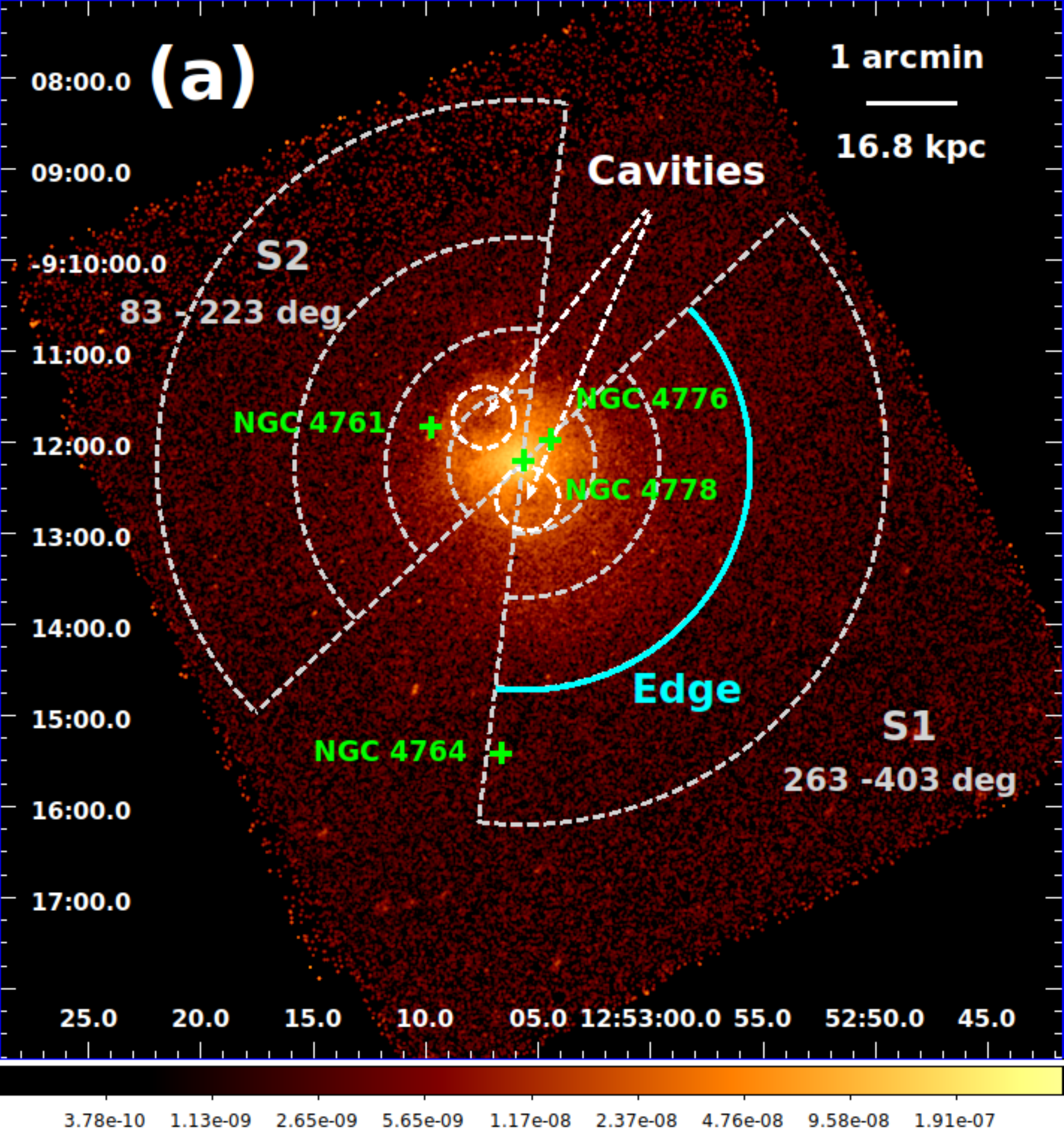}
\includegraphics[scale=.5]{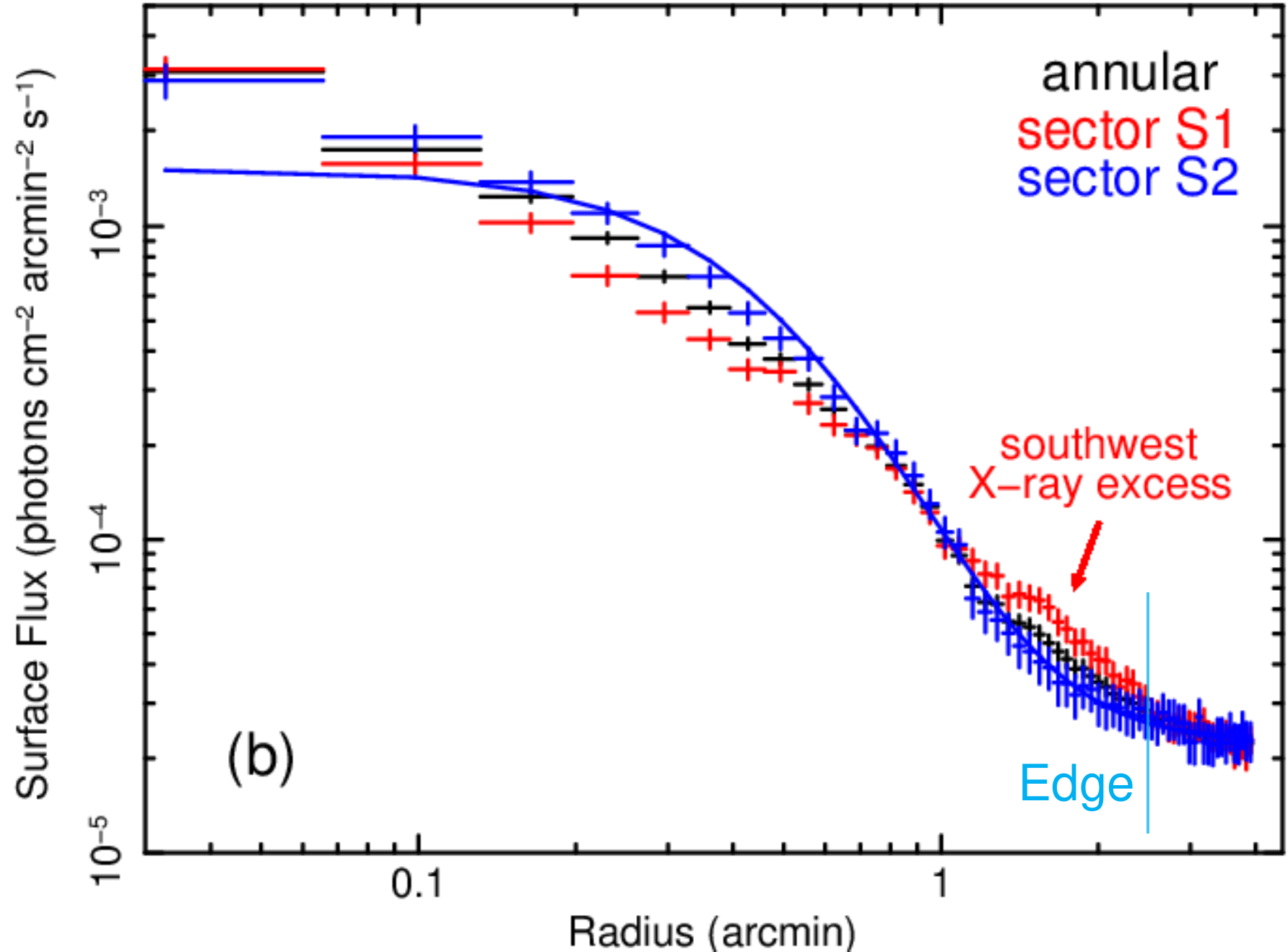}
\caption{($\it a$) Combined \Chandra\ ACIS S3 image of HCG 62 in $0.5-7.0$ keV, which is plotted in a log scale. The image has been exposure-corrected and smoothed with a Gaussian of 3\arcsec. Positions of surface brightness jump (edge), cavities, four bright member galaxies and pie-region defined in two sectors (S1 and S2) are marked. 
($\it b$) X-ray surface brightness profiles extracted from two pie-region sets and a set of concentric annuli. Best-fit model for the profile extracted in sector S2 is also shown as a solid line. The corresponding surface brightness jump (edge) is marked by a cyan line at 2.5'.  \label{fig1}}
\end{figure}

%=======================================================
In Figure~\ref{fig1}a we show the combined \Chandra\ ACIS S3 image obtained in $0.5-7.0$~keV from the three observations, which has been corrected for exposure and smoothed with a Gaussian of 3\arcsec. Locations of the two X-ray cavities and the southwest surface brightness jump (approximately at the outer edge of the region showing high Fe abundance; \citealt{gu07}; \citealt{gitti10}; \citealt{rafferty13}), as well as the locations of four bright member galaxies (i.e., NGC~4761, NGC~4764, NGC~4776, and NGC~4778), are marked in the figure. In order to examine the X-ray morphology in a quantitative way, we have defined pie-region sets in two circular sectors that have the same opening angles, one (sector S1) extending toward the southwest to straddle the entire high-abundance region, where it also exhibits an excess X-ray emission, and the other (sector S2) extending toward the opposite direction, to extract radial X-ray surface brightness profiles ($S(R)$, where $R$ is the two-dimensional radius) in $0.5-7.0$ keV (Fig.~\ref{fig1}). The surface brightness profile calculated from a set of concentric annuli, which are centered at the X-ray peak, is also extracted and plotted for comparison. In sector S1 an emission excess can be clearly detected at about $1\arcmin - 2.5\arcmin$, and this has been attributed to the high metallicity therein in previous works. 

We find that, except for the innermost 0.2\arcmin\ region where it shows a significant central excess primarily due to a temperature decline (i.e., the second $\beta$ component in \citealt{morita06} and \citealt{rafferty13}), and the $0.3\arcmin - 0.7\arcmin$ region where an X-ray deficit due to the northeastern cavity, the surface brightness profile extracted from sector S2 can be well fitted by an analytic solution derived from the empirical $\beta$ profile for gas density \citep{CF76}, 
\begin{equation}
 S(R)=S_{\rm 0}\left[ 1+\left(\frac{R}{R_{\rm c}}\right)^2\right]^{0.5-3\beta}+S_{\rm bkg},
\end{equation}
where $R_{\rm c}=0.57\arcmin \pm 0.20 \arcmin$ is the core radius, $\beta=0.85 \pm 0.11$ is the slope, $S_{\rm bkg} = (2.30 \pm 0.21) \times 10^{-5}$~$\rm photons~cm^{-2}~arcmin^{-2}~s^{-1}$ is the background, and $S_{\rm 0} = (1.44\pm 0.31)\times10^{-3}$~$\rm photons~cm^{-2}~arcmin^{-2}~s^{-1}$ is the normalization (Fig.~\ref{fig1}b).
In \S4, this result is adopted to constrain the simulated X-ray surface flux profile, in order to avoid the biases caused by the central excess, the two cavities and the high-abundance region.

%=====================================================================
\subsection{Deprojected Spectral Results with \Chandra\ and \XMM\ } 
\begin{figure}
\centering
\graphicspath{{figures/}}
\includegraphics[scale=.3]{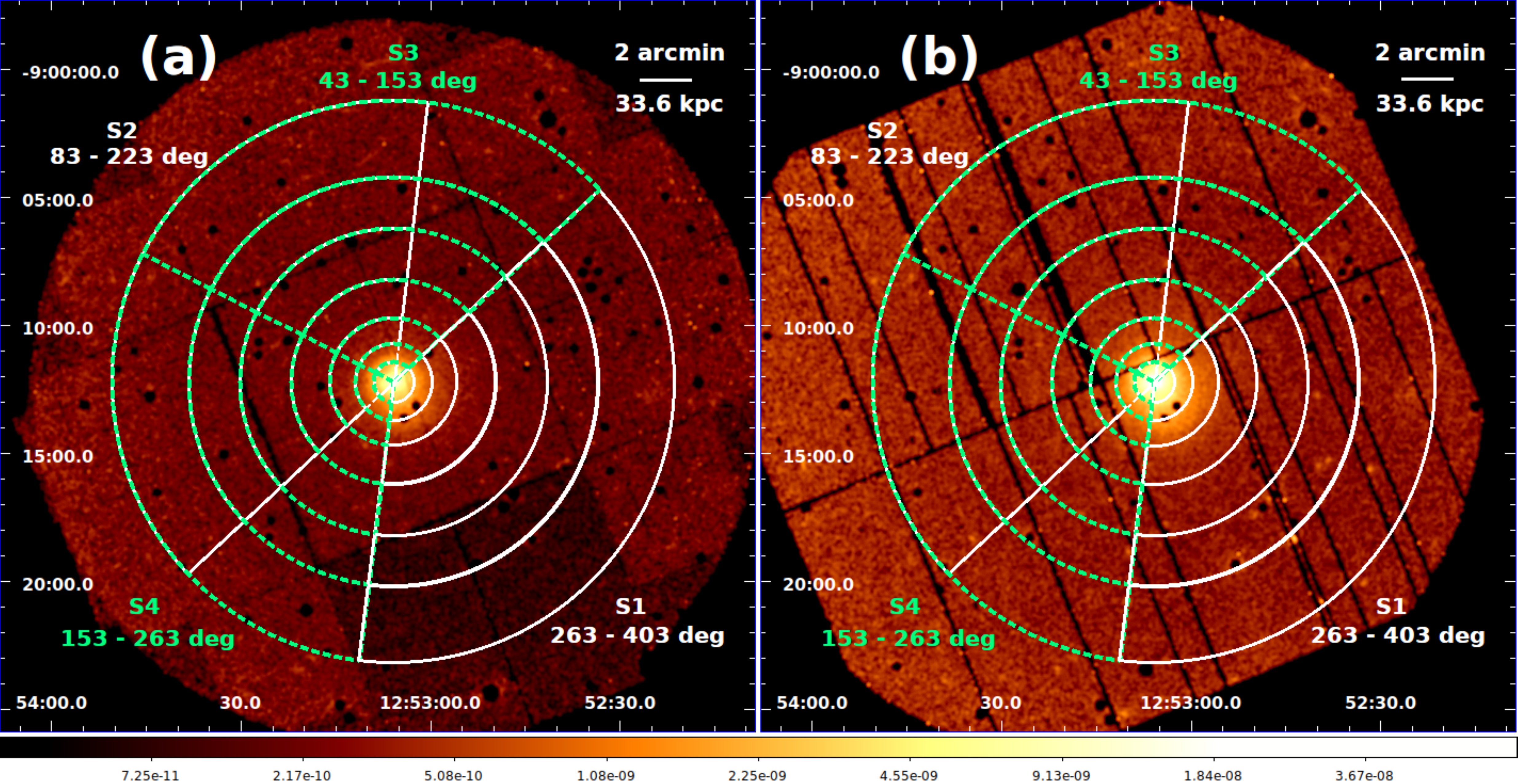}
\caption{ Combined \XMM\ EPIC-MOS (a) and EPIC-PN (b) images of HCG 62 in $0.5-7.0$ keV, which is also plotted in a log scale. Besides the pie-regions in sectors S1 and S2, those defined in two new sectors (S3 and S4) are also marked. \label{fig2}} 

\end{figure}

%=====================================================================
In order to study the spatial distributions of gas temperature and metal abundances in different directions, we extract \Chandra\ ACIS and \XMM\ EPIC MOS/PN spectra in $0.5-7.0$ keV from four pie-region sets, which are defined in sector S1, sector S2, and two new sectors pointing approximately toward north and southeast (i.e., S3 and S4 in Fig.~\ref{fig2}). For each pie-region set, the inner and outer radii of the pie regions are $0\arcmin-0.8\arcmin$, $0.8\arcmin-1.5\arcmin$, $1.5\arcmin-2.5\arcmin$, $2.5\arcmin-4\arcmin$, $4\arcmin-6\arcmin$, $6\arcmin-8\arcmin$, and $8\arcmin-11\arcmin$. Since the ACIS detector provides a narrower field of view than the MOS and PN detectors, the \Chandra\ spectra are extracted only from the inner 4\arcmin. In the \Chandra\ spectral analysis the appropriate Ancillary Response Files (ARFs) and Redistribution Matrix Files (RMFs) are created by using the CIAO tool \textit{specextract}; in the \XMM\ spectral analysis the ARFs and RMFs are created by using the SAS tasks \textit{mos-spectra} and \textit{pn-spectra} for the EPIC MOS and PN data, respectively. The extracted spectra are grouped with a minimum of 30 counts per bin, and for each pie region the spectra obtained from the three \Chandra\ observations and two \XMM\ observations are fitted jointly using the X-ray spectral fitting package XSPEC version 12.8.2 \citep{arnaud96} after the corresponding background templates are applied. In the fittings we adopt the collision ionization equilibrium plasma model VAPEC by fixing the abundances of He, C, and N at their solar values, and dividing other elements into five groups (i.e., O=Ne, Mg=Al, Si, S=Ar=Ca, and Fe=Ni), whose abundances are allowed to vary independently \citep{morita06,sasaki14}. Throughout the spectral analysis we fix the $N_{\rm H}$ at the Galactic value of $3.00 \times 10^{20}$~$\rm cm^{-2}$ \citep{DL90} unless otherwise stated.

To correct the projection effect in the spectral fittings we employ the code \textit{dsdeproj}\footnote{~https://www-xray.ast.cam.ac.uk/papers/dsdeproj/}, which is designed based on the Direct Spectral Deprojection approach of \citeauthor{SF07} (2007; see also \citealt{russell08}), to evaluate the influence of the spectra of the outer spherical shells on those of the inner ones. 

In the \XMM\ MOS and PN observations, a significant fraction of emission originating from one sky area may be scattered to the surrounding areas due to the point spread functions (PSFs) of the mirror assemblies, which have broad wings and vary as a function of both energy and off-axis angle. The treatment of this effect, i.e., the cross-talk, is crucial in the data analysis of extended sources like galaxy clusters and groups, especially those exhibiting substructures \citep{snowden08, SK14}. In this work we employ the SAS task \textit{arfgen} to calculate how many photons originating in one pie-region are eventually detected in another pie-region (see Table~\ref{a1} in Appendix B), and use the results to modify the corresponding ARFs.

%-----------------------------------------
\subsubsection{Single-Phase Plasma Model}
\begin{figure}
\centering
\epsscale{1}
\graphicspath{{figures/}}
\plottwo{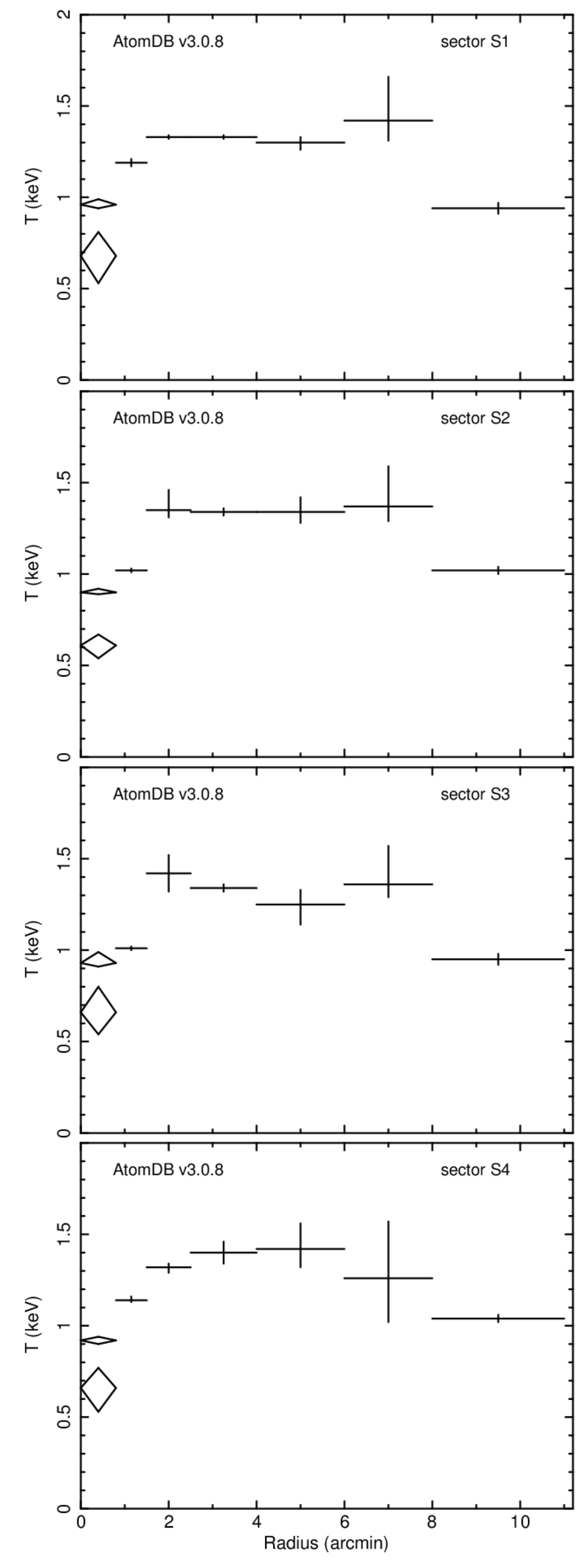}{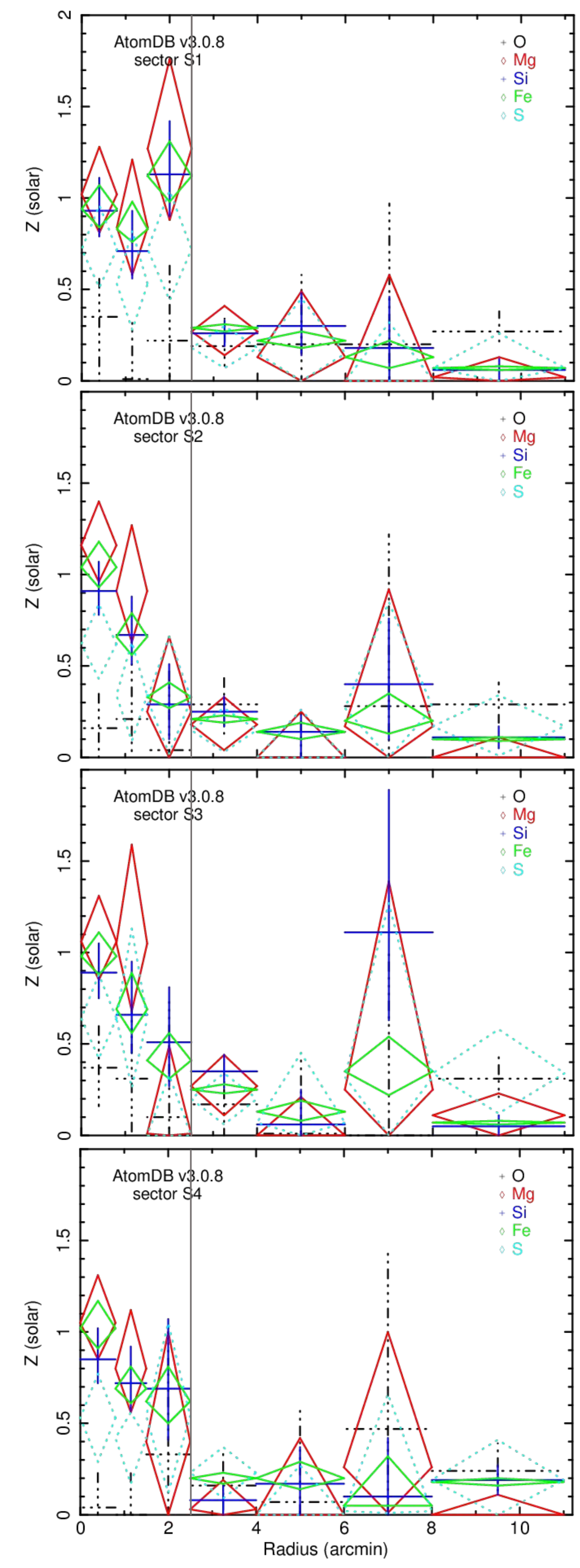}
\caption{ Deprojected radial temperature ($\it Left$) and metal abundance ($\it Right$) profiles  derived in the four sectors. Two-temperature model is used for the $< 0.8'$ region and single-temperature model is used for the $> 0.8'$ regions (\S 3.2.2). \Chandra\ data are used only in the spectral fittings in $<4\arcmin$ regions. Dark lines corresponding to the surface brightness jump (edge) are marked in Figure 1. \label{fig3}}
\end{figure}

%-----------------------------------------
First we fit the deprojected spectra using the absorbed thermal VAPEC model, and show the best-fit results in Table~\ref{tbl-2} and Figures~\ref{fig3}. We find that the emission measure-weighted gas temperature shows very similar radial variations in four directions; it rises from about 0.8~keV at the group center to about 1.4~keV in $1.5\arcmin-8\arcmin$, while in the outer regions it drops to about 1.0~keV or slightly below. The central gas temperature drop can be attributed to either an inward monotonous temperature decrease for a single-phase gas, or the appearance of a cooler component (\S3.2.2). 

The single-phase spectral analysis based on five high-quality \Chandra\ and \XMM\ observations shows that at the 90\% confidence level the deprojected abundances of nearly all the metals (O, Si, S, Mg, and Fe) exhibit a flat distribution in $>2.5\arcmin$ (i.e., $>42$~kpc or $>0.07r_{200}$\footnote{~$r_{200}$ is defined as the radius within which the mean enclosed mass density of the target is 200 times the critical density of the universe at the target's redshift.}), except that in sector S1 the Fe abundance shows a mild outward decrease. The average Fe abundance in $>2.5\arcmin$ is found to be about 0.2 solar, which agrees well with the previous results \citep{morita06,tokoi08,gitti10,sasaki14}. Note that the value is slightly lower than those typically found in $>0.3r_{200}$ in galaxy clusters, where a uniform metallicity is seen \citep[e.g.,][]{werner13,urban17}, but consistent with the abundances measured at the outskirts of the Virgo cluster \citep{simionescu15,simionescu17} and many galaxy groups \citep{RP07}.

The high-quality X-ray data reveals that in the three inner regions of sector S1 the abundances of Si, S, Mg, and Fe are significantly higher than those in $>2.5\arcmin$, all showing an abrupt jump at about $2.5\arcmin$. In the inner regions the Si, Mg, and Fe abundances reach about 1 solar, while the S abundance is relatively lower. Note that the abundance jumps approximately coexist with the X-ray surface brightness jump shown in Figure~\ref{fig1}b, and such metal abundance distributions actually form a high-abundance plateau spanning a region (i.e., $0\arcmin-2.5\arcmin$ in sector S1) broader than the `high-abundance arc' (a region approximately covering $1.7\arcmin-2.2\arcmin$ from south to northwest) identified in previous works \citep[e.g.,][]{gu07,gitti10,rafferty13}, in which only part of the data analyzed in this work were used. In order to further confirm the existence of the high-abundance plateau we have applied different region partitions by reducing the sizes of the pie-regions by half either azimuthally and/or radially, and obtained consistent results.

The high Mg abundances observed in the central regions, raise a possibility that these regions might have been polluted by additional SN II yields after the epoch of early enrichment. This can be ascribed to recent star formation triggered by mergers (see \S1 and references therein), the possibility of which will be investigated in \S4 and \S5 via numerical simulations. Besides, it is also interesting to note that the O abundance in the high-abundance plateau is rather low and is consistent with those of the outer regions. This results in a high Mg/O ratio ($\sim 2$) in the central regions in agreement with the results of \citet{morita06} and \citet{tokoi08}. We will probe into the cause of this phenomenon in \S5.4.

In other directions (sectors S2, S3, and S4) unambiguously higher abundances are detected only in the innermost ($<0.8\arcmin$) region for Si (about $0.7-0.8$ solar), Mg (about 1 solar), and Fe (about 1 solar). In $0.8\arcmin-2.5\arcmin$ all the Si, Mg, and Fe abundances have a tendency to decrease outward. Unlike the high-abundance plateau in sector S1, this type of cuspy central abundance increase is not rare in galaxy clusters and groups.

%-------------------------------------
\subsubsection{Multi-Phase Plasma Model}
%------------------------------------
If there exists a multi-phase gas or a cooling flow in the core region, the measurement of the central abundances may be biased by the use of a single-phase plasma model \citep{buote001,buote002}.
In order to examine whether or not this is true, especially in the inner regions ($<2.5\arcmin$) where both a significant inward gas temperature drop and a high-abundance plateau are found, we attempt to add an additional cool component (APEC) in the deprojected spectral fittings. The temperature and normalization of the cool component are left free, and the abundance is assumed to be the same as the coexisting hot component. We find that in sector S1 the cool component cannot be well determined in $< 0.8\arcmin$ when fitting the \Chandra\ ACIS spectra, whereas the inclusion of the cool component can slightly improve the fitting of the \XMM\ EPIC spectra. The best-fit temperatures derived in the joint \Chandra$-$\XMM\ fittings ($T_{\rm cool}=0.68_{-0.15}^{+0.13}$~keV and $T_{\rm hot}=0.96_{-0.02}^{+0.03}$~keV) are very close to those obtained by \citet{morita06} in their deprojected fittings of the \Chandra\ spectra for the central $0.6'$, and the cool component accounts for up to 8\% of the $0.5-7.0$~keV luminosity with a volume filling factor of only $0.04\pm0.01$, which is calculated by following \citet{gu12}. In the $> 0.8\arcmin$ regions, however, we find that the two-temperature model is not well constrained since either the temperature or the normalization of the cool component, sometimes both, cannot be determined in the fittings. When we set the temperatures of the cool component in $0.8'-1.5'$ and $1.5'-2.5'$ to be 0.68 keV, which is the same as what is found in $< 0.8'$ by jointly fitting the \Chandra\ and \XMM\ data, we find that the cool components account for only $\lesssim$ 2\% of the $0.5-7.0$~keV luminosity, meanwhile the variations of the best-fit metal abundances are negligible when comparing with the results obtained in the single-phase fitting. Note that, when the single-phase plasma model is applied to fit the spectra of these regions the goodness-of-fit is already acceptable (Table~\ref{tbl-2}). These results indicate that in the outer regions the possible cool component is too weak and the spectra are actually dominated by the hot gas.

In order to quantitatively estimate the scale of the weak cool gas component, we alternatively add an isobaric cooling flow component (MKCFLOW) to the spectral model. The low temperature of the cooling flow is fixed to 0.08~keV, and the high temperature is tied to that of the VAPEC component \citep[e.g.,][]{rafferty13}. The results for sector S1 (Table~\ref{tbl-3}) show that, except for the central 0.8\arcmin\ region where the mass deposition rate is $<0.034~\rm M_{\sun}~yr^{-1}$ (sector S1), in all other regions the mass deposition rate is very low ($< 0.005~\rm M_{\sun}~yr^{-1}$). The derived low mass deposition rates are consistent with those obtained by \citet{rafferty13}, and are very similar to what have been observed in many giant elliptical galaxies \citep{xu02,bregman05}.

The above results show that the cool components can be ignored except for the innermost region ($ < 0.8\arcmin$). Similar results are found for other three sectors (Fig.~\ref{fig3}). Therefore in the data analyses, calculations, and simulations that follow we will adopt the two-temperature best-fits for the $ < 0.8\arcmin$ regions, and the single-temperature best-fits for outer regions.

%--------------------------------------------------------------
\subsubsection{Consistency between \Chandra\ and \XMM\ Results}
\begin{figure}
\centering
\epsscale{1.1}
\graphicspath{{figures/}}
\plottwo{f4a}{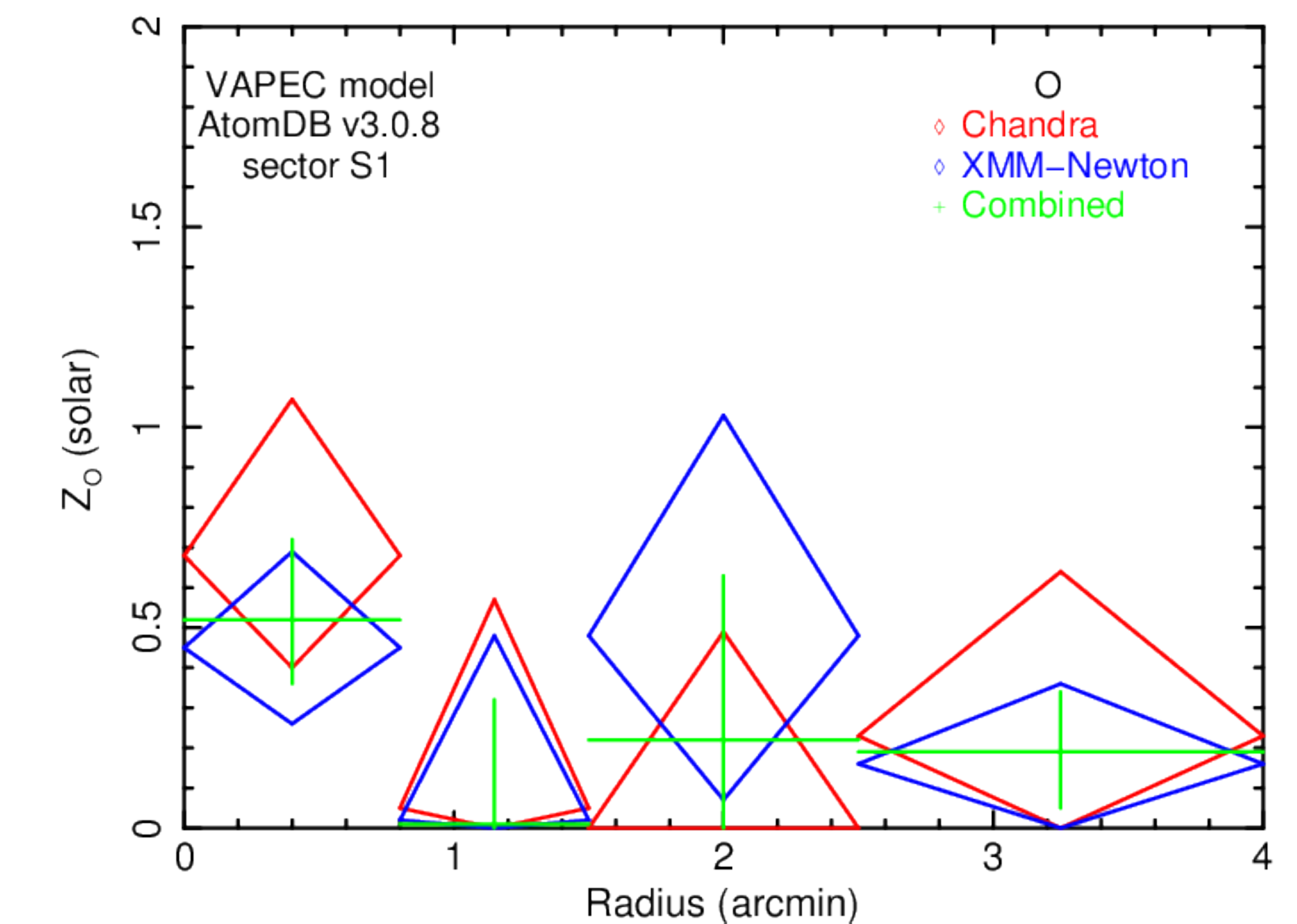}
\plottwo{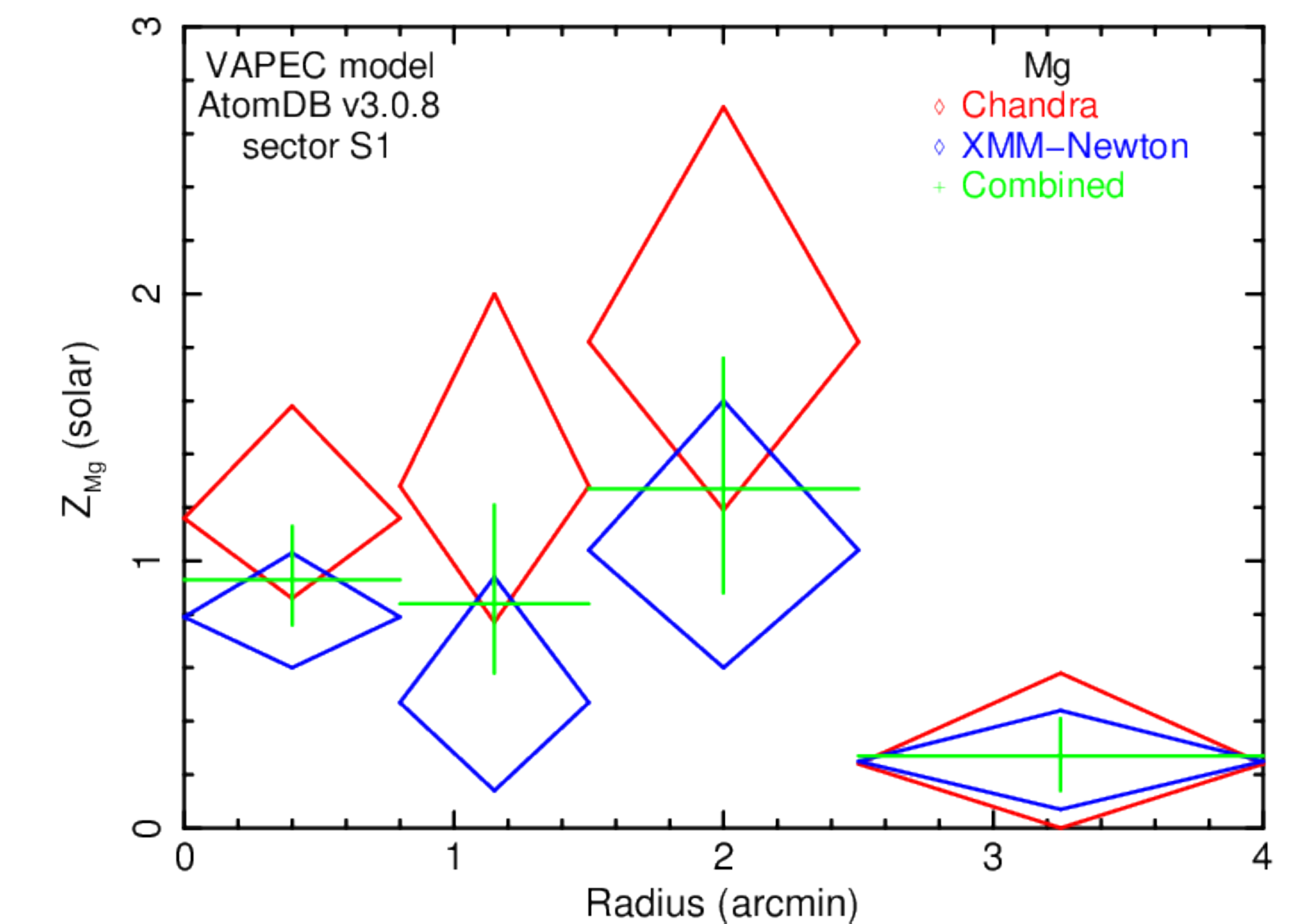}{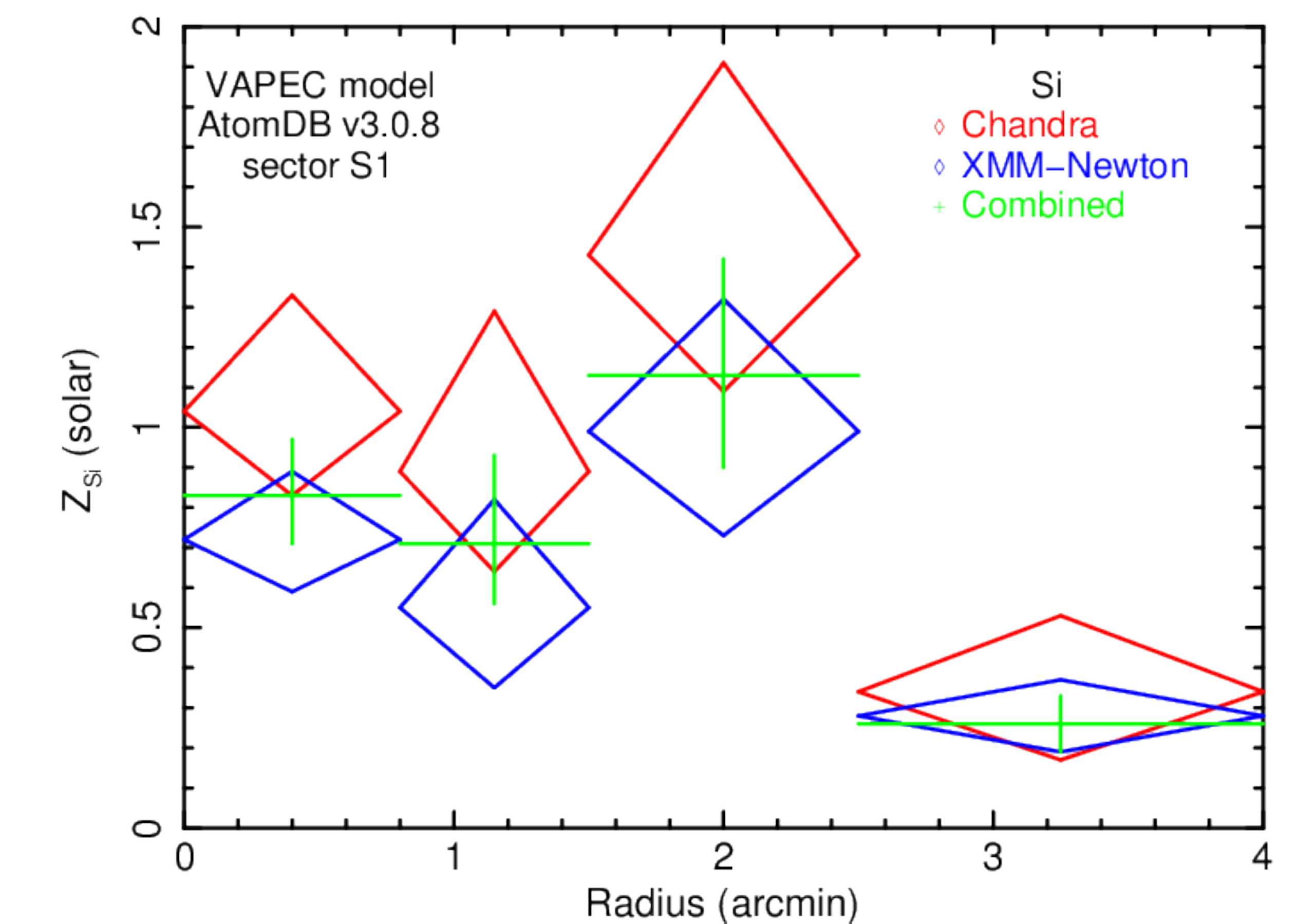}
\plottwo{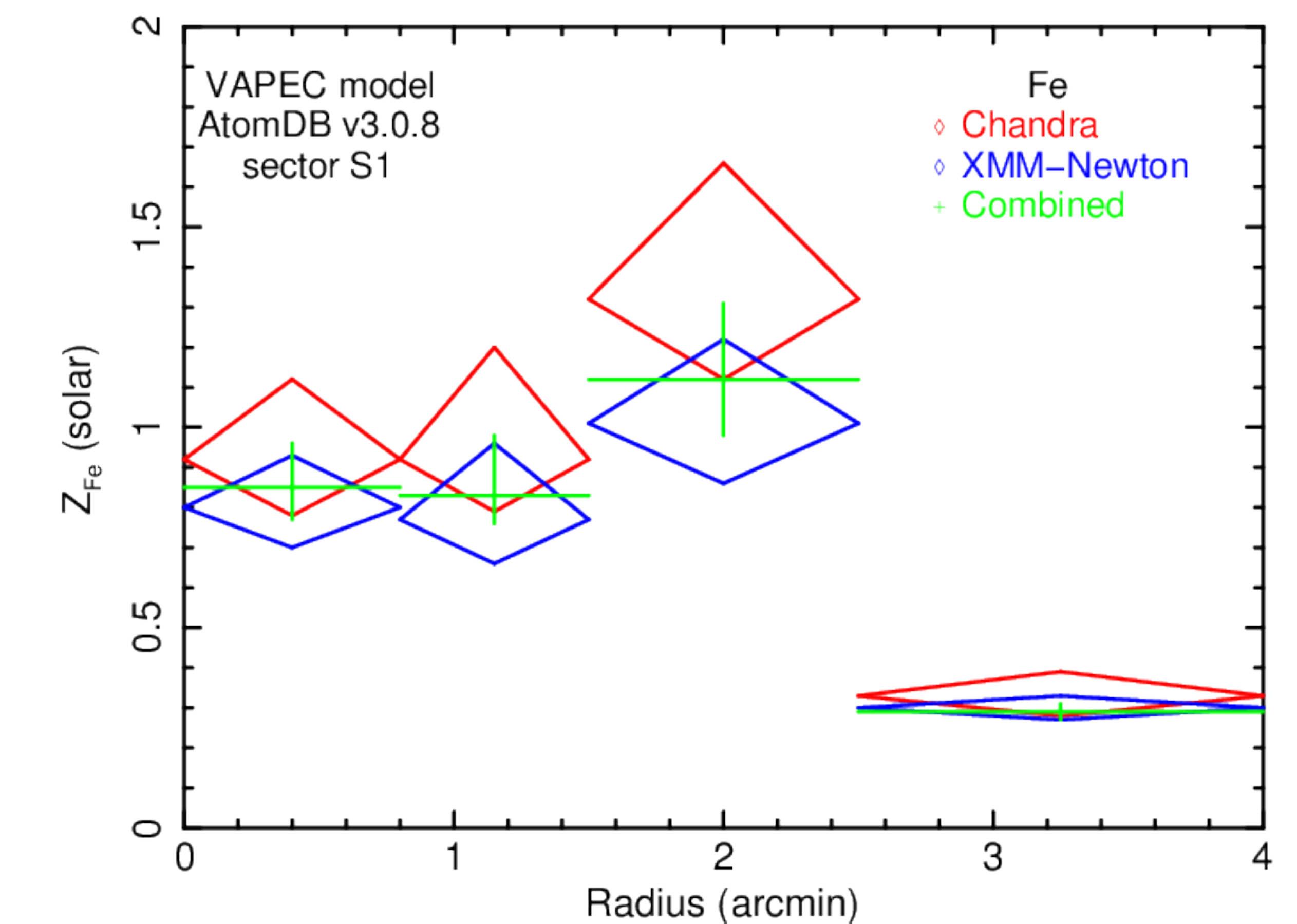}{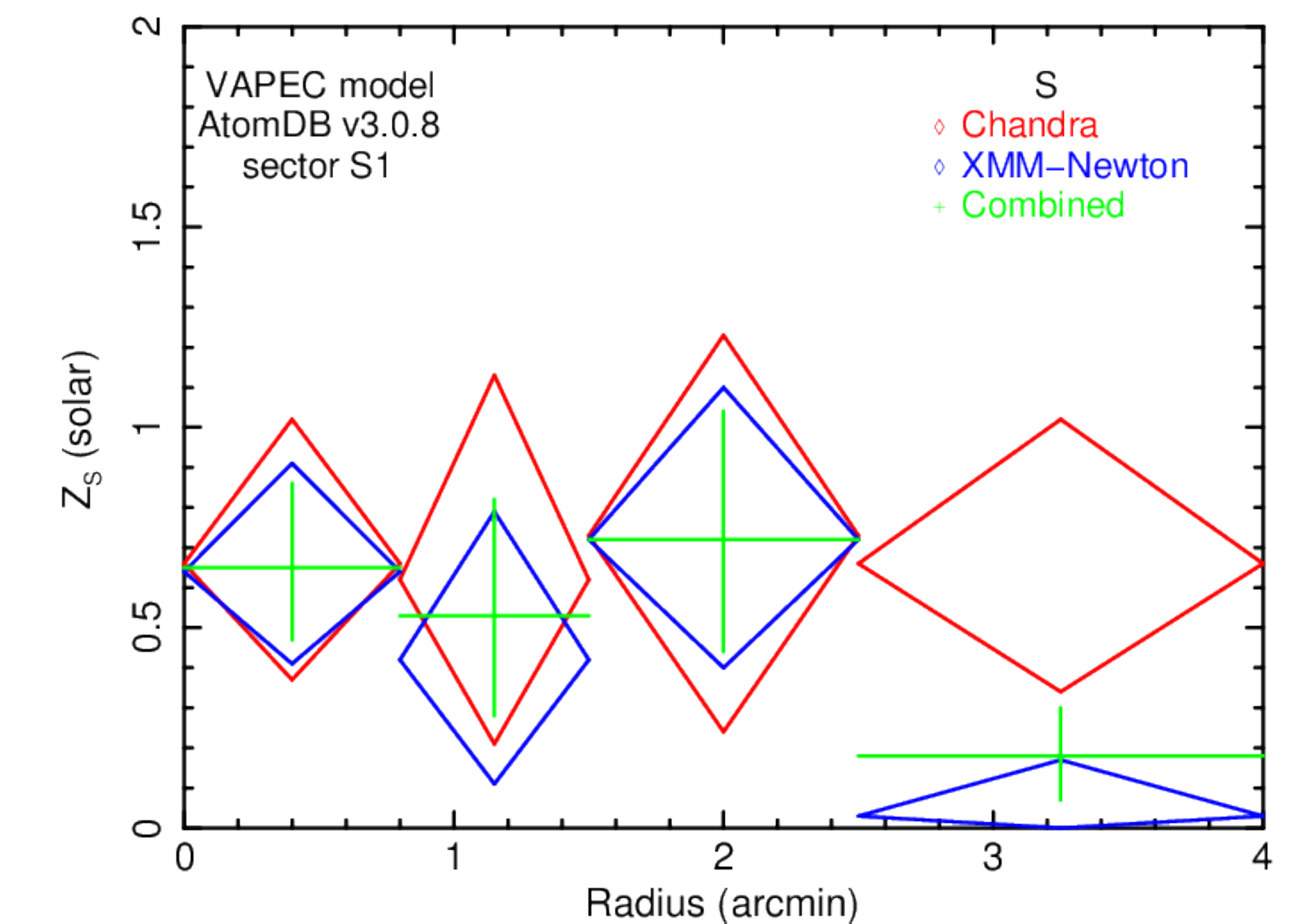}
\caption{ Deprojected temperature and metal abundance profiles measured in sector S1 with different instruments (\Chandra\ ACIS, \XMM\ EPIC, and Combined) by applying single-phase model (\S3.2.2). \label{fig4}}
\end{figure}

\begin{figure}
\centering
\epsscale{1.1}
\graphicspath{{figures/}}
\plottwo{f5a}{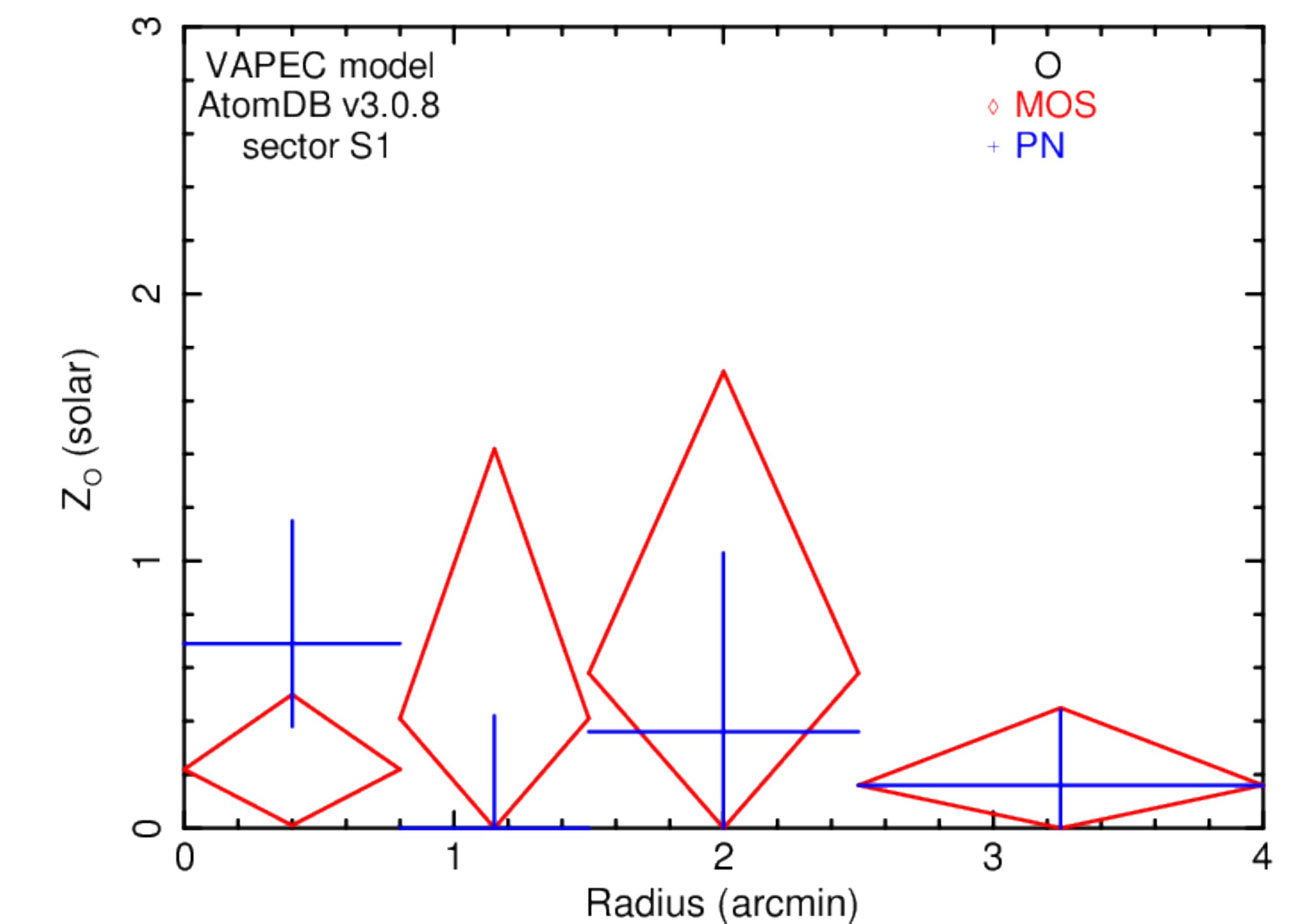}
\plottwo{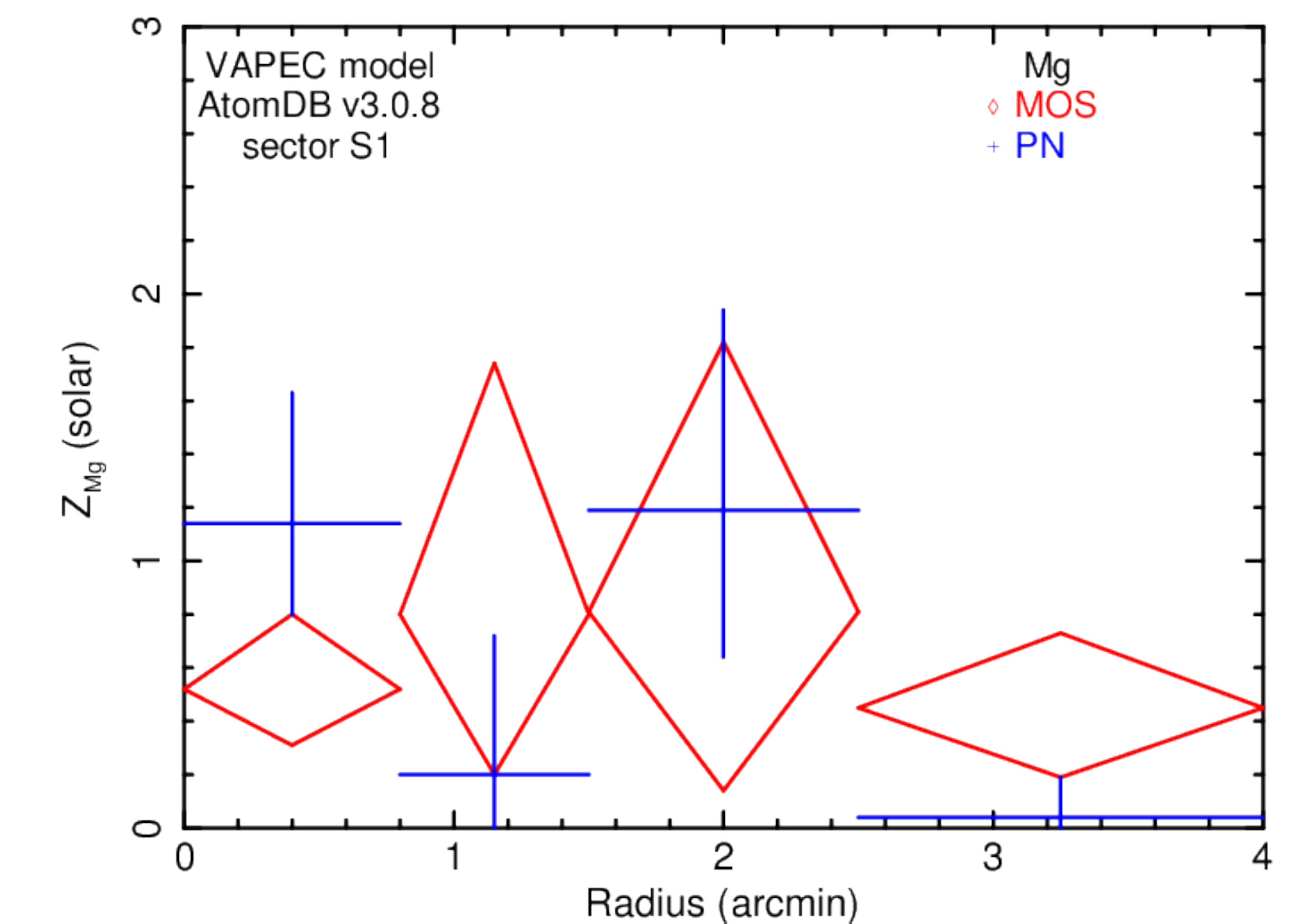}{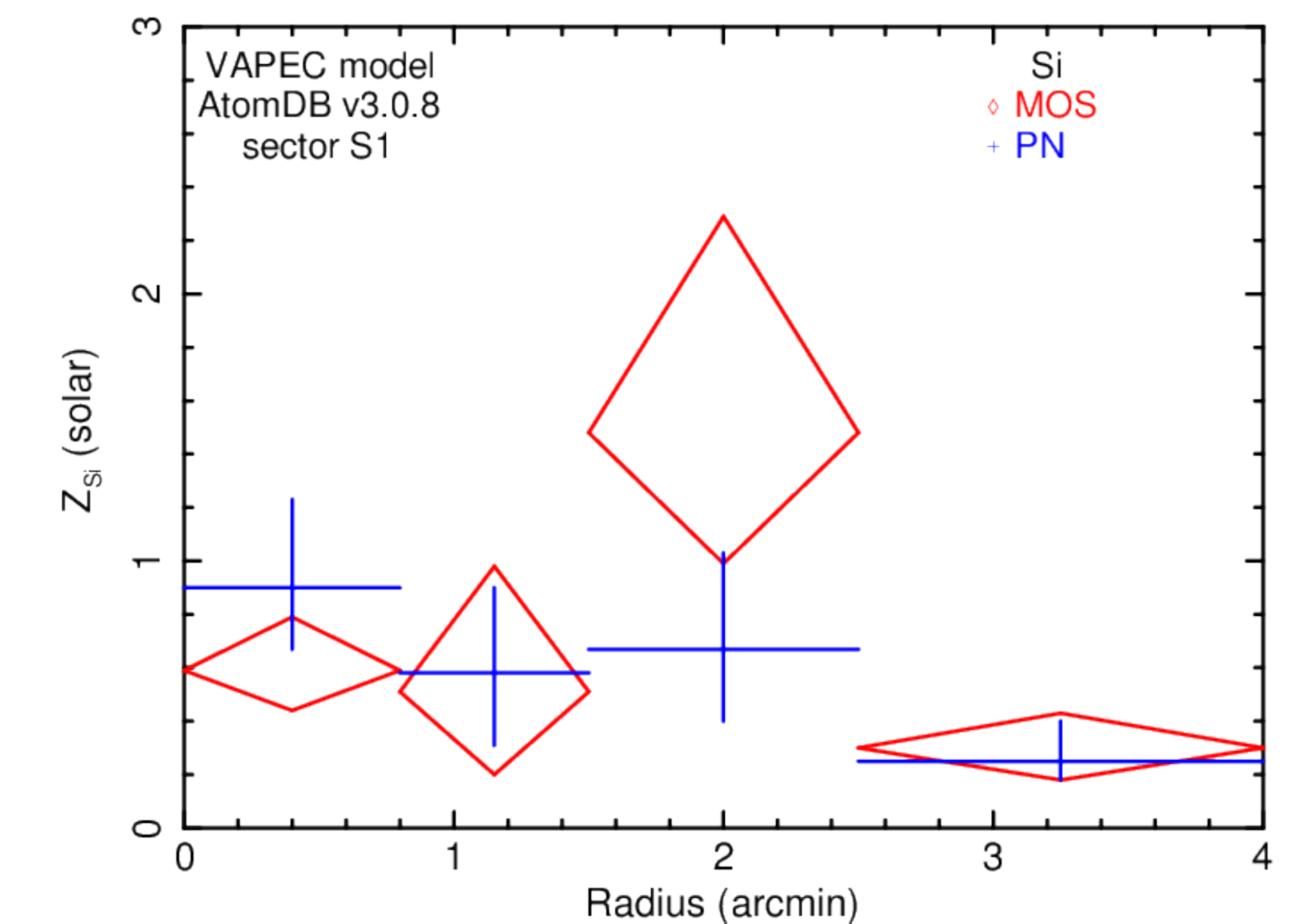}
\plottwo{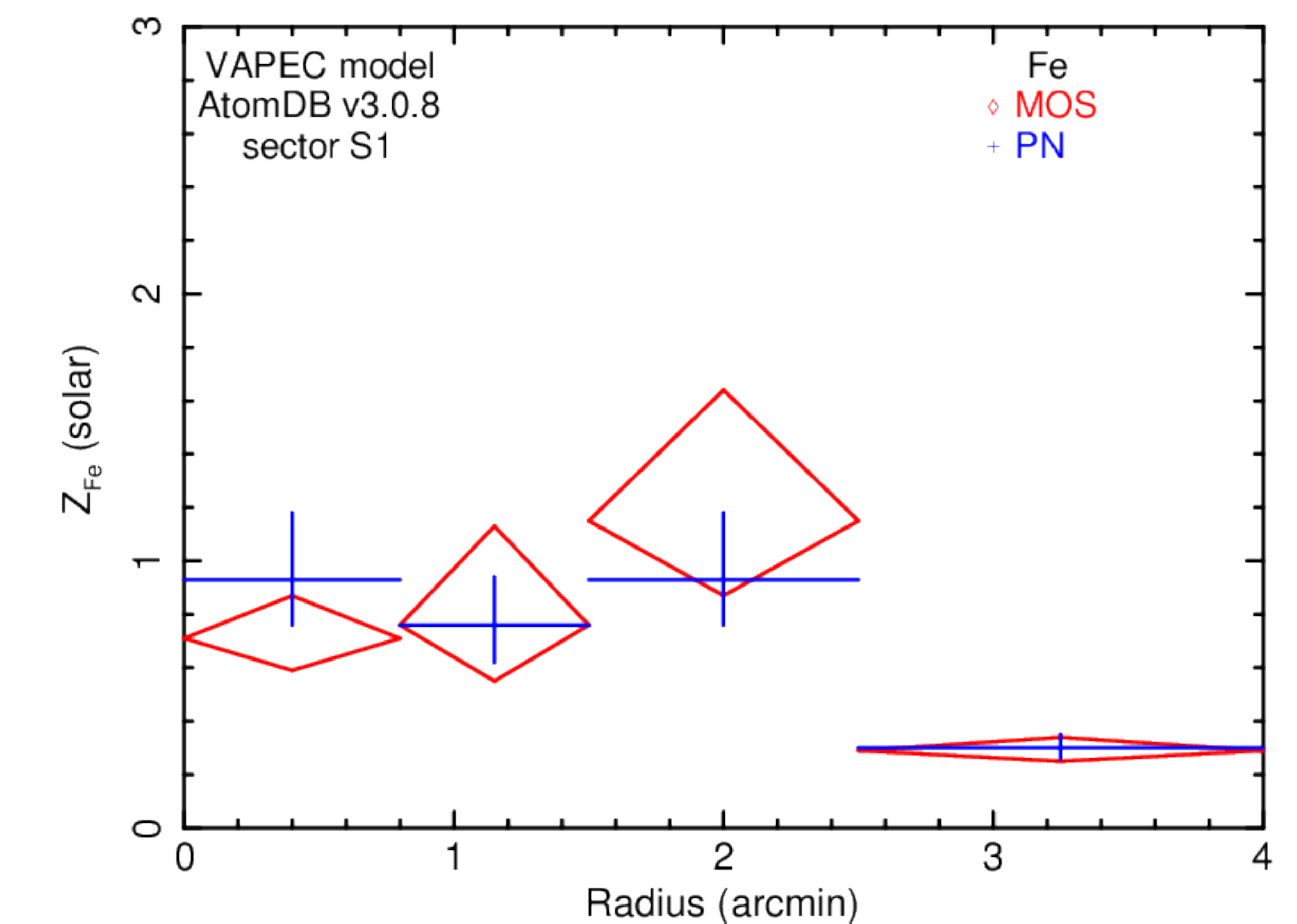}{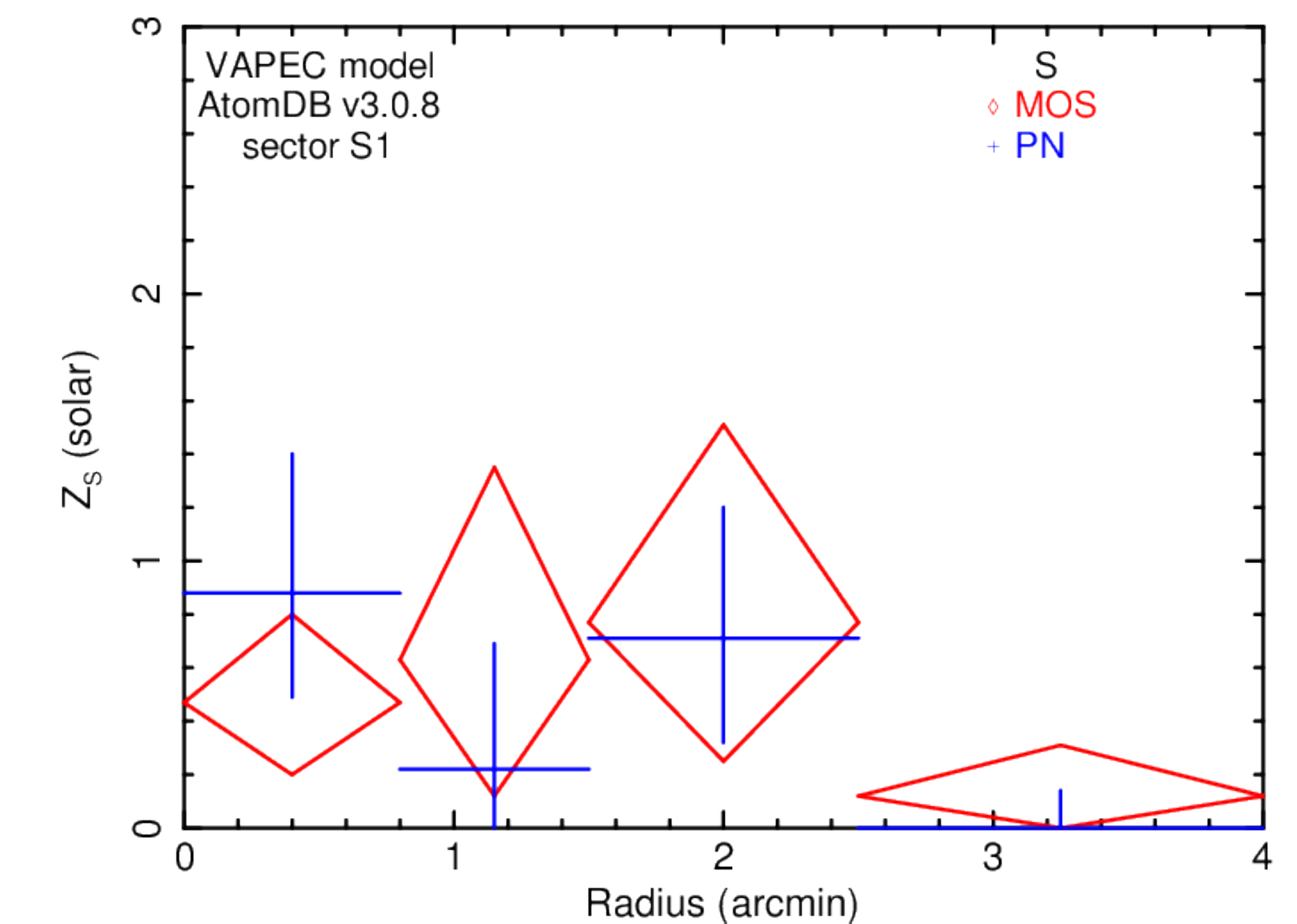}
\caption{ Deprojected temperature and metal abundance profiles measured in sector S1 with different instruments on board \XMM\ by applying single-phase model (\S3.2.2). \label{fig5}}
\end{figure}

%--------------------------------------------------------------
The energy-dependent difference between the effective areas of the \Chandra\ ACIS and \XMM\ MOS or PN instruments, i.e., the energy dependence of the stacked residuals ratios \citep[e.g.,][]{schellenberger15}, exists even after careful calibrations. It may cause differences between the temperatures, and then the metal abundances due to the temperature-abundance dependency, measured with \Chandra\ and \XMM. However, this effect turns to be minor when gas temperature is not high, and eventually becomes negligible in systems in which gas temperature is $<2.0$~keV \citep{schellenberger15}. This is confirmed in our case, as shown in Table~\ref{tbl-2}, Figures~\ref{fig4} and \ref{fig5} where we compare the gas temperature and abundance distributions obtained with the ACIS, MOS, and PN instruments in sector S1. We find that ACIS and EPIC results, or MOS and PN results, agree with each other at the 90\% confidence level.

%----------------------------------------------------
\subsubsection{AtomDB v3.0.8 vs v1.3.1}
\begin{figure}
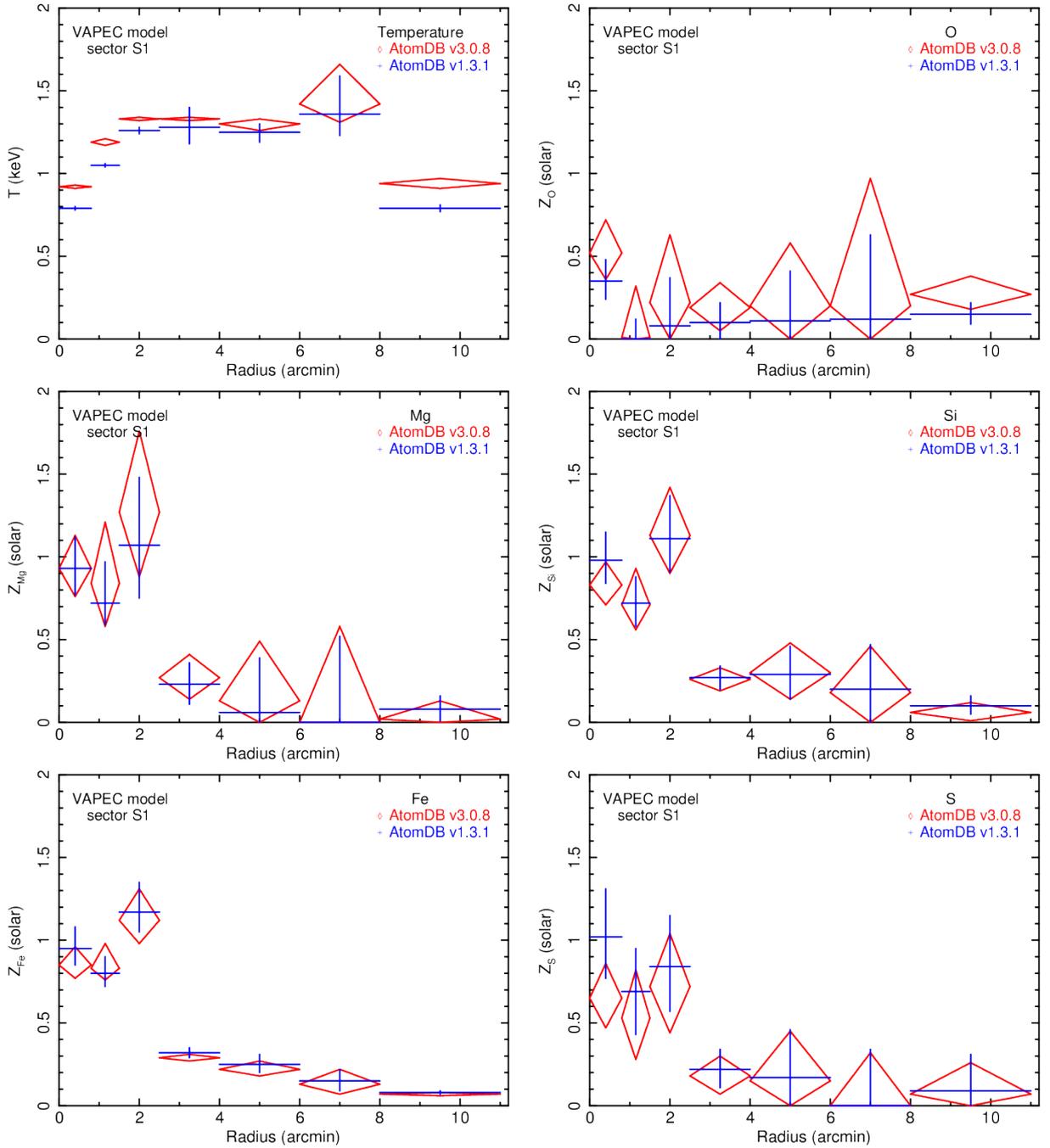

\centering
\epsscale{1.1}
\graphicspath{{figures/}}
\plottwo{f6a}{f6b}
\plottwo{f6c}{f6d}
\plottwo{f6e}{f6f}
\caption{ Deprojected temperature and metal abundance profiles measured in sector S1 with different AtomDB versions (v3.0.8 and v1.3.1) by applying single-phase model. \Chandra\ data are used only in the spectral fittings in $<4\arcmin$ regions.  \label{fig6}}
\end{figure}

%----------------------------------------------------
In this work we apply the latest astrophysical atomic database for collisional plasma AtomDB v3.0.8. Compared with the old versions AtomDB used in previous works on HCG~62 \citep{morita06,gu07,tokoi08,gitti10,rafferty13}, the new AtomDB may leads to a temperature higher by about $10\%-20\%$ for a plasma at about 1.0~keV \citep[e.g.,][]{foster12,sasaki14}, because significant updates of the Fe L-shell complex have been included in AtomDB since version 2.0.1. We find that the temperatures obtained with AtomDB v3.0.8 are indeed higher than those obtained with AtomDB v1.3.1 by about $0.1-0.2$~keV in $<1.5\arcmin$ and $>8\arcmin$. In other regions the temperature difference caused by the use of different versions of atomic database is nearly negligible. Meanwhile the abundance difference caused by the update of AtomDB, is always smaller than the abundance error ranges (Fig.~\ref{fig6}).

%==================================================================
\subsection{Spatial Distributions of Gas and Dark Matter Densities}
\begin{figure}
\centering
\epsscale{1.1}
\graphicspath{{figures/}}
\plottwo{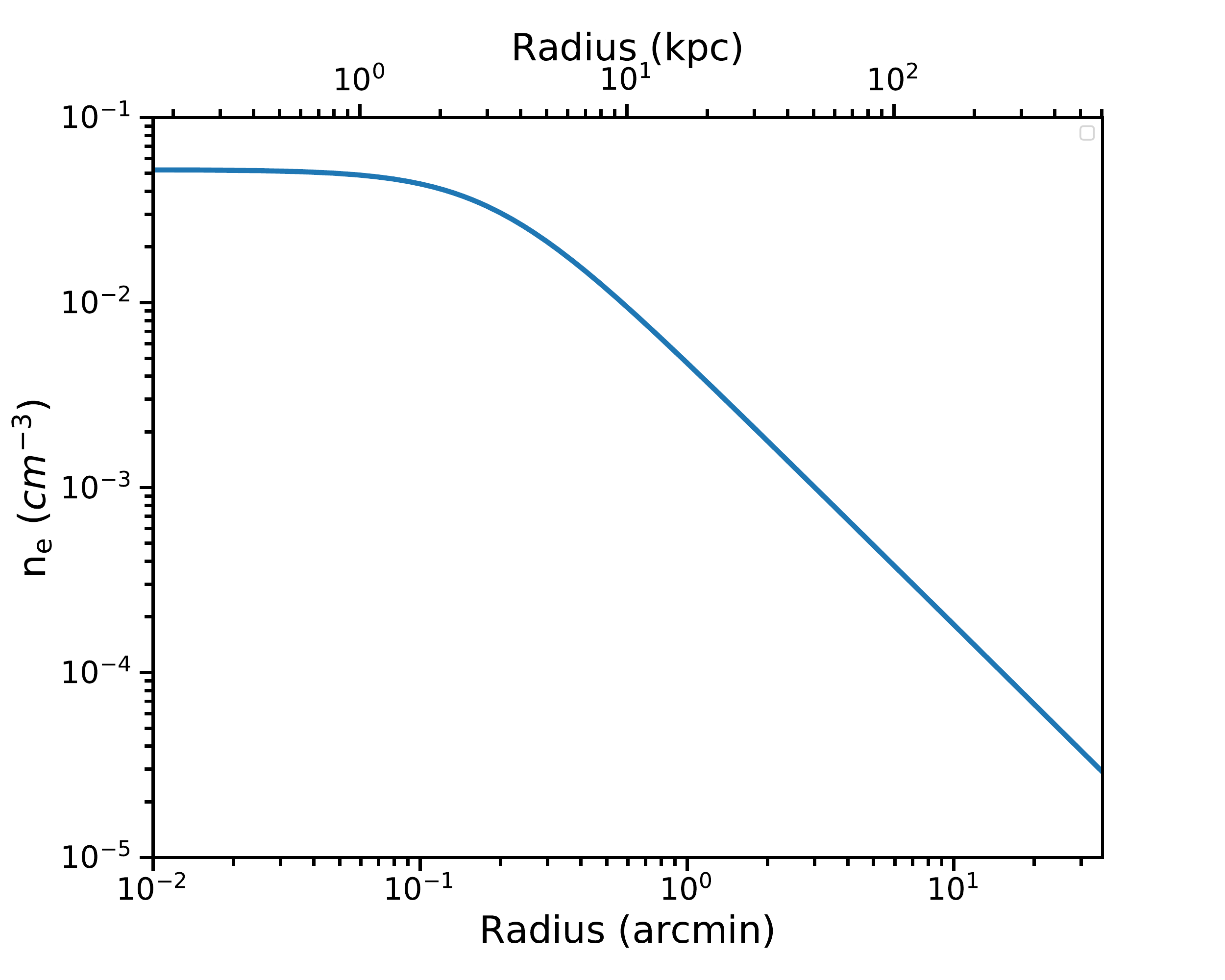}{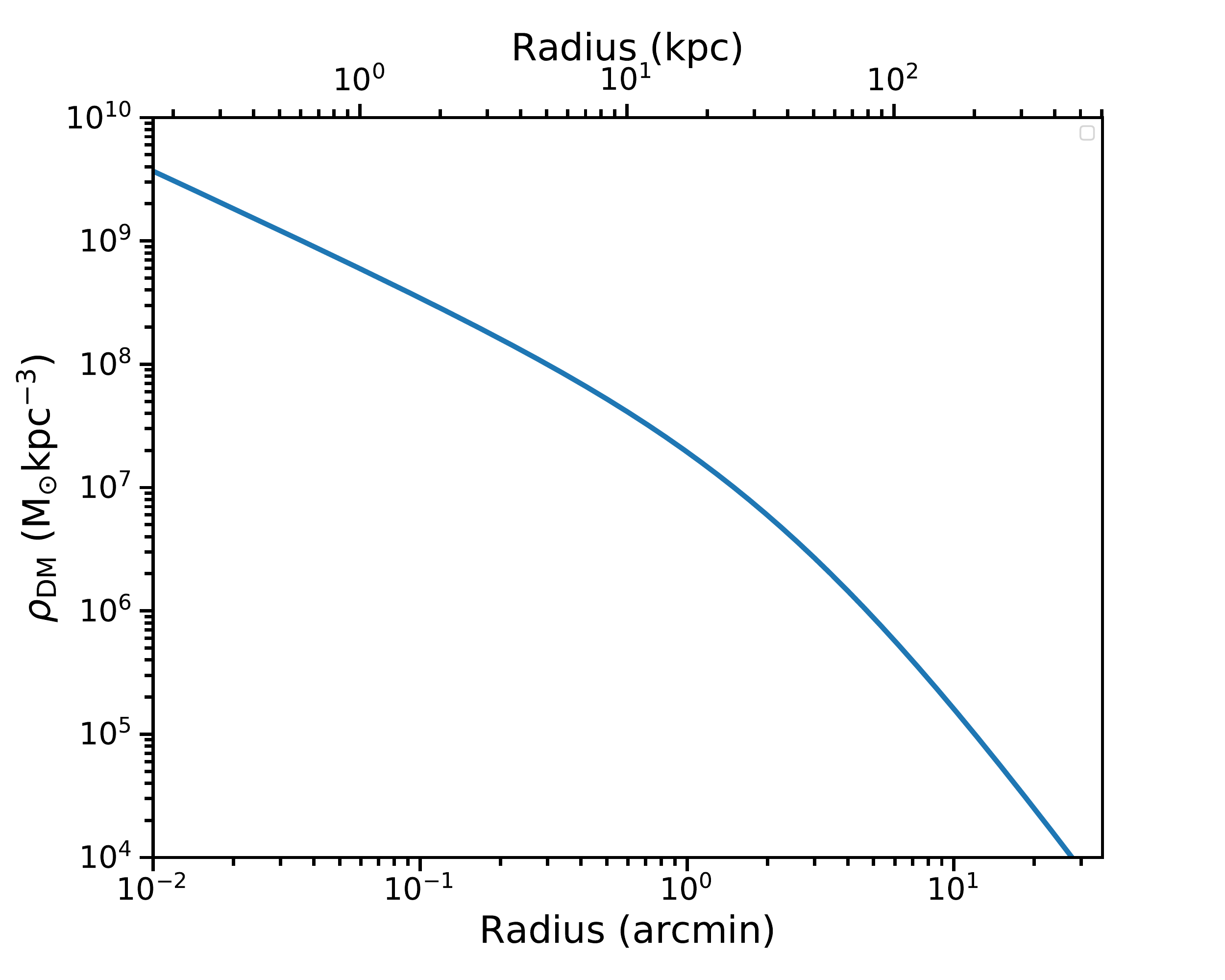}
\caption{ Electron number density profile (left) and dark matter density profile (right) are calculated by applying the best-fit X-ray surface brightness and spectral models. \label{fig7}}
\end{figure}

\begin{figure}
\centering
\graphicspath{{figures/}}
\includegraphics[scale=.33]{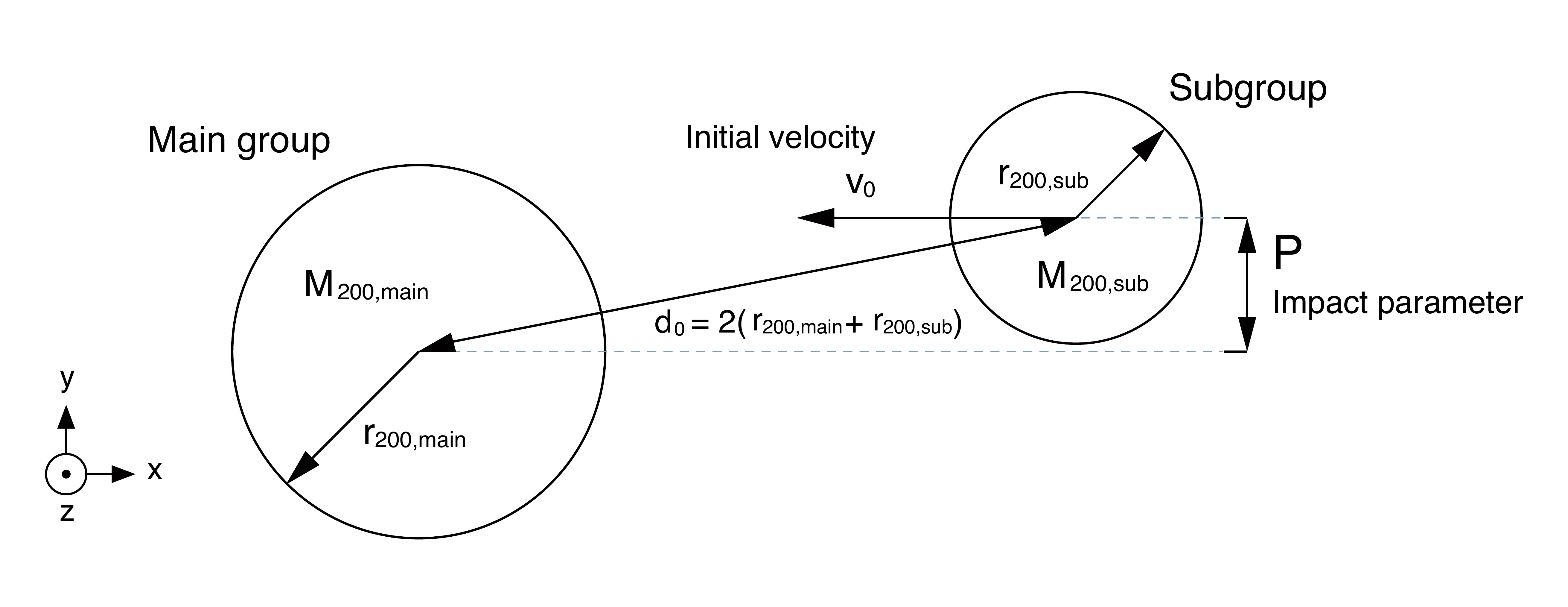}
\caption{ Schematic diagram of the initial settings of the simulation. \label{fig8}}
\end{figure}

%==================================================================
In order to constrain the simulations carried out in this work, we also calculated the spatial distributions of gas and dark matter densities, the results of which will be fed into the GADGET-3 code. We assume that the three-dimensional spatial distribution of the electron density ($n_e(r)$, where $r$ is the three-dimensional radius) satisfies a spherically symmetric $\beta$-model \citep{CF76} 
	\begin{equation}
	n_e(r)=n_0\left[1+(\frac{r}{r_c})^2\right]^{-\frac{3}{2}\beta},
	\end{equation}  
where $r_c$ and $\beta$ are the core radius and slope parameter, respectively. Given Eq.(2) and the best-fit radial distributions of gas temperature and metal abundances in sector S2 (Fig.~\ref{fig3} and \ref{fig4}), the surface brightness profile is calculated as a function of the two-dimensional radius $R$ by integrating the gas emission along line of sight 
	{\begin{equation}\label{s_fit}
	S_X(R)=\int_{R}^{\infty}\Lambda(T,A)n_e(r) n_p(r)\frac{rdr}{\sqrt{r^2-R^2}}+S_{\rm bkg},
	\end{equation}
where $n_p(r)$ ($\thickapprox n_e(r)/1.2$) is the proton density for a fully ionized plasma with one solar abundance \citep{cavagnolo09}, and $\Lambda(T,A)$ is the cooling function calculated by using the best-fit spectral parameters \citep[e.g.,][]{cavagnolo08}. Using Eq.(3) to fit the observed surface brightness profile extracted from sector S2, the electron density $n_e(r)$ can be determined when the best-fit is achieved ($r_{\rm c} = 3.2$~kpc, $\beta = 0.47$, and $n_{\rm 0} = 0.05~\rm cm^{-3}$; Figure~\ref{fig7}). Thus the gas mass distribution in the group can be calculated by integrating the best-fit gas densities (see \citealt{ettori13} for a review).  

Assuming that the group is in a hydrodynamic equilibrium state, the total gravitating mass, including the contributions of dark matter, gas, and stellar components, within radius $r$ can be calculated as
 	 \begin{equation}
 	 M_{\rm tot}(<r)=-\frac{r^2k_bT(r)}{G\mu m_p}\left[\frac{1}{T(r)}\frac{dT(r)}{dr}+\frac{1}{n_e(r)}\frac{dn_e(r)}{dr}\right],
 	 \end{equation}
where $\mu$ $=$ $0.61$ is the mean molecular weight per hydrogen atom, $k_b$ is the Boltzmann constant, and $m_p$ is the proton mass \citep[e.g.,][]{sarazin88}. By subtracting the gas mass and stellar mass, which is given in \citet{morita06}, the dark matter mass distribution can be determined (Figure~\ref{fig7}), based on which we derive a virial radius of $r_{200}= 610 \pm 30$~kpc and a total mass within $r_{200}$ of $M_{\rm tot}(r_{200})=(2.82 \pm 0.35) \times 10^{13}$~$\rm M_{\sun}$.

%###################################
\section{SIMULATION OF THE MERGER}
%###################################
In order to investigate the possible recent merger event and its impact on the redistribution of supernova yields, we perform a series of idealized numerical simulations to model the collision processes of two initially isolated galaxy groups (Fig.~\ref{fig8}) by using the TreePM-SPH GADGET-3 code \citep{springel05}. The code treats both gas and collisionless dark matter as particles and has been widely used to study merger processes of galaxy clusters \citep[e.g.,][]{springel07,ML13,zhang14,ML15,machado15}. In this work we focus on an ideal case in which the gas is assumed to be adiabatic ($\gamma = 5/3$; Case A), i.e., neither radiative cooling nor energetic feedback is involved (e.g., \citealt{springel07,ML13,zhang14}), and a case in which only gas cooling is included (Case B) for comparison.

Following the previous works of, e.g., \citet{springel07}, \citet{ML13}, and \citet{zhang14}, we apply several simplifications in the simulations. Firstly, the stellar component is not involved in the simulations. To study the metal enrichment caused by the star formation process we alternatively employ a toy model as described in \S 4.1.4. Secondly, magnetic field is ignored, which is expected to have little impact on the thermodynamics of gas \citep[e.g.,][]{lagana10}. In addition to these, cosmological expansion is ignored due to the small spatial extent considered here.

%================================================
\subsection{Model Settings and Analysis Methods}
%================================================
%------------------------------------------------
\subsubsection{Dark Matter and Gas Distributions}
%------------------------------------------------
We assume that both of the groups are initially isolated and stay in a hydrodynamic equilibrium state before the collision. Based on a series of simulations we find that, unless a head-on merger with an extremely high speed occurs, a nearly merged system like HCG~62 should possess global dark matter and gas distributions very similar to those initially possessed by the main colliding group \citep[main group hereafter;][]{roediger11,ML13,ML15,machado15}. Therefore, as a fair approximation, we set the initial dark matter and gas density distributions of the main group by referring to the best-fit results for HCG~62 (\S3.3).

To be specific, we assume that within the virial radius $r_{200}$ the initial dark matter distribution of the main group follows the NFW profile \citep{NFW96,NFW97}, whereas outside $r_{200}$ an exponential truncation is introduced to avoid the divergence of total mass \citep{KMM04}, 
\begin{equation} \label{eq:5}
   {\rho_{\rm DM}}(r) = 
	\begin{cases}
		\frac{\rho_{\rm s}}{(r/r_{\rm s})(1+r/r_{\rm s})^{2}}       & \quad \text{if } r \leqslant r_{200}, \\
		{\rho_{\rm DM}}(r_{200}) \left(\frac{r}{r_{200}} \right)^{\delta} {\rm exp}\left(- \frac{r-r_{200}}{r_{\rm decay}}\right)   & \quad \text{if } r > r_{200},         
	\end{cases}
\end{equation}
where $r_{\rm s}$ is the scale radius, $\rho_{\rm s}$ is the critical density, $r_{\rm decay} =  0.3r_{200}$ is the truncation scale, and $\delta$ is a parameter used to ensure a smooth transition at $r_{200}$. By fitting this profile to the observed dark matter density distribution (i.e., using the observation to constrain the model; \S3.3), we impose the following best-fit parameters on the main group as initial conditions in order to ensure that the corresponding parameters of the simulated merged system are similar to those of HCG 62: $r_{\rm s,main} = 44$~kpc, $\rho_{\rm s,main} = 1.42 \times 10^{7}$~$\rm M_{\sun}~kpc^{-3}$, $r_{\rm 200,main} = 610$~kpc, and $M_{\rm 200,main} = 2.67 \times 10^{13}$~$\rm M_{\sun}$ \footnote{~We use subscripts ``main'' and ``sub'' to denote parameters of the main group and subgroup, respectively.}. In each run of the simulations, these initial settings are always the same for the main dark matter halo, and the initial profile of the dark matter distribution in the infalling subgroup is scaled down according to the mass ratio between the two groups. Based on test simulations we choose to focus on mass ratios of $M_{\rm 200,main} / M_{\rm 200,sub} = 1.5,~3,~5$ and $7$. Also we define major and minor mergers as $M_{\rm 200,main} / M_{\rm 200,sub}$ $\le 3$ and $> 3$, respectively.

Similarly, we assume that within $r_{200}$ the initial gas distribution in the main group follows the $\beta$-model that best fits the observation, and outside $r_{200}$ the gas fraction is fixed at the value observationally derived at $r_{200}$
\begin{equation} \label{eq:6}
   \rho_{\rm gas}(r) = 
	\begin{cases}
		\rho_{\rm 0} \left[1+(\frac{r}{r_{\rm c}})^{\rm 2}\right]^{-\frac{3}{2}\beta}     & \quad \text{if } r \leqslant r_{200}, \\
		{\rho_{\rm DM}}(r)\frac{\rho_{\rm gas}(r_{200})}{\rho_{\rm DM}(r_{200})}    & \quad \text{if } r > r_{200},  	
	\end{cases}
\end{equation}
where $r_{\rm c}$, $\beta$, $\rho_{\rm 0}$, and $\rho_{\rm gas}(r_{200})$ can be found in \S3.3 and are imposed on the main group here. The initial gas distribution of the subgroup is also scaled down accordingly.

%-----------------------------------------------------
\subsubsection{Relative Velocity and Impact Parameter}
%-----------------------------------------------------
N-body simulations \citep[e.g.,][]{poole06, DS13} show that, when two cluster- or group-sized halos collide, the mean infall velocity of the minor object at the virial radius of the main object is $\bar{v}(r_{\rm 200,main}) = (1.1 \pm 0.1)V_{\rm c}(r_{\rm 200,main})$, where $V_{\rm c}$ is the circular velocity at $r_{\rm 200,main}$. 
The tangential component of $\bar{v}(r_{\rm 200,main}$) is typically $v_{\bot}$ $\thickapprox$ $(0-0.5)V_{\rm c}$ \citep[][] {RS01, poole06}, depending on the impact parameter. According to these we derive that the initial relative velocity $v_{0}$ ranges from 200 to 800~$\rm km~s^{-1}$ and the initial impact parameter $P$ from 0 to 600~kpc.

%-----------------------------
\subsubsection{Other Settings}
%----------------------------- 
The initial separation between the centers of the two merging groups is $d_0 = 2(r_{\rm 200,main}+r_{\rm 200,sub})$ \citep{ML13,zhang14}. The mass resolutions for the dark matter and gas components are $4.68 \times \rm 10^{6}$~$\rm M_{\sun}$ and $3.71 \times \rm 10^{5}$~$\rm M_{\sun}$, respectively. The evolution of each merger is followed for 10~Gyr and the gravitational softening length $\epsilon$ is fixed at 2~kpc.

%------------------------------------------------
\subsubsection{Analysis Methods}
%------------------------------------------------
In order to study the star formation activity and the related central metal enrichment of the X-ray gas caused by the SN Ia and SN II events after the merger begins, we employ a toy model, on which several assumptions have been made as follows. 
(1) Sharing the same center of mass, the stellar component (group dominating galaxy or GDG) co-exists with the dark matter halo \citep{zitrin12,harvey15}. 
The stellar mass of the main GDG is set to be $M_{\rm star} = (7.6 \pm 1.5) \times 10^{10}$~$\rm M_{\sun}$, i.e., the stellar mass of NGC 4778 along with an uncertainty of 20\%, as given in \citealt{morita06}. The stellar mass of the sub-GDG is scaled down according to the mass ratio $M_{\rm 200,main} / M_{\rm 200,sub}$ in use.
(2) The interacting GDGs are rich in gas, and the molecular gas fraction $f_{\rm mol}$\footnote{~ $f_{\rm mol} = \frac{M_{\rm mol}}{M_{\rm mol} + M_{\rm star}}$, where $M_{\rm mol}$ is the molecular gas mass and $M_{\rm star}$ is the total mass of stars.} is between 10\% and 30\% (\citealt{dimatteo08}). 
(3) Once the two GDGs approach each other within a threshold distance of $d_{\rm SF} \lesssim 20$~kpc \citep{ellison08,cao16}, intensive star forming activity is assumed to be triggered in their core regions.
(4) Based on numerical simulations \citep[e.g.,][]{dimatteo07,dimatteo08} and observational measurements \citep[e.g.,][]{mcnamara06,woods10}, two types of merger-induced starburst durations (i.e., the molecular gas depletion time scale) are assumed. The first one (SFR1) is related to continuous starbursts, for which a typical duration of $4\times10^{8}$~yr is considered and the corresponding peak value of the linearly decelerating star formation rate is in a range of $50 - 200$~$\rm M_{\sun}~yr^{-1}$. The second is related to instantaneous starbursts, for which a constant high star formation rate (SFR2; $500 - 2000$~$\rm M_{\sun}~yr^{-1}$) is assumed for an extremely short duration of $ 2\times10^{7}$~yr. {Note that these SFRs are among the typical values given in the starburst sample study of \citet{violino18}.}
(5) Additional metal enrichment of the X-ray gas will be contributed by the newborn stars during their final evolutionary stages. Since massive stars are short-lived (less than a few tens of Myr) and the time interval of the simulations is 20~Myr, the new supply of SN II yields can be assumed to begin to accumulate immediately after the merger-induced starburst begins. For new-born low-mass stars, however, a longer time delay ($> 1$~Gyr) between the star formation and SN Ia explosions is required \citep [e.g.,][]{montuori10}. On the other hand, the old stellar components of the GDGs are assumed to be contributing SN Ia yields continuously. 
(6) Since the X-ray gas contained within about $20-30$~kpc, which is approximately the scales of the GDGs, is expected to be enriched via superwind and galactic wind \citep{strickland04,strickland09,chisholm15}, two enrichment scales (i.e., $r_{\rm enrich} =$ 20~kpc and 30~kpc) are concerned in this work.  
With the assumptions described above, the toy model is calculated by applying the Salpeter initial mass function (IMF; \citealt{salpeter55}), the stellar evolutionary model of \citet{schaller92}, and the SN II model with solar-metallicity ($Z=0.02$; \citealt{nomoto06}). The metal abundances, together with their 90\% confidence intervals, are derived by running 1000 runs of Monte Carlo simulations with the toy model to take into account the ranges and/or uncertainties of model parameters that are described above.

We take the following steps to evaluate the simulation results. First, we carry out visual inspection of the snapshots of the X-ray maps obtained in each run of the simulations to search for a morphology similar to that observed in the X-ray observations. Once a snapshot shows both a relatively relaxed appearance and an off-center excess, the corresponding simulated X-ray surface flux distributions are calculated by applying the same pie-region sets and a set of concentric annuli used in \S3.2, which are compared with the observations. If the simulated X-ray surface flux profiles, including the location and the significance of the excess, agree with observations, the simulated abundance distributions of Mg, which is regarded as a good tracer of the SN II yields \citep[e.g.,][]{montuori10}, are calculated to examine whether or not they show the same behavior as shown in the observations. If the simulated Mg abundance profiles also match the observed ones, the last step is carried out by examining whether or not the relative positions of the simulated GDGs, which are indicated by the dark matter halos \citep{zitrin12,harvey15}, match those of NGC~4778 and NGC~4776 with reference to the X-ray peak.

%==============================
\subsection{Simulation Results}
%==============================
For each run of the simulations, which is assigned with a different configuration of the initial conditions, we create projected maps of the simulated X-ray emission under inclination angles ($i_{\rm in}$) of 0\degr, 45\degr, and 90\degr at intervals of 0.02~Gyr.

%----------------------------------
\subsubsection{Differences between Merger Configurations}
%----------------------------------
We find that the relative motion between the main group and the subgroup, as well as the motion of the merged gas halo, significantly depends on the initial orbital angular momentum of the infalling subgroup $L_{\rm orb,init} = M_{\rm 200,sub}v_{0}P$. 
In a typical head-on or a nearly head-on ($L_{\rm orb,init}\simeq0$, $P\simeq0$) collision the resulting gas halo always possesses a relatively smooth and axisymmetrical appearance, which shows no particular local metal substructure and is too diffuse to account for the X-ray halo observed in HCG~62.
In a typical off-axis merger, the cores of the two interacting groups do not collide directly with each other at the first pericentric passage and the asymmetrical interaction between the two groups usually causes a global rotation in both gas halos with a certain degree of gas sloshing. We notice that, although in the mergers occurring with either small $L_{\rm orb,init}$ (e.g., when initial condition configurations such as $P = 300$~kpc and $v_{0} = 200$~km~s$^{-1}$ are applied) or large $L_{\rm orb,init}$ (e.g., when $P = 600$~kpc and $v_{0} = 500$~km~s$^{-1}$ are applied) asymmetric gas substructures can be formed at certain stages, neither their morphologies, nor the locations, nor the profiles of the gas halo are favored by observations. Features similar to the observed ones can only be found with intermediate $L_{\rm orb,init}$. In fact, using the initial condition configuration $M_{\rm 200,main} / M_{\rm 200,sub} = 3$, $P = 300$~kpc, and $v_{\rm 0} = 500$~km~s$^{-1}$ we have successfully reproduced the observed off-center gas substructure under inclination angle $i_{\rm in} \simeq 0\degr$, together with the observed high-abundance plateau. Hereafter we refer to the model with such an initial condition configuration as the `best-match' model and will focus our discussions on it.

%-----------------------------------
\subsubsection{`Best-match' Model}
%-----------------------------------
In Figures~\ref{fig9} and \ref{fig10} we illustrate the X-ray maps and dark matter distributions simulated with the `best-match' model, respectively. The maps are projected under an inclination angle of 0\degr\ and are displayed in frames of $0.8\times0.8$ Mpc, corresponding to the following six stages: the first pericentric passage ($t_{\rm 1}$), the first apogee ($t_{\rm 2}$), the second pericentric passage ($t_{\rm 3}$), the second apogee ($t_{\rm 4}$), the third pericentric passage ($t_{\rm 5}$), and the instant when the `best-match' substructure is found ($t_{\rm 6}$).

\paragraph{\it Case A vs. Case B}\

In order to compare with the observations in a straightforward way, we rotate the maps obtained at the `best-match' instant ($t_{\rm 6}$) to match the orientation of the observations, and calculate the simulated X-ray surface flux profiles, as well as the distribution of the surface flux ratio between sectors S1 and S2, by applying the same pie-region sets and a set of concentric annuli as used in \S3.1 (Fig.~\ref{fig11}). In Case A we find that in sector S2 the simulated gas emission distribution can be well modeled by the $\beta-$model ($R_{\rm c} = 7.0 \pm 0.5$~kpc and $\beta = 0.6 \pm 0.04$), and in sector S1 there exists a remarkable emission excess beyond the $\beta-$model at $\sim 10-40$~kpc, which is very similar to the observations.

In Case B we find that the `best-match' model also shows an emission excess in sector S1 at nearly the same instant as in Case A. However, this gas substructure exhibits a relatively smaller spatial extent (about $10 - 20$~kpc) than in Case A (Fig.~\ref{fig11}), i.e., the gas substructure does not develop to the extent that can account for the observed one. This can be attributed to the fact that when the gas cools down, a denser and more compact core will appear quickly after the second pericentric passage, which becomes the primary impediment to gas sloshing. In addition, an extra plume of gas is seen in Case B to the north-western direction of the gas halo. By tracing the evolution of the plume, we conclude that it is formed due to quick cooling and concentrating of a dense infalling gas stream. Consequently, Case B is not favored by the observations.

\paragraph{\it Gas Sloshing in Case A}\

By examining the simulation snapshots we determine that the emission excess in Case A is caused by weak central gas sloshing from the south to the northwest, which occurs after the third pericentric passage ($t_{\rm 5}$; the direction of the sloshing is marked on Figure~\ref{fig11}). A similar picture was also described in \citet{AM06}, who performed hydrodynamical simulations and confirmed that sloshing is capable of uplifting the gas from the core region and creating low-temperature spiral or bow-like features on scales of about a few to 100~kpc. In order to further testify this, we plot the simulated projected temperature map obtained at the `best-match' instant in Figure~\ref{fig12}. We find that the gas within the region showing emission excess is systematically cooler than the surrounding regions by about 0.3~keV, which is expected as a feature of the sloshing. This also agrees with \citet{rafferty13} who identified a cold front approximately at the outer edge of the emission excess. We note that, however, the simulated gas temperatures of both the substructure and its surrounding regions show a bias of about $\Delta T_{\rm X} \sim 0.3$~keV when compared with the observation \citep{rafferty13}. This has been discussed by \citet{ML13} and \citet{ML15}, who argued that a similar systematic gas temperature bias can be found in the simulations due to the fact that the simulations are idealized. When physical processes, such as the feedbacks from supernovae and AGNs, are included in the calculation, gas temperatures may be increased by $\sim 1$~keV or even more (see \citealt{MN07} for a review). It is true that the AGN activity can possibly destroy such a sloshing cold gas clump when heating the environment. However, the case of the Fornax cluster \citep{su171} does show that the features associated with the gas sloshing can survive in some cases.

\paragraph{\it Mg Abundance Profiles}\

We calculate the simulated deprojected distributions of Mg abundance in Case A (Figure~\ref{fig12}), which are assumed to have been affected by the merger-induced starburst that occurs at the second pericentric passage ($t_{\rm 3}$) when the linear distance between the two GDGs is less than the threshold distance $d_{\rm SF}$. In the calculation we assume a uniform initial Mg abundance distribution (0.3 solar; \citealt{simionescu15}), which is consistent with the early enrichment scenario. We find that, when the linearly decelerating star formation rate (SFR1) and an enrichment scale of $r_{\rm enrich} = 20$~kpc or 30~kpc are employed, the Mg abundances of both the inner regions ($< 1.5\arcmin$ or $25.2$~kpc in sectors S1 and S2) and the region observationally showing emission excess ($\simeq 1.5\arcmin-2.5\arcmin$ or $25.2-42$~kpc in sector S1) turn to be higher than those of the outer regions by about 0.6 solar. Also, the Mg abundance of the region showing the emission excess is about three times higher than that obtained at the same radius in sector S2. In $> 2.5'$ (or $42$~kpc) the simulated Mg abundances are in accordance with the assumed initial values (0.3 solar) since very few enriched gas particles can reach such radii.

Alternatively, we also assume a constant high star formation rate (SFR2), which lasts for only $2 \times 10^{7}$~yr, and repeat the calculation of the simulated Mg abundance profiles. The results (Fig.~\ref{fig13}) are similar to those obtained with the linearly decelerating star formation rate (SFR1). Based on these we may conclude that the merger-induced starburst and gas sloshing can be responsible for the observed Mg abundance substructures, i.e., a central abundance peak and a high-abundance plateau in sector S1.

As a further test we assume that the initial Mg abundance profile is also centrally peaked before the second pericentric passage, since in some sources the spatial distributions of the SN II products are found to possess an increase towards the center of the GDGs \citep{simionescu09,million11,sato09,lovisari11}. We tentatively adopt an initial Mg abundance profile that linearly decreases from 1 solar at the center to 0.3 solar at $0.2r_{200}$, which keeps uniform at 0.3 solar outwards. Using the star formation rate SFR1 and an enrichment scale of $r_{\rm enrich} = 20$~kpc, we find that in the region observationally showing the emission excess the simulated Mg abundances (Fig.~\ref{fig13}) are consistent with the observed ones within the 90\% confidence level, although the simulated abundances are higher by about $0.2-0.4$ solar than those derived with a uniform initial Mg abundance distribution. In $2.5'-4'$ (or $42-67.2$~kpc) the simulated Mg abundances are higher than the observed ones, while the simulated results in $> 4'$ are in accordance with the observations. Based on the results presented above, we conclude that the observed Mg abundance profiles can be reproduced by assuming either a uniform or a centrally peaked initial Mg abundance profile; the former provides a sightly better description of abundance distribution in $2.5'-4'$.

\paragraph{\it Fe Abundance Profiles}\

Furthermore, we attempt to calculate the Fe abundance profiles using the `best-match' model in Case A, and compare the results with the observations. Since there exists a long time delay ($>1$~Gyr) between the SN Ia explosions and the merger-induced starburst triggered at the second pericentric passage (about 0.78~Gyr before the `best-match' instant), we deduce that the enrichment of Fe that can be currently observed is attributed to both the SN II explosions occurring immediately after the starburst and the SN Ia explosions of the original stellar component in the GDGs. We assume that before the merger-induced starburst each of the gas halos possesses either a centrally peaked initial Fe abundance profile, which linearly decreases from 1 solar at the center to 0.2 solar at $0.2r_{200}$ and keeps uniform at 0.2 solar outwards, as often observed in galaxy groups \citep[e.g.,][]{RP07}, or a uniform (0.2 solar) initial Fe abundance profile.

We calculate the SN II enrichment of Fe using the same approach as described above for the Mg enrichment. As for the SN Ia enrichment, we take into account both the iron blown out directly into the gas during SN Ia explosions and the iron lost in the stellar winds (\citealt{bohringer04,wang05} and references therein). The former is characterized by the direct SN Ia iron-enriching rate 
\begin{equation}
R_{\rm SN Ia} = SR 10^{-12} L_{B,\odot}^{-1} \eta_{\rm Fe}
\end{equation}
in units of $\rm M_{\odot}$ yr$^{-1}$ $L_{\odot}$, where $\eta_{\rm Fe} = 0.7~M_{\odot}$ is the iron yield per SN Ia event and $SR = 0.18$ is the SN Ia rate in units of SNu [1 SNu = 1 supernova $(10^{10}~L_{B,\odot})^{-1}$ century$^{-1}$]. In the calculations the luminosity of the main GDG is approximated by adopting the $B$-band luminosity of NGC~4778 ($1.6\times10^{10}$~$L_{B,\sun}$; \citealt{gu07}), and the luminosity of the sub-GDG is scaled down simply according to the mass ratio $M_{\rm 200,main} / M_{\rm 200,sub}$. The iron loss rate of the stellar winds is calculated as
\begin{equation}
R_{\rm wind} = 1.5 \times 10^{-11} L_{B,\odot}^{-1} t_{15}^{-1.3}\gamma_{\rm Fe},
\end{equation}
where $\gamma_{\rm Fe}$ is the iron mass fraction in stellar winds and $t_{15}=0.9$ is the age in units of 15~Gyr. By adding the Fe produced in both SN Ia and SN II explosions, the latter of which is calculated with SFR1, we obtain the Fe abundance profiles at the `best-match' instant and show them in Figure~\ref{fig14}. 
We find that when the initial Fe profile is assumed centrally peaked, the obtained Fe abundances in the central $1.5'$ of both sector S1 and S2 and in $1.5'-2.5'$ of sector S1 match the observed ones, showing that an high abundance plateau is formed. When a uniform initial Fe abundance profile is applied, the obtained Fe abundance in $1.5'-2.5'$ of sector S1 is lower than the observed values by about 0.3 solar, although the Fe abundances in the inner regions of both sector S1 and S2 agree with the observed ones. In either case, the simulated Fe abundances in $> 2.5'$ are in accordance with the corresponding initial values. Since the Fe abundance profiles obtained with the assumptions of a centrally peaked and a uniform initial Fe abundance profile turn to provide a better fit to the observed profiles in $1.5'-2.5'$ and $> 2.5'$, respectively, we propose an alternative, initial Fe abundance profile, i.e., a flat-topped profile, \begin{equation} Z_{\rm Fe}(r)= \begin{cases} -0.075r^{2}+1 & \quad \text{if } r \leqslant 3.25', \\ 0.2 & \quad \text{if } r > 3.25',\end{cases} \end{equation} where $Z_{\rm Fe}$ is Fe abundance and $r$ is the radius in units of arcmin. It seems that the simulated results with the flat-topped initial Fe abundance profile best match the observations (Fig.~\ref{fig14}).

\paragraph{\it Locations of two GDGs}\

The locations of the centers of the colliding dark matter halos, which can be used to represent the positions of the GDGs, are also marked in Figure~\ref{fig11} for the `best-match' instant. With this we conclude that at the `best-match' instant the positions of the two simulated GDGs (the distance between them is about 10~kpc) relative to both the X-ray peak and the region showing X-ray emission excess are consistent with those of NGC 4776 and NGC 4778 (the distance between them is about 8~kpc).

%===================
\section{Discussion}
%===================

%========================================================
\subsection{A Comparison with Other Observational Works }
%========================================================
By reanalyzing all available high-quality \Chandra\ (165~ks) and \XMM\ (401.9~ks) data, we have revealed that there exists a high-abundance (Fe, Si, and Mg) plateau in about $0'-2.5'$ southwest of the HCG 62's X-ray peak, the spatial extension of which is broader than the `high-abundance arc' identified in previous works \citep{gu07,gitti10,rafferty13}. Since the authors of the previous works used only part of the data analyzed in this work, and we are able to reproduce the previous results precisely by applying our approach to the corresponding datasets, we speculate that the differences in the results should be attributed to data quality.

\citet{tokoi08} analyzed the \Suzaku\ XIS spectra extracted from an annulus (r $< 1.1'$), a northeast arc located in $1.1'-3.3'$ and a southwest arc located in $1.1'-3.3'$, and failed to find the evidence for the high-abundance substructure. The main reasons for this may be twofold. First, the region partition used in the spectral analysis of \citet{tokoi08} is relatively large, spanning a radial extent of $\sim 2'$, which is bound to smear out any metal abundance features that are not sufficiently significant. Second, the \Suzaku\ XIS spectra extracted from a certain region are inevitably contaminated by the photons scattered from the adjacent bright regions due to \Suzaku's broad PSF (about 2.5\arcmin), which makes it hard to resolve such a metal abundance substructure.  
In order to verify this point, we use the \textit{xissim} tool provided by the \Suzaku\ team to create a fake \Suzaku\ observation, using the best-fit gas temperature profiles and Fe abundance profiles obtained in \S3.2 and the same observational conditions (e.g., normal clocking mode, exposure time, and offset) as applied in the \Suzaku\ observation analyzed by \citet{tokoi08}. Then we follow the same approach of \citet{tokoi08} to extract and analyze the simulated XIS spectra. The results show that the Fe abundance measured in r $< 1.1'$, northeast arc, and southwest arc are $1.60 \pm 0.52$ solar, $0.67 \pm 0.15$ solar, and $0.85 \pm 0.26$ solar, respectively, which agree with those derived by \citet{tokoi08}. This confirms that the metal abundance substructure revealed in our work cannot be identified in such a \Suzaku\ observation.}

%=======================================
\subsection{The Southwest Shock }
%=======================================
By analyzing the X-ray surface brightness profile extracted along a sector ($260\degr - 330\degr$) with a broken power-law model, \citet{gitti10} identified an outer edge at about 36~kpc southwest to the center of HCG~62, where a temperature jump of about 0.16~keV is simultaneously found (see their figure 6, left panel). Since the outer edge roughly coincides with the southern radio lobe, the existence of a weak shock related to an AGN outburst was suggested. In this section we examine the gas temperature distributions near the edge and investigate whether or not there exists such a shock.

By using the same region partition scheme, energy band ($0.7-3$~keV), abundance standard \citep{AG89}, and blank-sky background as employed by \citet{gitti10}, we extract and analyze the ACIS spectra obtained in the three \Chandra\ observations with the use of both the newest and earlier versions of AtomDB. We find that, the same results, including the temperature jump across the edge (the temperatures inside and outside the edge derived by us are $T_{\rm in} = 1.50 \pm 0.08$~keV and $T_{\rm out} = 1.34 \pm 0.06$~keV, respectively, corresponding to the temperatures measured in \citealt{gitti10}. See their figure~6), can be reproduced only with the data obtained in the observation ObsID 921 and an early version of AtomDB which were used by \citet{gitti10}. However, when newer AtomDB and/or data are in use, the evidence for the existence of a temperature jump cannot be confirmed. For example, when the spectra obtained in observations ObsID 921, 10462, and 10874 and AtomDB version 3.0.8 are used, we obtained $T_{\rm in} = 1.35 \pm 0.03$~keV and $T_{\rm out} = 1.36 \pm 0.05$~keV. 

Furthermore, we attempt to add a power-law component in the spectral model to approximate the non-thermal emission radiated from the possible shock region \citep{million09}, i.e., $1.5'-2.5'$ in sector S1. As an appropriate assumption we set the mean photon index $\Gamma$ of the power-law component free or fixed at 2. The results (Table~\ref{tbl-3}) show that the power-law component is only able to account for less than 1\% of the total $0.5-7.0$~keV flux, and the best-fit spectral parameters (gas temperatures and abundances) essentially remain unchanged.

These results agree with the negative detection of the temperature jump in \citet{gu07} and \citet{rafferty13}, who jointly analyzed the three \Chandra\ observations and did not find any sudden temperature change in their temperature map. Apparently the existence of the southwest shock needs to be examined further with high-quality data in the future.

%================================================================
\subsection{Is the Merger-induced Starburst Necessary?}
%================================================================

In order to examine whether or not the merger-induced starburst is really indispensable, we re-run the simulations by switching off the metal enrichment process via SN II explosions, which are expected to be triggered at the second pericentric passage (\S4.1). We study this process by applying a centrally peaked Mg abundance profile and a flat-topped Fe abundance profile as initial distributions. The uniform initial abundance profiles are not included here, because we find that if the uniform initial abundance profiles are assumed, the abundance distributions are invariant during the merger because of the gas mixture/redistribution with the equal metallicity. 

We find that a large amount of the metals initially residing inside the core are scattered to outer regions (about tens to hundreds of kpc) during/after the second pericentric passage (an intense, nearly head-on collision), and diluted by inflows of the outer gas with low-metallicity simultaneously before the third pericentric passage. Thus at the `best-match' instant the metal concentrations in both the inner region and the southwest substructure, which is formed by gas sloshing after the third pericentric passage, are significantly weakened (Fig.~\ref{fig15}). Although the SN Ia explosions of the original stellar component can contribute additional Fe (the contribution of Mg is so little that can be ignored), the amount of the supplied metals is not adequate to account for the observed features. Therefore, we may conclude that the contribution of a starburst is necessary to explain not only the high-abundance substructure that formed by gas sloshing, but also the central peak of Fe and Mg after the second pericentric passage.

In order to confirm the conclusion made above, the contributions of the SN Ia and SN II events to the gas enrichment in the southwest abundance substructure are calculated for different cases. To be specific, when the linearly decelerating star formation rate (SFR1) and the uniform initial Mg abundance profile are assumed, it is estimated that about $52\% \pm 17\%$ of Mg in the southwest substructure is produced by SN II explosions after the merger-induced starburst (we define this fraction as $f_{\rm Mg,SN II}$), while the rest Mg is primarily produced during early enrichment ($f_{\rm Mg,initial}$). When the SFR1 and the centrally peaked initial Mg profile are assumed, the derived $f_{\rm Mg,SN II}$ is $39\% \pm 11\%$. As for Fe, when the SFR1 and the flat-topped initial abundance profile are assumed, we find that about $21\% \pm 8\%$ of Fe is produced by SN II explosions during the merger-induced starburst ($f_{\rm Fe,SN II}$), about $13\% \pm 3\%$ of Fe is contributed by the continuous SN Ia explosions ($f_{\rm Fe,SN Ia}$) of the old stellar component during the interval between the second pericentric passage and the `best-match' instant, while the rest Fe is primarily produced during early enrichment ($f_{\rm Fe,initial}$). These results confirm again that the contribution of the merger-induced starburst to the gas enrichment cannot be ignored.

%==============================
\subsection{The O Abundance  }
%==============================

Comparing with magnesium which is found centrally concentrated, oxygen shows a relative low abundance in the central region. Thus, in contrast with the ratio Mg/O $\sim 1$ predicted by the standard SN II models \citep{nomoto06}, in the innermost 13.44~kpc the Mg/O ratio reaches about 2, which, however, is consistent with the values (about $1.5-3$) measured in some other poor systems, such as Abell 1060 \citep{sato07}, AWM 7 \citep{sato08}, and NGC 507 group \citep{sato09}.
This phenomenon may indicate a possibility that the overestimation of O yield in standard SN II models, as suggested in previous studies of elliptical galaxies \citep[e.g.,][]{HB06, ji09, LD10}. Another possibility is that there exits a pre-enrichment by Population III hypernovae \citep{loewenstein01,HB06}, which may account for anomalous O abundance since the O-burning region of hypernovae is generally more expanded than that of supernovea (thus, more O is burnt into heavier elements). However, this is questioned by, e.g., \citet{yoshida04} and \citet{MP05}, who calculated that the yields of Population III hypernova might not be responsible for the observed anomalous O abundance.

As for the southwest high-abundance substructure, by using the method described in \S4.1.4 we estimate that the amount of O enriched by SN II during the starburst will account for about 0.3 solar, which will in turn result in an abundance of about 0.6 solar after adding the initial O abundance ($\sim$ 0.3 solar, which is expected to be consistent with the O abundance observed in the outer regions). With current data quality, however, it is difficult to detect both such an increase in O abundance at the 90\% confidence level and the O abundance plateau. Since the measurement of O abundance suffers from both systematic uncertainties caused by low sensitivity at the low energies and contamination on the ACIS optical blocking filters, the measurement of Mg abundance is more reliable.

%=========================================================================
\subsection{Possible Impacts of AGN Activity and Ram-Pressure Stripping }
%=========================================================================

Although the merger scenario is interesting, there still exists other physical processes that can help form the observed southwest high-abundance substructure. Here we discuss the possible contributions of AGN activity and ram-pressure stripping occurring when a gas-rich galaxy moves fast though the intragroup medium. 

%==========================
\subsubsection{AGN activity}
%==========================

By studying the two-dimensional metal distributions in Abell 262, Abell 1835, and Hydra A, the central dominating galaxies of which show clear evidence (i.e., radio structure, such as, double lobes, jets, and a mini-halo) for recent AGN activity, \citet{kirkpatrick09,kirkpatrick11} found that in these clusters the X-ray plasma tends to exhibit filamentous high-abundance regions along the directions of the X-ray cavities detected on both sides. This infers that the metals have been transported outwards by large-scale outflows during multiple outbursts. As argued by \citet{roediger07}, large bubbles are more efficient at transporting metals outwards, resulting in elongated metal concentrations along the directions of the bubbles. We note that, however, such metal distribution patterns differ significantly from that of HCG 62, which hosts one off-center high-abundance gas clump spanning a linear scale ($\sim 60$ kpc in the south to the northwest direction; see also \citealt{rafferty13}) much larger than that of the southwest X-ray cavity ($\sim 10$~kpc). In fact such a small cavity size suggests that the AGN activity might be able to lift the metals only to about 12 kpc from the center \citep{roediger07}.

Furthermore, the radio power and the radio spectral index of the central AGN in the dominating galaxy NGC 4778 (a weak FR I; \citealt{gitti10}) in HCG 62 measured at 235/610~MHz are $P_{\rm 610~MHz} = (5.7\pm0.3) \times 10^{21}$~$\rm W~Hz^{-1}$ and $\alpha_{\rm 235~MHz}^{\rm 610~MHz} = 1.2\pm0.1$, respectively \citep{gitti10}, corresponding to a radio power of $P_{\rm 1.4~GHz} = (1400/610)^{-\alpha} \times P_{\rm 610~MHz} = 2.1\pm0.1 \times 10^{21}$~$\rm W~Hz^{-1}$ at 1.4~GHz. These infer that the AGN hosted by NGC 4778 is a weak radio source. In fact, the AGN in NGC 4778 is about two orders of magnitude fainter than the weakest source (J132814.81+320159.4) listed in the large radio AGN sample of \citet{BH12}. It is not very clear whether or not such an AGN outburst can blow out a sufficiently large amount of metals to explain the observed high-abundance substructure in HCG~62, because the efficiency of metal transport via AGN-induced outflows is still under debate as demonstrated in recent observational works \citep[e.g.,][]{kirkpatrick15,su172} and a series of hydrodynamical simulations \citep[e.g.,][]{churazov01,heath07,roediger07,barai11}.

%=====================================
\subsubsection{Ram-Pressure Stripping}
%=====================================

Some clusters and groups, such as Abell 4059 \citep{reynolds08} and RGH 80 \citep{cui10}, simultaneously show signatures of high-abundance substructures and ram-pressure stripping. In Abell 4059 \citet{reynolds08} found that there exists a high-abundance ridge extending for about 30~kpc in the azimuthal direction, containing about $\sim 5 \times 10^{9}$~$M_{\sun}$ of gas. The substructure is located at about 25~kpc southwest of the center, the appearance of which is similar to that of the southwest high-abundance substructure in HCG 62. Since one of the bright member galaxies of Abell 4059, the spiral ESO 349–G009 (which sits at about $266$~kpc north to the cluster central galaxy ESO 349–G010) exhibits a high radial velocity difference relative to ESO 349–G010 ($\Delta v \thickapprox 2100~\rm km~s^{-1}$), meanwhile it displays ongoing starburst signatures in both X-ray and near-ultraviolet bands, the authors speculated that ESO 349–G009 may have been plunging through the core of Abell 4059 and is responsible for the observed comet-like metal-rich gas ridge via intense ram-pressure stripping. 
A similar case is found in RGH 80 by \citet{cui10}, who argued that the high-abundance arc detected in this group is possibly formed due to the metals blown out from the early-type galaxy PGC 046529 via ram-pressure stripping. This is likely, since the galaxy is located at one end of the high-abundance substructure.

As for HCG 62, however, after using the optical data provided in \citet{ZM98} and \citet{sasaki14} to cross-check the coordinates and line-of-sight velocities of the three bright member galaxies (NGC 4776, NGC 4761, and NGC 4764, all located within the central 50~kpc), we find that none of these galaxies possesses a high radial velocity ($\Delta v > 1000$~$\rm km~s^{-1}$) relative to the group central galaxy (NGC 4778). The radial velocities of the four galaxies are actually very close to each other to within $3500-4500$~$\rm km~s^{-1}$. In addition to this, unlike the case of Abell 4059, no X-ray, near-ultraviolet, or infrared evidence for ongoing starburst has been reported for the member galaxies in HCG 62 \citep{johnson07}. 
As suggested by \citet{rafferty13}, the metal in the high-abundance substructure are more expected to originate in the core, where it is more likely to experience efficient enrichment, meanwhile there is no evidence that the location of the high-abundance substructure can be associated with that of any member galaxy. In the context of this, a scenario including merger-induced starburst and gas sloshing appears to be more favored by the high-quality \Chandra\ and \XMM\ data used in this work (see also \citealt{rafferty13}).

%==============================
\section{SUMMARY}
%==============================
By reanalyzing the high-quality \Chandra\ and \XMM\ data we reveal that, both Fe and Mg abundances show a high value in the innermost region and in the southwest gas substructure, forming a high-abundance plateau that is more extended than the `high-abundance arc' studied in previous works \citep{gu07,gitti10,rafferty13}. In the outer regions flat metal distributions are found in favor of the prediction of the early enrichment scenario. By employing the TreePM-SPH GADGET-3 code to model the collision between two galaxy groups, we find that the observed X-ray emission and metal distributions, as well as the relative positions of two central dominating galaxies with reference to the X-ray peak, can be well reproduced in a major merger with a mass ratio of 3, if the infalling subgroup possesses an intermediate initial orbital angular momentum and if the gas is assumed to be adiabatic. We also find that both the Fe and Mg abundance profiles could be reproduced only when merger-induced starburst was switched on in the toy model. By following the evolution of the simulated merging system, we also demonstrate that the effects of such a major merger on the chemical enrichment in the intragroup medium is mostly restricted within the core region when the final relaxed state is reached.

%============================
\acknowledgments
%============================
We sincerely thank the referee for providing valuable comments.
We would like to thank Congyao Zhang for sufficient help in simulations and extensive contributions to this manuscript.
We are grateful to Volker Springel for his kind offer of the developer version of GADGET code.
We also thank Yuying Zhang and Francois Mernier for helpful discussions. 
This work was supported by the Ministry of Science and Technology of China (grant No.
2018YFA0404601), the National Science Foundation of China (grant Nos. 11203017, 11333008, 11433002, 11621303, 11673017, 11835009 and 61371147), and the National Key Research and Discovery Plan (grant No. 2017YFF0210903).
C.L. acknowledges the National Key Basic Research Program of China (2015CB857002). C.L. is supported by Key Laboratory for Particle Physics, Astrophysics and Cosmology, Ministry of Education, and Shanghai Key Laboratory for Particle Physics and Cosmology (SKLPPC).

\clearpage

%===============================================================================================

% Table 1
\begin{table}
\begin{center}
\centering
\tablewidth{1pt}
\tablecolumns{4}
\caption{Summary of \Chandra\ and \XMM\ observations of HCG~62. \label{tbl-1}}
\setlength{\tabcolsep}{0.5pt}
\renewcommand{\arraystretch}{1.5}
\begin{tabular}{c@{\hspace{2em}}c@{\hspace{2em}}cc}
\tableline\tableline\
Observation     &   Detector   &    ObsID    &         Raw/Clean Exposure  \\
    date        &              &             &              (ks)        \\
\tableline    
2000 Jan 25  &  \Chandra\ ACIS-S  &     921       &   48.5/47.5   \\
2009 Mar 02  &  \Chandra\ ACIS-S  &    10462      &   67.1/66.6   \\
2009 Mar 03  &  \Chandra\ ACIS-S  &    10874      &   51.4/50.9   \\
2007 Jun 27  &  \XMM\ EPIC-MOS1   &  0504780501   &   127.3/104.1   \\
2007 Jun 27  &  \XMM\ EPIC-MOS2   &  0504780501   &   127.3/105.1   \\
2007 Jun 27  &  \XMM\ EPIC-PN     &  0504780501   &   123.9/96.8   \\
2007 Jun 29  &  \XMM\ EPIC-MOS1   &  0504780601   &   37.8/33.5   \\
2007 Jun 29  &  \XMM\ EPIC-MOS2   &  0504780601   &   37.8/33.7   \\ 
2007 Jun 29  &  \XMM\ EPIC-PN     &  0504780601   &   33.9/28.7   \\
\tableline\tableline 
\end{tabular}
\end{center}
\end{table}

%---------------------------------------------------------------------------------------------------

% Table 2
\begin{table}
\centering
\caption{Deprojected gas temperature and metal abundances with the single-phase model measured in sectors S1 and S2. \label{tbl-2} }
\scriptsize
\tablecolumns{8}
\tablewidth{1pc}
\setlength{\tabcolsep}{8pt}
\renewcommand{\arraystretch}{0.9}
\begin{tabular}{lccccccc}
\tableline\tableline\
 Radius &  kT  &   O   &  Mg  &  Si &  S  &   Fe & $\chi^{2}$/dof  \\
(arcmin) &  (keV)  &   (solar)   &   (solar)   &  (solar) &  (solar)  &   (solar)  &  \\
\tableline\tableline 
\multicolumn{8}{c}{sector S1} \\
\tableline
\multicolumn{4}{l}{\Chandra\ ACIS} \\
\tableline
$0.0 - 0.8$ & $0.94 \pm 0.01$ & $0.68_{-0.28}^{+0.39}$ & $1.16_{-0.30}^{+0.42}$ & $1.04_{-0.21}^{+0.29}$ & $0.66_{-0.29}^{+0.36}$ & $0.92_{-0.14}^{+0.20}$ & 330/221(1.49) \\

$0.8 - 1.5$ & $1.16 \pm 0.03$ & $0.05_{-0.05}^{+0.52}$ & $1.28_{-0.51}^{+0.72}$ & $0.89_{-0.25}^{+0.40}$ & $0.62_{-0.41}^{+0.51}$ & $0.92_{-0.13}^{+0.28}$ & 224/221(1.02) \\

$1.5 - 2.5$ & $1.34 \pm 0.01$ & $0.00_{-0.00}^{+0.49}$ & $1.82_{-0.63}^{+0.88}$ & $1.43_{-0.34}^{+0.48}$ & $0.73_{-0.49}^{+0.50}$ & $1.32_{-0.20}^{+0.34}$ & 249/221(1.13) \\

$2.5 - 4.0$ & $1.34 \pm 0.03$ & $0.23_{-0.23}^{+0.41}$ & $0.24_{-0.24}^{+0.34}$ & $0.34_{-0.17}^{+0.19}$ & $0.66_{-0.32}^{+0.36}$ & $0.33_{-0.05}^{+0.06}$ & 256/221(1.16) \\
\tableline

\multicolumn{4}{l}{\XMM\ EPIC} \\
\tableline
$0.0 - 0.8$ & $0.91 \pm 0.01$ & $0.45_{-0.19}^{+0.24}$ & $0.79_{-0.19}^{+0.24}$ & $0.72_{-0.13}^{+0.17}$ & $0.64_{-0.23}^{+0.27}$ & $0.80_{-0.10}^{+0.13}$ & 672/627(1.07) \\

$0.8 - 1.5$ & $1.21 \pm 0.03$ & $0.02_{-0.02}^{+0.46}$ & $0.47_{-0.33}^{+0.47}$ & $0.55_{-0.20}^{+0.27}$ & $0.42_{-0.31}^{+0.37}$ & $0.77_{-0.11}^{+0.19}$ & 625/627(1.00) \\

$1.5 - 2.5$ & $1.32 \pm 0.01$ & $0.48_{-0.41}^{+0.55}$ & $1.04_{-0.44}^{+0.56}$ & $0.99_{-0.26}^{+0.33}$ & $0.72_{-0.32}^{+0.38}$ & $1.01_{-0.15}^{+0.21}$ & 605/627(0.96) \\

$2.5 - 4.0$ & $1.33 \pm 0.02$ & $0.16_{-0.16}^{+0.20}$ & $0.25_{-0.18}^{+0.19}$ & $0.28_{-0.09}^{+0.09}$ & $0.03_{-0.03}^{+0.14}$ & $0.30_{-0.03}^{+0.03}$ & 560/627(0.89) \\

$4.0 - 6.0$ & $1.28 \pm 0.06$ & $0.17_{-0.17}^{+0.42}$ & $0.13_{-0.13}^{+0.40}$ & $0.36_{-0.20}^{+0.22}$ & $0.17_{-0.17}^{+0.35}$ & $0.20_{-0.05}^{+0.07}$ & 577/627(0.92) \\

$6.0 - 8.0$ & $1.46_{-0.19}^{+0.20} $ & $0.28_{-0.28}^{+0.80}$ & $0.00_{-0.00}^{+0.61}$ & $0.14_{-0.14}^{+0.31}$ & $0.00_{-0.00}^{+0.36}$ & $0.10_{-0.06}^{+0.10}$ & 609/627(0.97) \\

$8.0 - 11.0$ & $0.94 \pm 0.03$ & $0.26_{-0.11}^{+0.13}$ & $0.03_{-0.03}^{+0.12}$ & $0.05_{-0.05}^{+0.07}$ & $0.04_{-0.04}^{+0.20}$ & $0.07_{-0.01}^{+0.01}$ & 633/627(1.01) \\

\tableline

\multicolumn{4}{l}{Combined} \\
\tableline
$0.0 - 0.8$ & $0.92 \pm 0.01$ & $0.52_{-0.16}^{+0.20}$ & $0.93_{-0.17}^{+0.20}$ & $0.83_{-0.12}^{+0.14}$ & $0.65_{-0.18}^{+0.21}$ & $0.85_{-0.08}^{+0.11}$ & 1088/893(1.22) \\

$0.8 - 1.5$ & $1.19 \pm 0.02$ & $0.01_{-0.01}^{+0.31}$ & $0.84_{-0.26}^{+0.37}$ & $0.71_{-0.15}^{+0.22}$ & $0.53_{-0.25}^{+0.29}$ & $0.83_{-0.07}^{+0.15}$ & 896/893(1.00) \\

$1.5 - 2.5$ & $1.33 \pm 0.01$ & $0.22_{-0.22}^{+0.41}$ & $1.27_{-0.39}^{+0.49}$ & $1.13_{-0.23}^{+0.29}$ & $0.72_{-0.28}^{+0.32}$ & $1.12_{-0.14}^{+0.19}$ & 901/893(1.01) \\

$2.5 - 4.0$ & $1.33 \pm 0.01$ & $0.19_{-0.14}^{+0.15}$ & $0.27_{-0.13}^{+0.14}$ & $0.26_{-0.07}^{+0.07}$ & $0.18_{-0.11}^{+0.12}$ & $0.29_{-0.02}^{+0.02}$ & 849/893(0.95) \\
\tableline\tableline

\multicolumn{8}{c}{sector S2} \\
\tableline

\multicolumn{4}{l}{\Chandra\ ACIS} \\
\tableline
$0.0 - 0.8$ & $0.83 \pm 0.01$ & $0.67_{-0.25}^{+0.36}$ & $1.37_{-0.31}^{+0.44}$ & $1.12_{-0.23}^{+0.32}$ & $0.76_{-0.31}^{+0.39}$ & $1.06_{-0.17}^{+0.25}$ & 373/238(1.57) \\

$0.8 - 1.5$ & $1.00 \pm 0.01$ & $0.20_{-0.20}^{+0.40}$ & $1.22_{-0.39}^{+0.57}$ & $0.67_{-0.21}^{+0.29}$ & $0.40_{-0.35}^{+0.43}$ & $0.66_{-0.12}^{+0.19}$ & 268/238(1.13) \\

$1.5 - 2.5$ & $1.34_{-0.05}^{+0.13}$ & $0.01_{-0.01}^{+0.62}$ & $0.17_{-0.17}^{+0.54}$ & $0.26_{-0.24}^{+0.30}$ & $0.38_{-0.38}^{+0.49}$ & $0.33_{-0.07}^{+0.11}$ & 230/238(0.97) \\

$2.5 - 4.0$ & $1.45_{-0.10}^{+0.08}$ & $0.36_{-0.30}^{+0.38}$ & $0.14_{-0.14}^{+0.25}$ & $0.29_{-0.14}^{+0.15}$ & $0.31_{-0.23}^{+0.23}$ & $0.26_{-0.05}^{+0.06}$ & 278/238(1.17) \\
\tableline

\multicolumn{4}{l}{\XMM\ EPIC} \\
\tableline
$0.0 - 0.8$ & $0.83 \pm 0.01$ & $0.64_{-0.19}^{+0.25}$ & $1.07_{-0.21}^{+0.28}$ & $0.76_{-0.14}^{+0.17}$ & $0.55_{-0.22}^{+0.26}$ & $0.94_{-0.12}^{+0.16}$ & 516/401(1.29) \\

$0.8 - 1.5$ & $1.05 \pm 0.01$ & $0.17_{-0.17}^{+0.50}$ & $0.37_{-0.33}^{+0.47}$ & $0.64_{-0.22}^{+0.33}$ & $0.25_{-0.25}^{+0.40}$ & $0.64_{-0.13}^{+0.21}$ & 404/401(1.01) \\

$1.5 - 2.5$ & $1.41_{-0.11}^{+0.16}$ & $0.11_{-0.11}^{+0.93}$ & $0.32_{-0.32}^{+0.74}$ & $0.31_{-0.30}^{+0.37}$ & $0.20_{-0.20}^{+0.49}$ & $0.37_{-0.11}^{+0.21}$ & 358/401(0.89) \\

$2.5 - 4.0$ & $1.33 \pm 0.02$ & $0.31_{-0.19}^{+0.22}$ & $0.17_{-0.17}^{+0.20}$ & $0.22_{-0.09}^{+0.10}$ & $0.06_{-0.06}^{+0.15}$ & $0.20_{-0.02}^{+0.03}$ & 427/401(1.06) \\

$4.0 - 6.0$ & $1.32_{-0.08}^{+0.10}$ & $0.00_{-0.00}^{+0.31}$ & $0.00_{-0.00}^{+0.28}$ & $0.16_{-0.16}^{+0.13}$ & $0.00_{-0.00}^{+0.30}$ & $0.12_{-0.06}^{+0.08}$ & 329/401(0.82) \\

$6.0 - 8.0$ & $1.40_{-0.09}^{+0.24} $ & $0.33_{-0.33}^{+0.98}$ & $0.20_{-0.20}^{+0.85}$ & $0.37_{-0.32}^{+0.38}$ & $0.24_{-0.24}^{+0.61}$ & $0.16_{-0.09}^{+0.17}$ & 353/401(0.88) \\

$8.0 - 11.0$ & $1.02 \pm 0.02$ & $0.27_{-0.10}^{+0.13}$ & $0.00_{-0.00}^{+0.11}$ & $0.12_{-0.07}^{+0.08}$ & $0.15_{-0.17}^{+0.19}$ & $0.10_{-0.01}^{+0.01}$ & 393/401(0.98) \\

\tableline

\multicolumn{4}{l}{Combined} \\
\tableline
$0.0 - 0.8$ & $0.83 \pm 0.01$ & $0.66_{-0.16}^{+0.19}$ & $1.19_{-0.18}^{+0.23}$ & $0.89_{-0.12}^{+0.15}$ & $0.63_{-0.18}^{+0.21}$ & $0.99_{-0.10}^{+0.13}$ & 902/645(1.40) \\

$0.8 - 1.5$ & $1.02 \pm 0.01$ & $0.21_{-0.21}^{+0.30}$ & $0.91_{-0.28}^{+0.36}$ & $0.67_{-0.16}^{+0.21}$ & $0.34_{-0.25}^{+0.29}$ & $0.66_{-0.10}^{+0.13}$ & 697/645(1.08) \\

$1.5 - 2.5$ & $1.35_{-0.04}^{+0.11}$ & $0.04_{-0.04}^{+0.45}$ & $0.25_{-0.25}^{+0.41}$ & $0.29_{-0.19}^{+0.22}$ & $0.32_{-0.30}^{+0.34}$ & $0.33_{-0.06}^{+0.08}$ & 589/645(0.91) \\

$2.5 - 4.0$ & $1.34 \pm 0.02$ & $0.29_{-0.16}^{+0.17}$ & $0.18_{-0.14}^{+0.15}$ & $0.25_{-0.08}^{+0.08}$ & $0.15_{-0.12}^{+0.12}$ & $0.21_{-0.02}^{+0.02}$ & 717/645(1.11) \\
\tableline\tableline\
\end{tabular}
\end{table}

%---------------------------------------------------------------------------------------------------

% Table 3
\begin{table}
\centering
\caption{Deprojected gas temperatures and metal abundances with different models measured in sectors S1. \label{tbl-3} }
\scriptsize
\tablecolumns{10}
\setlength{\tabcolsep}{5pt}
\renewcommand{\arraystretch}{1}
\begin{tabular}{lccccccccccc}
\tableline\tableline\
 Radius\tablenotemark{a} &  $\rm kT_{1} $  &   $\rm kT_{2}$  & O   &  Mg  &  Si &  S  &   Fe &  $\dot{M}$\tablenotemark{b} & $\chi^{2}$/dof \\
(arcmin) &  (keV)   &  (keV)  &  (solar)   &   (solar)   &  (solar) &  (solar)  &  (solar) & $\rm M_{\sun}yr^{-1}$ &  \\
\tableline\tableline 

\multicolumn{4}{l}{1T+ CF} \\
\tableline

$0.0 - 0.8$ &  $0.08$ & $0.92\pm0.01$ & $0.52_{-0.15}^{+0.10}$ & $0.93_{-0.17}^{+0.15}$ & $0.83_{-0.07}^{+0.07}$ & $0.65_{-0.10}^{+0.16}$ & $0.85_{-0.08}^{+0.11}$ & $< 0.0340$ & 1088/892(1.22)  \\

$0.8 - 1.5$ &  $0.08$ & $1.18\pm0.02$ & $0.00_{-0.00}^{+0.16}$ & $0.84_{-0.26}^{+0.18}$ & $0.71_{-0.07}^{+0.18}$ & $0.53_{-0.24}^{+0.25}$ & $0.83_{-0.13}^{+0.08}$ & $< 0.0050$  & 896/892(1.00)    \\

$1.5 - 2.5$ &  $0.08$ & $1.33\pm0.01$ & $0.22_{-0.22}^{+0.37}$ & $1.27_{-0.40}^{+0.34}$ & $1.13_{-0.24}^{+0.29}$ & $0.72_{-0.25}^{+0.25}$ & $1.12_{-0.14}^{+0.14}$ & $< 0.0008$  & 901/892(1.01)   \\

$2.5 - 4.0$ &  $0.08$ & $1.49_{-0.08}^{+0.07}$ & $0.13_{-0.13}^{+0.26}$ & $0.58_{-0.25}^{+0.31}$ & $0.40_{-0.12}^{+0.14}$ & $0.23_{-0.16}^{+0.17}$ & $0.46_{-0.08}^{+0.10}$ & $< 0.0006$&  829/892(0.93)    \\ 

$4.0 - 6.0$ &  $0.08$ & $2.82_{-0.81}^{+0.88}$ & $3.55_{-3.55}^{+56.5}$ & $2.85_{-2.85}^{+28.7}$ & $0.10_{-0.10}^{+4.22}$ & $0.00_{-0.00}^{+4.63}$ & $1.61_{-0.70}^{+1.07}$ & $< 0.0001$&  570/626(0.91)   \\ 

$6.0 - 8.0$ &  $0.08$ & $1.39_{-0.09}^{+0.53}$ & $0.00_{-0.00}^{+1.64}$ & $0.00_{-0.00}^{+0.69}$ & $0.17_{-0.17}^{+0.37}$ & $0.00_{-0.00}^{+0.82}$ & $0.12_{-0.05}^{+0.22}$ & $<0.0003$ &  607/626(0.97)  \\

$8.0 - 11.0$ & $0.08$ & $1.14_{-0.06}^{+0.03}$ & $0.09_{-0.09}^{+0.31}$ & $0.00_{-0.08}^{+0.28}$ & $0.05_{-0.05}^{+0.14}$ & $0.00_{-0.00}^{+0.34}$ & $0.12_{-0.02}^{+0.02}$ & $<0.0001$&  638/626(1.02)   \\

\tableline\tableline\
 Radius &  $\Gamma$\tablenotemark{c}  &   $\rm kT$  & O   &  Mg  &  Si &  S  &   Fe & $R_{\rm flux}$\tablenotemark{d}  & $\chi^{2}$/dof  \\
(arcmin) &    &  (keV)  &  (solar)   &   (solar)   &  (solar) &  (solar)  &  (solar)  & & \\

\tableline\tableline
\multicolumn{4}{l}{1T+power-law} \\
\tableline

$1.5 - 2.5$ & $<1.75$ & $1.33 \pm 0.01$ & $0.25_{-0.25}^{+0.42}$ & $1.30_{-0.40}^{+0.51}$ & $1.14_{-0.23}^{+0.30}$ & $0.67_{-0.28}^{+0.33}$ & $1.14_{-0.15}^{+0.20}$ &  $<1\%$ &  898/891(1.01) 
\\
$1.5 - 2.5$ & $2.00$ & $1.33 \pm 0.01$ & $0.26_{-0.26}^{+0.44}$ & $1.31_{-0.41}^{+0.52}$ & $1.15_{-0.24}^{+0.31}$ & $0.68_{-0.29}^{+0.33}$ & $1.16_{-0.16}^{+0.22}$ &  $<1\%$ &  900/891(1.01)  
\\
\tableline

\tablenotetext{a}{~Combined \Chandra\ and \XMM\ data are used in the spectral fittings in $<4\arcmin$ regions, and only the \XMM\ data are used in $>4\arcmin$ regions.}
\tablenotetext{b}{~mass deposition rate: calculated for the corresponding pie-region.}
\tablenotetext{c}{~photon index of power-low component.}
\tablenotetext{d}{~Flux ratio of the power-law component to the VAPEC component between $0.5-7.0$~keV.}

\end{tabular}
\end{table}

\clearpage
%---------------------------------------------------------------------------------------------------

% Figure 1

%--------------------------------------------------------------------------------------------------

% Figure 2

%--------------------------------------------------------------------------------------------------

% Figure 3

%--------------------------------------------------------------------------------------------------

% Figure 4

%--------------------------------------------------------------------------------------------------

% Figure 5

%--------------------------------------------------------------------------------------------------

% Figure 6

%--------------------------------------------------------------------------------------------------

% Figure 7

%--------------------------------------------------------------------------------------------------

% Figure 8

%--------------------------------------------------------------------------------------------------

% Figure 9
\begin{figure}
\centering
\epsscale{1.2}
\graphicspath{{figures/}}
\includegraphics[scale=.3]{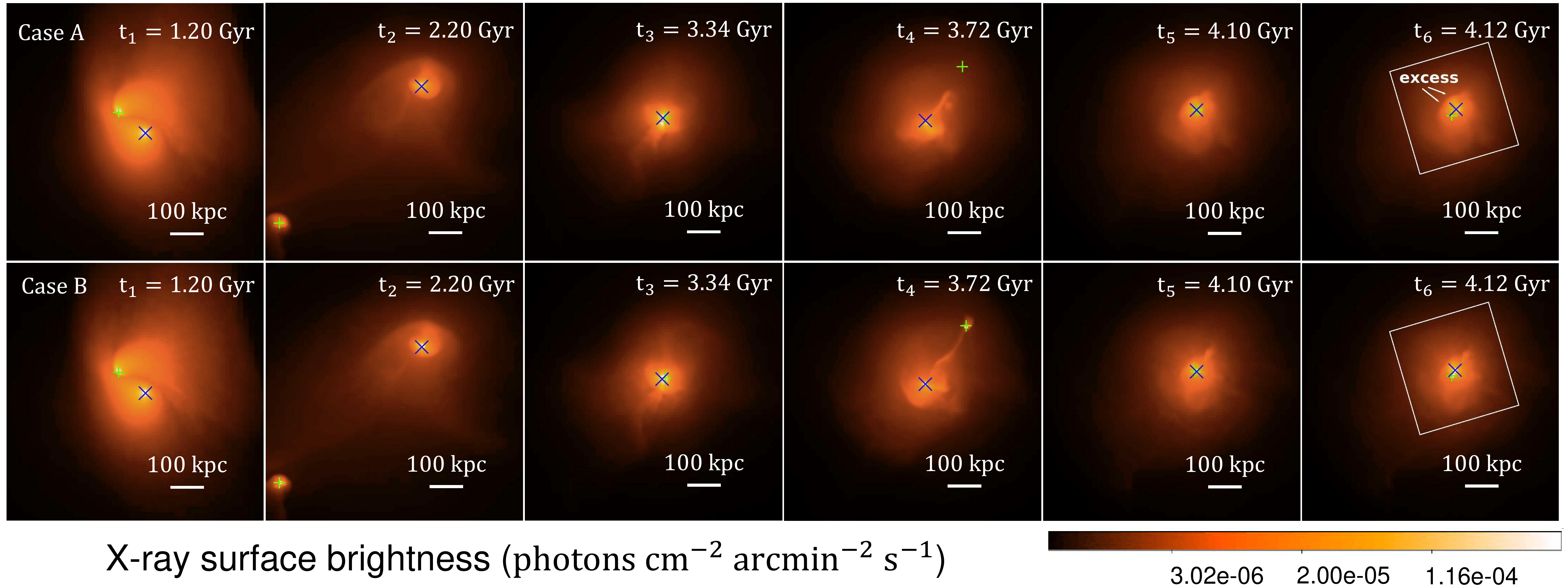}
\caption{ ($\it Upper$) Case A shows project X-ray images simulated with the `best-match' model at the snapshots of different stages (\S4.2.2). ($\it Lower$) Corresponding results for Case B (gas cooling only) is also plotted for comparison. All images are projected under an inclination angle of $i_{\rm in} = 0$. The marks of `X' and `+' are used to represented the centers of mass of the main and sub dark matter halos, respectively. Details of the excess emission in white box are shown in Figure~\ref{fig11}. \label{fig9}}
\end{figure}

%--------------------------------------------------------------------------------------------------

% Figure 10
\begin{figure}
\centering
\epsscale{2}
\graphicspath{{figures/}}
\includegraphics[scale=.3]{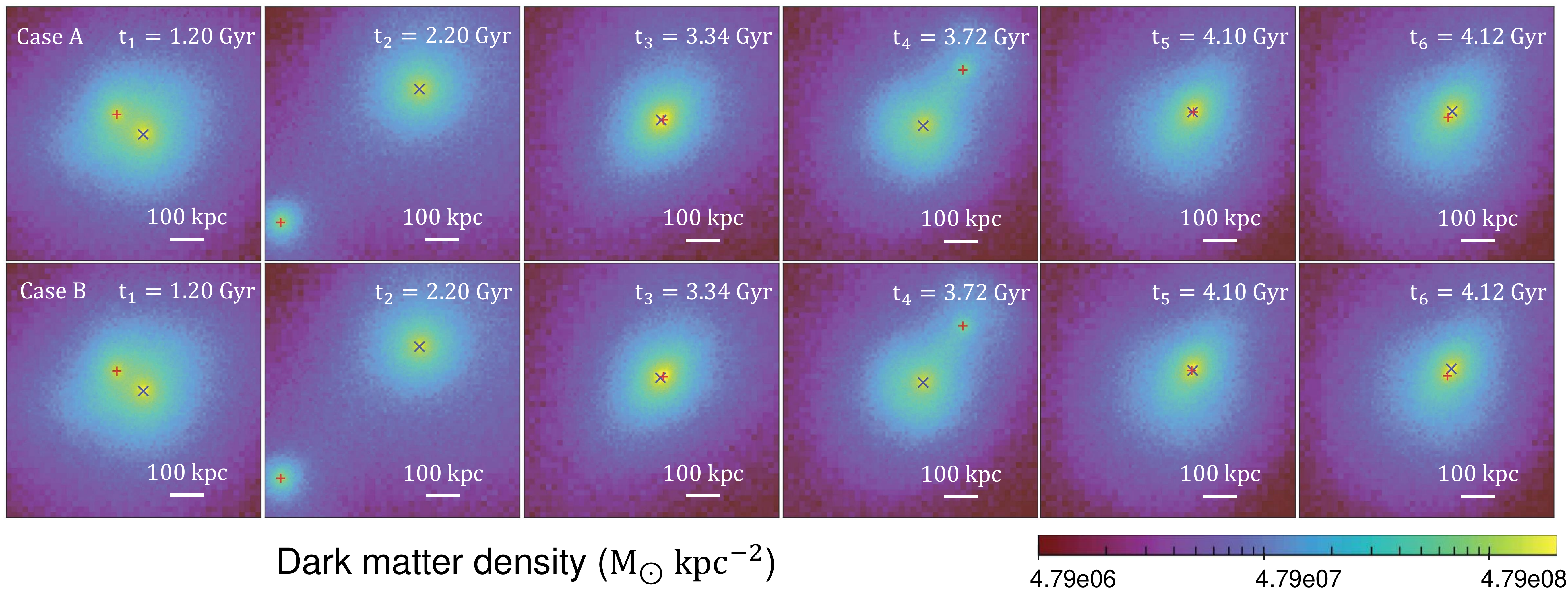}
\caption{ Same as Figure~\ref{fig9}, but for dark matter density distributions. \label{fig10}}
\end{figure}

%--------------------------------------------------------------------------------------------------

% Figure 11
\begin{figure}
\centering
\graphicspath{{figures/}}
\includegraphics[scale=.42]{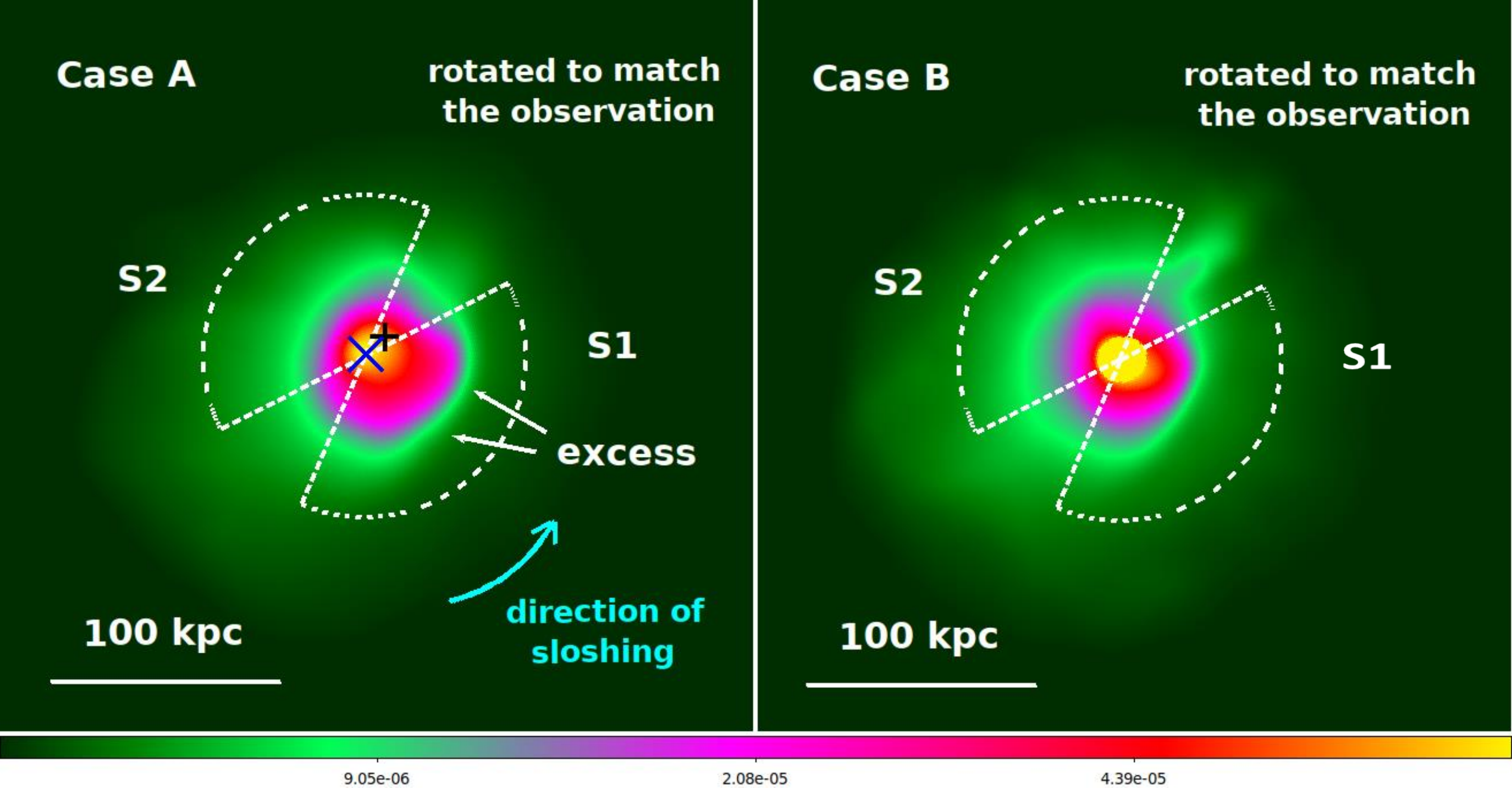}
\includegraphics[scale=.27]{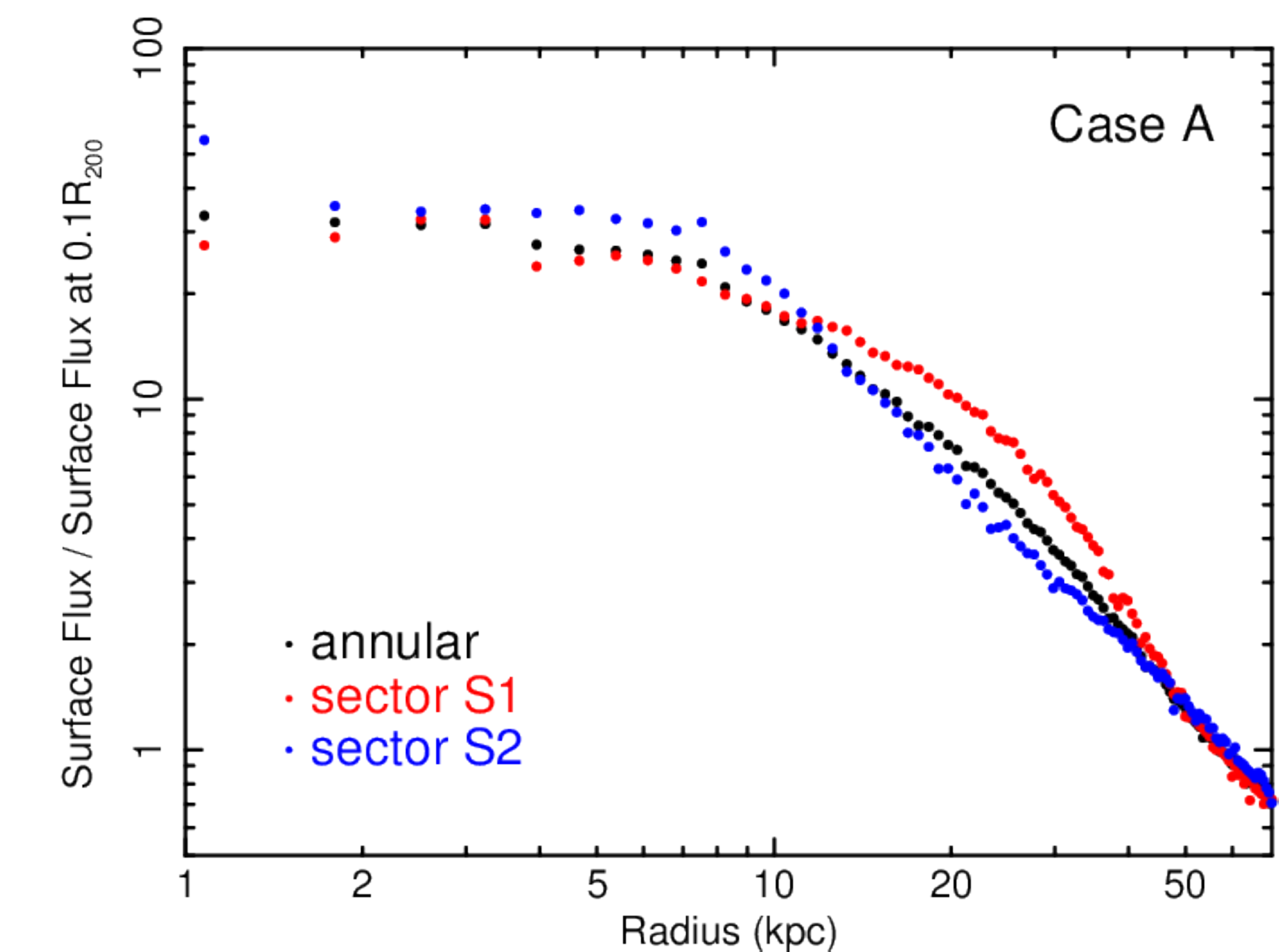}
\includegraphics[scale=.27]{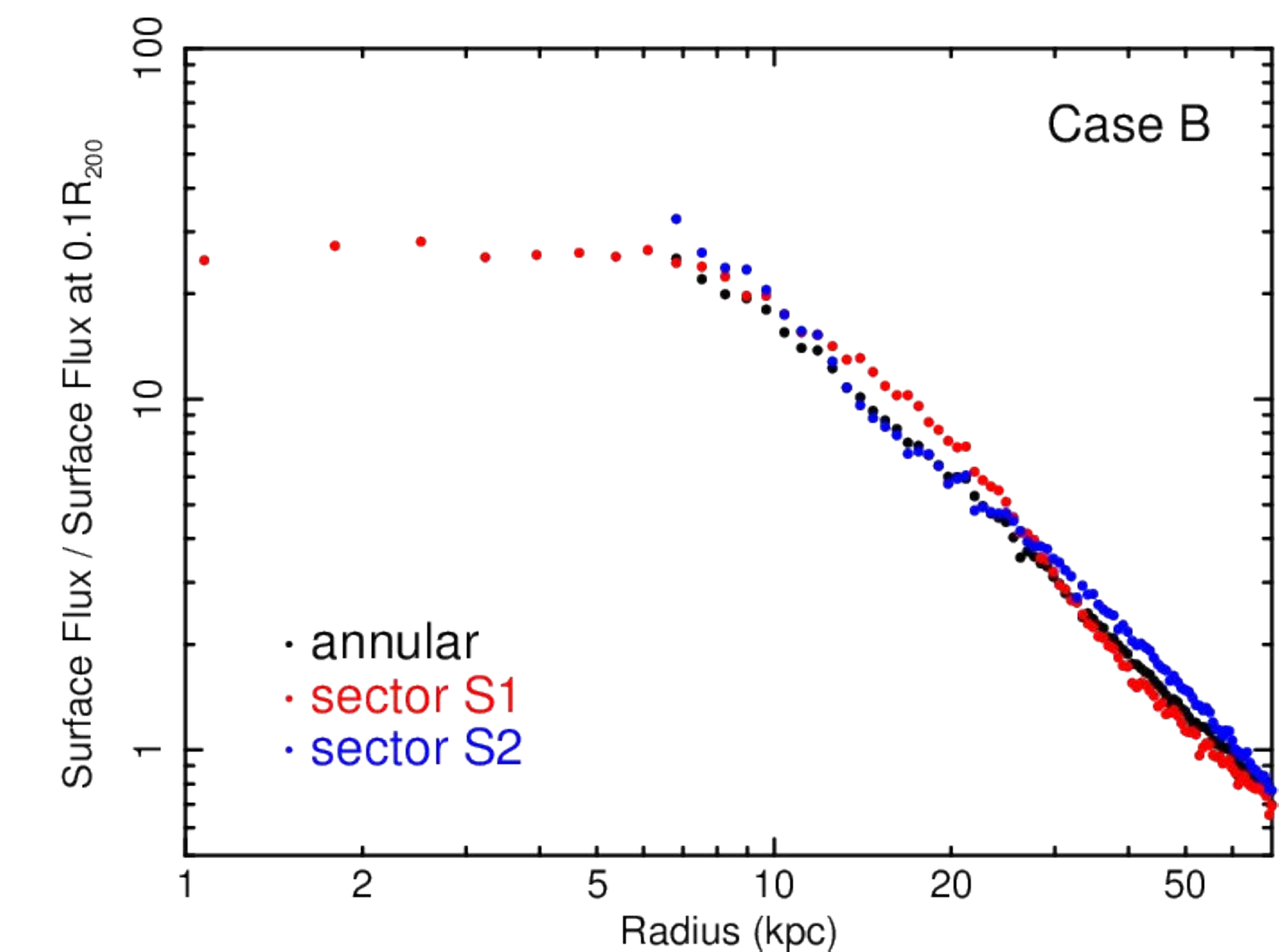}
\includegraphics[scale=.27]{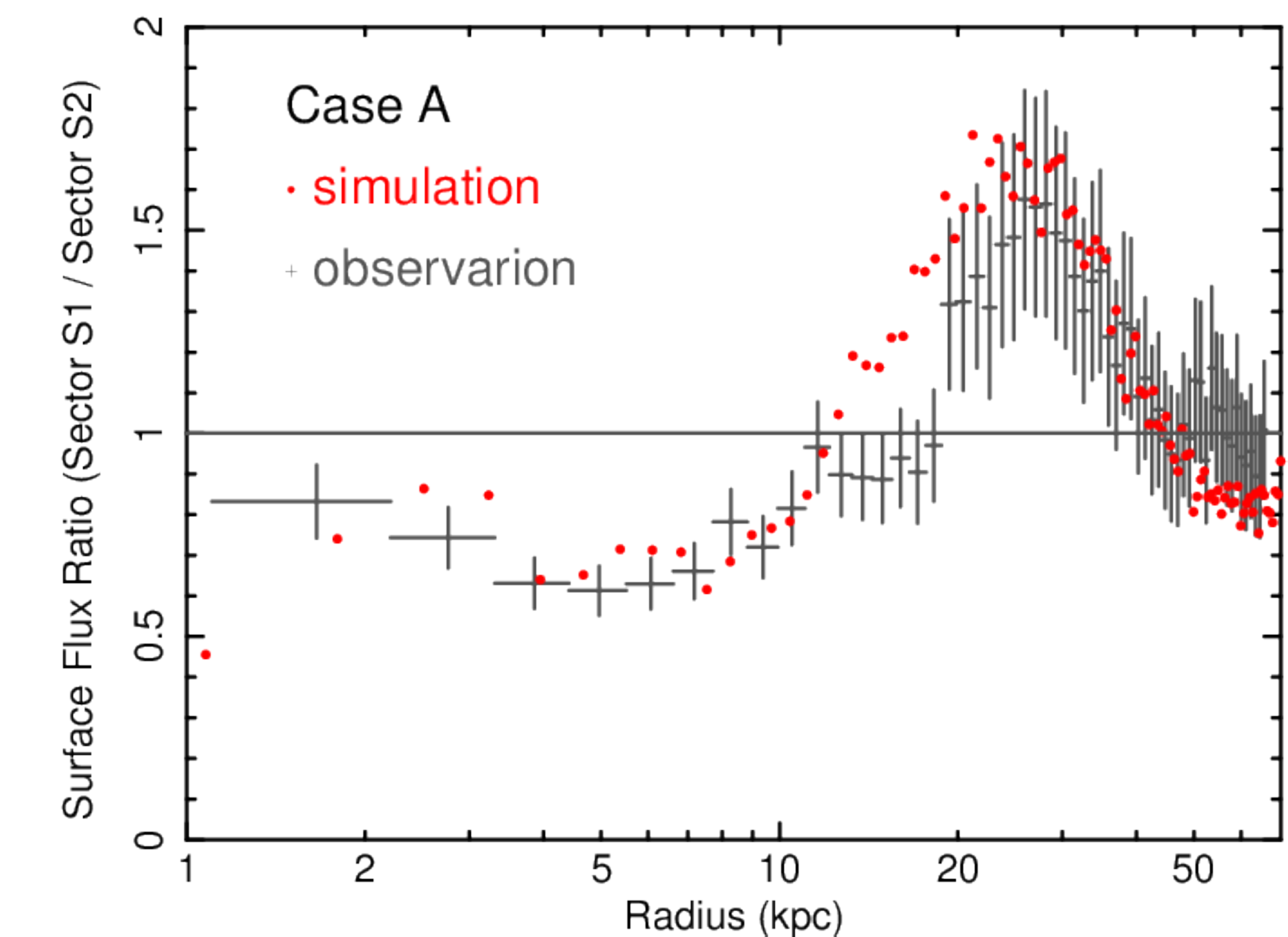}
\includegraphics[scale=.27]{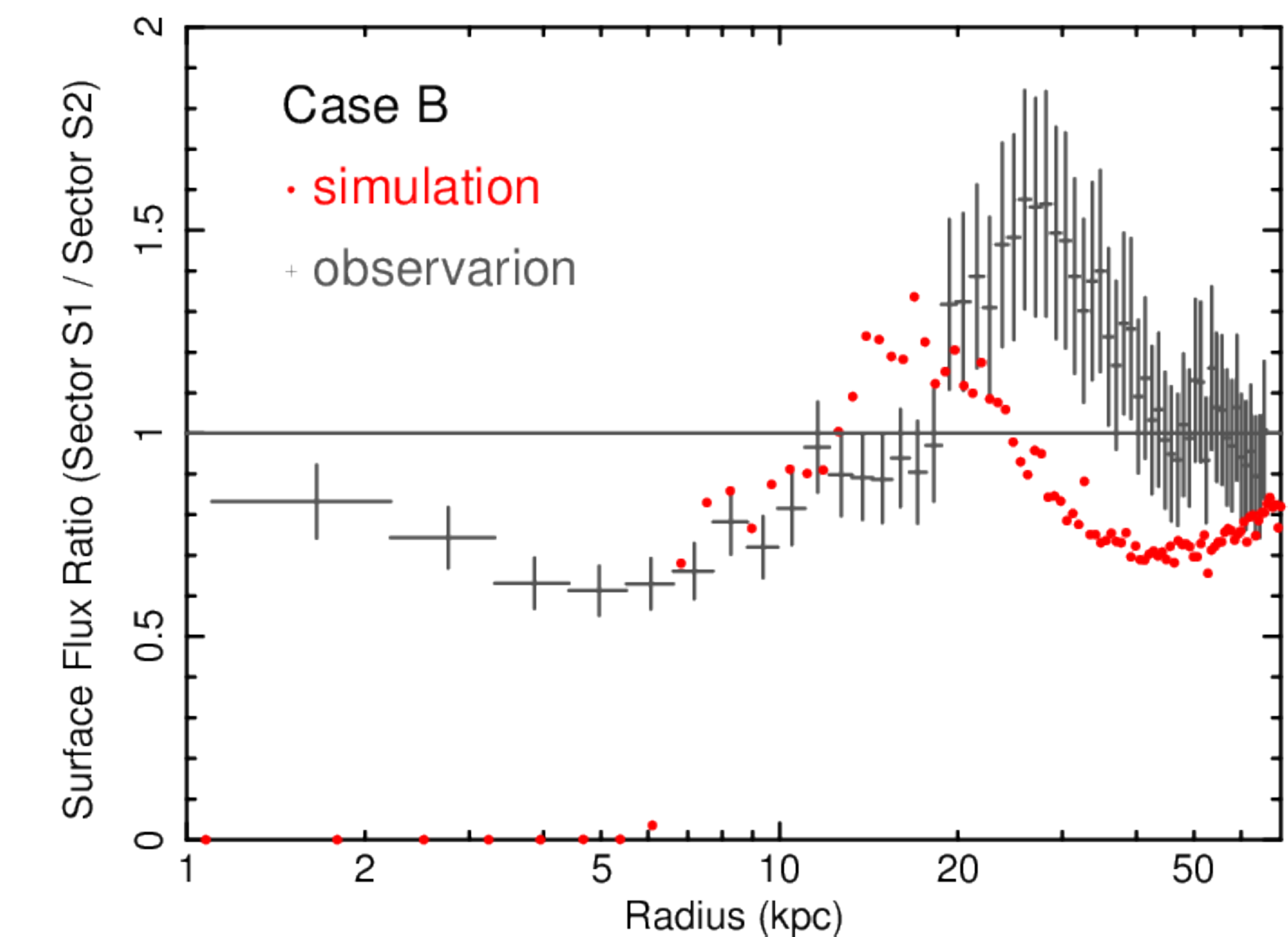}
\caption{ ($\it Upper$) Simulated X-ray image of Case A at the `best-match' snapshot with the `best-match' model ($i_{\rm in}$ = 0). Simulated X-ray image of Case B is also shown for comparison.
($\it Middle$) Corresponding X-ray surface flux distributions extracted from sector S1, sector S2 and annular regions. 
($\it Lower$) The flux ratio between sector S1 and S2 for the simulation and observation. \label{fig11}}
\end{figure}

%--------------------------------------------------------------------------------------------------

% Figure 12
\begin{figure}
\centering
\epsscale{1.0}
\graphicspath{{figures/}}
\includegraphics[scale=.21]{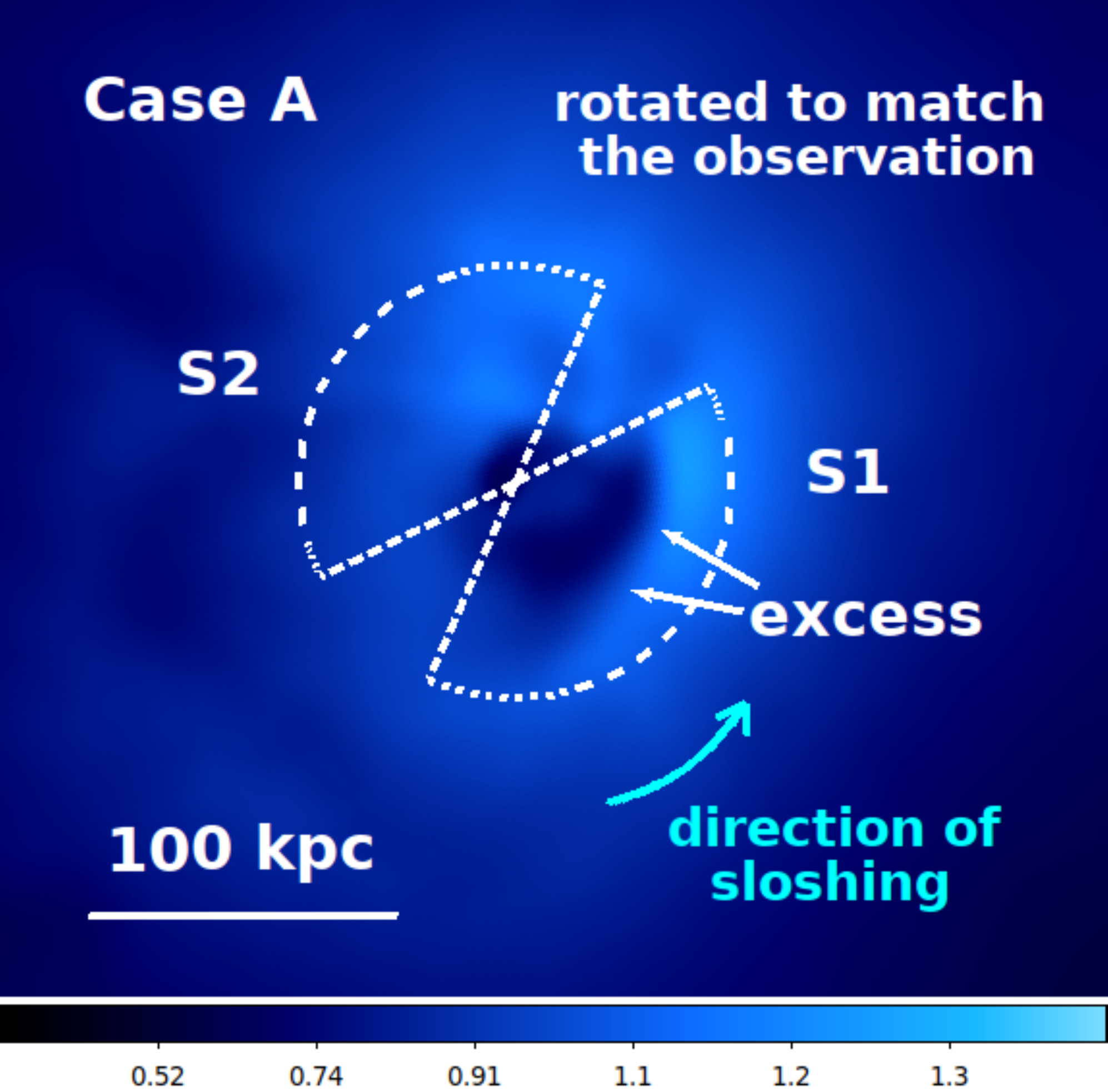}
\includegraphics[scale=.23]{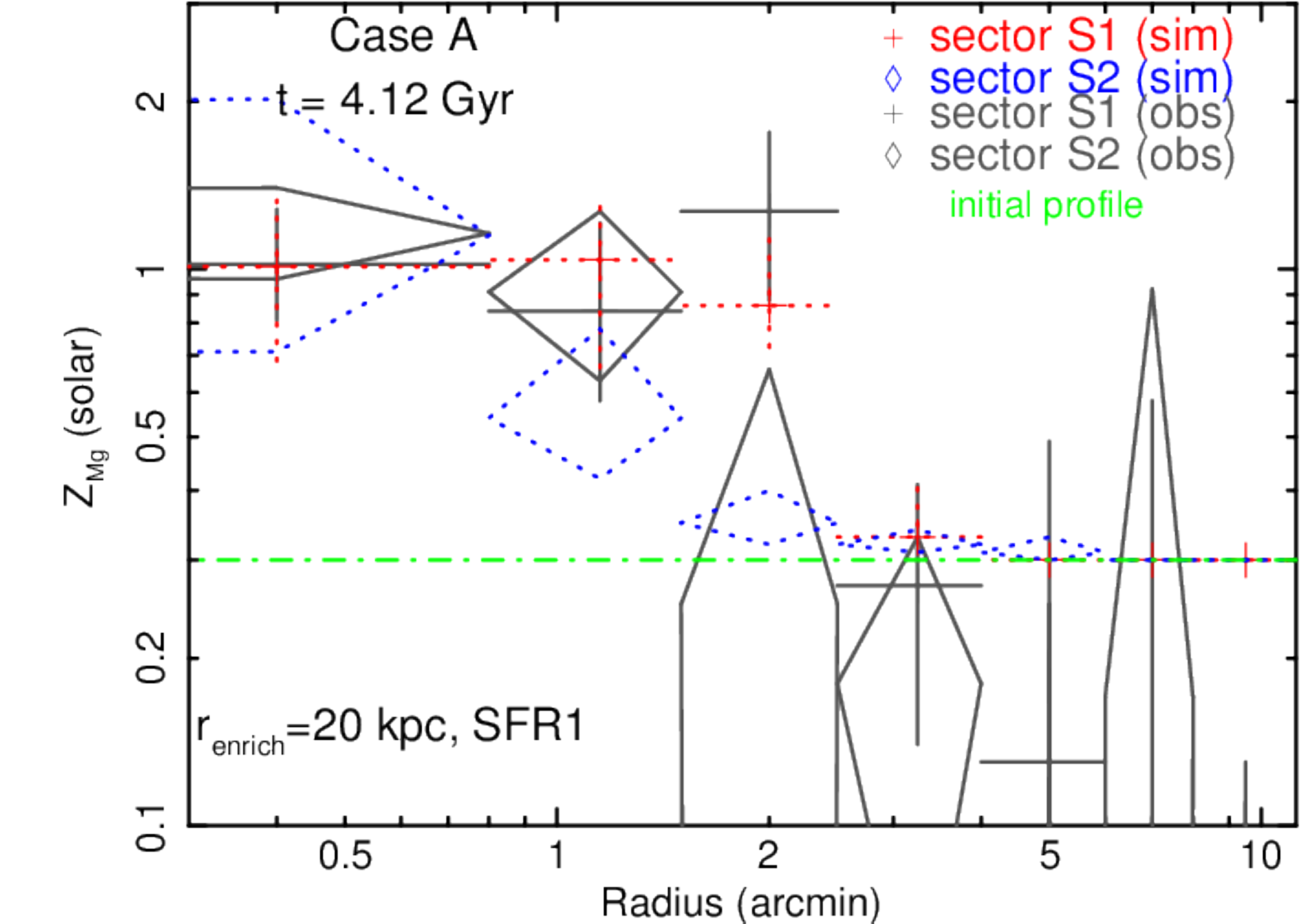}
\includegraphics[scale=.23]{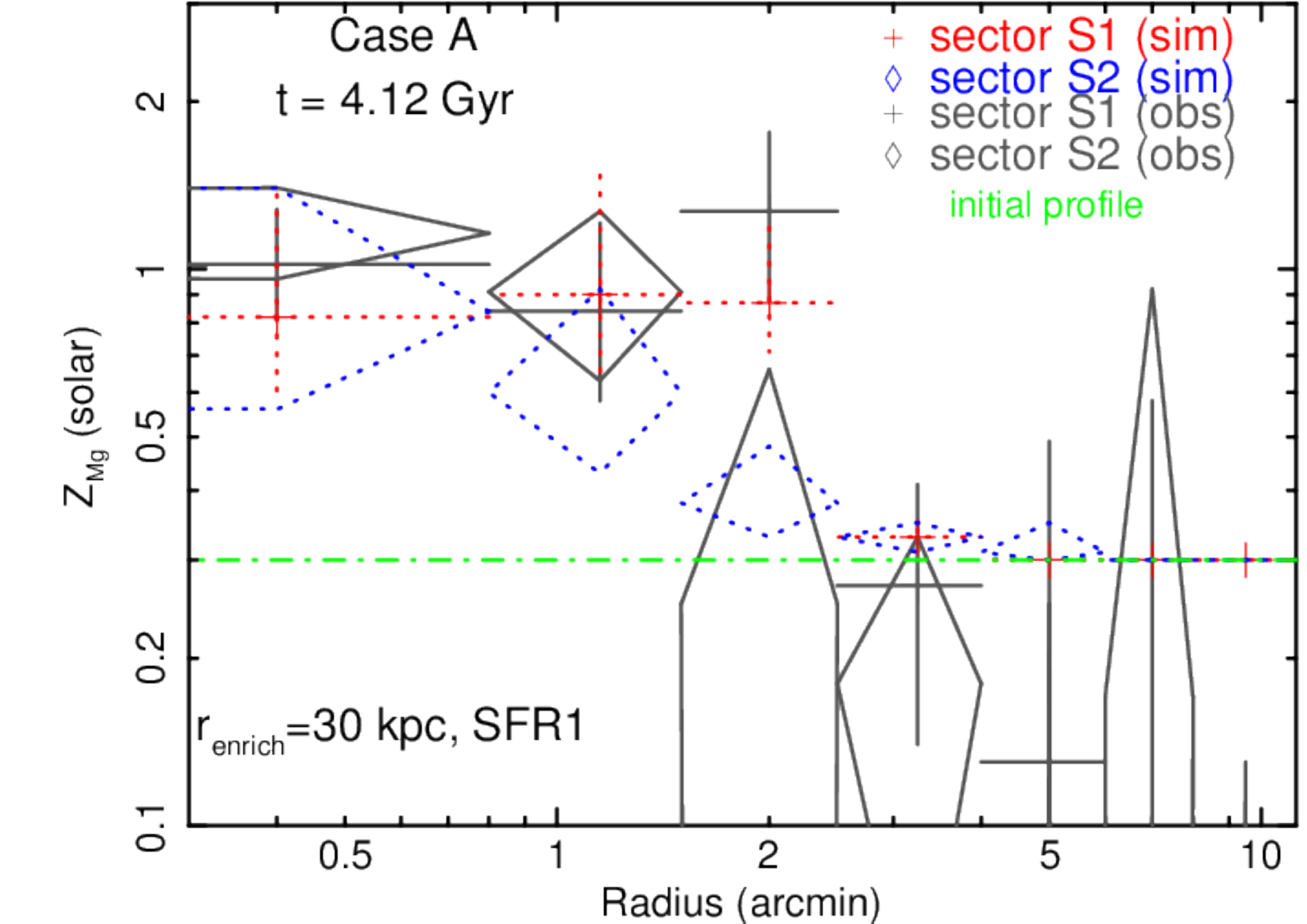}
\caption{ Temperature map and Mg abundance distributions simulated with the `best-match' model (Case A) at the `best-match' snapshot ($i_{\rm in}$ = 0). An initial uniform Mg abundance distribution due to early enrichment (0.3 solar; \citealt{simionescu15}), a linearly decelerating star formation rate (SFR1), and two kinds of enriched GDG's scale (20~kpc and 30~kpc) are assumed in the calculation (\S4.2.2). Corresponding Mg abundance profiles of observation are also plotted by dark-grey line. \label{fig12}}
\end{figure}

%--------------------------------------------------------------------------------------------------

% Figure 13
\begin{figure}
\centering
\graphicspath{{figures/}}
\includegraphics[scale=.3]{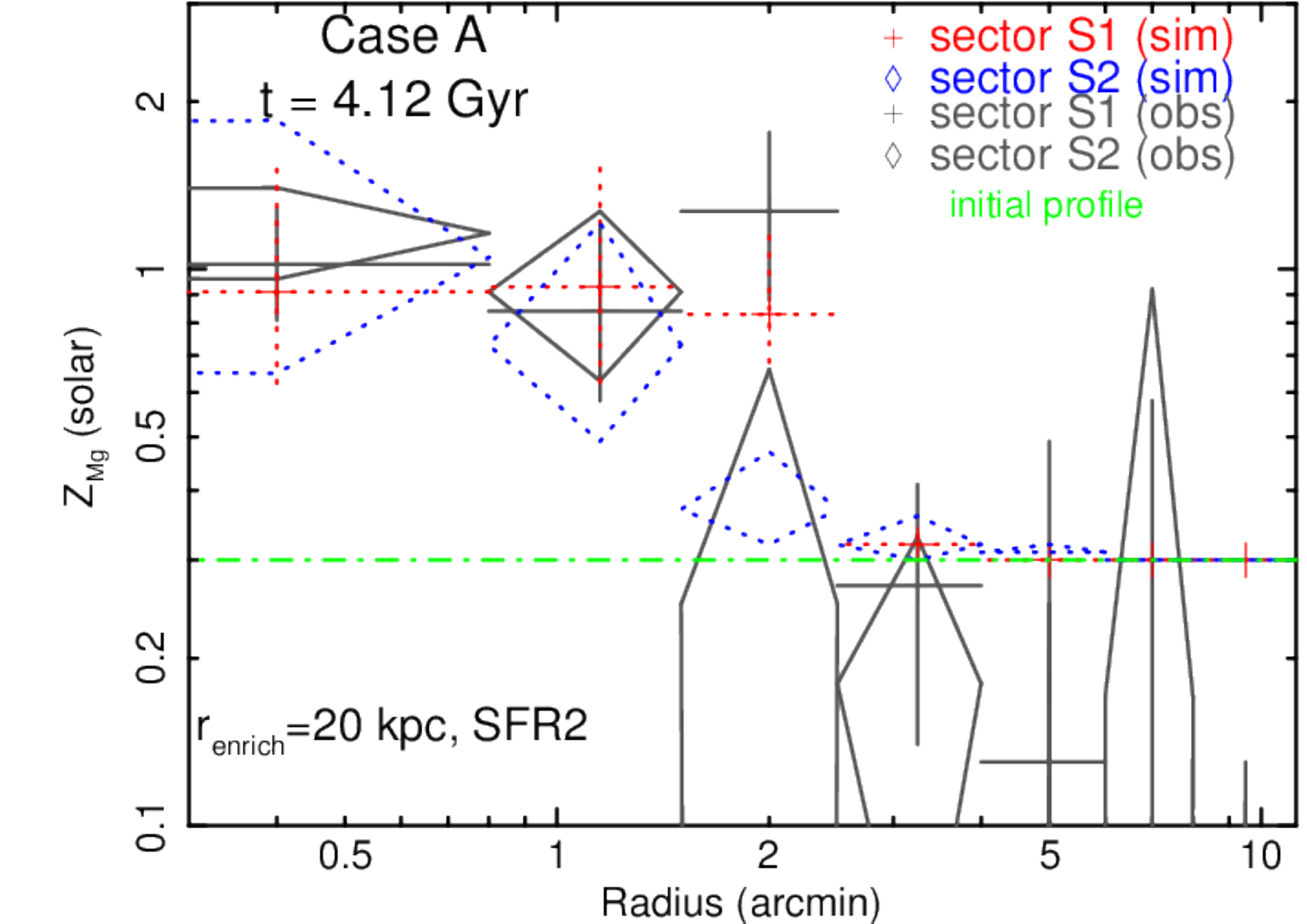}
\includegraphics[scale=.3]{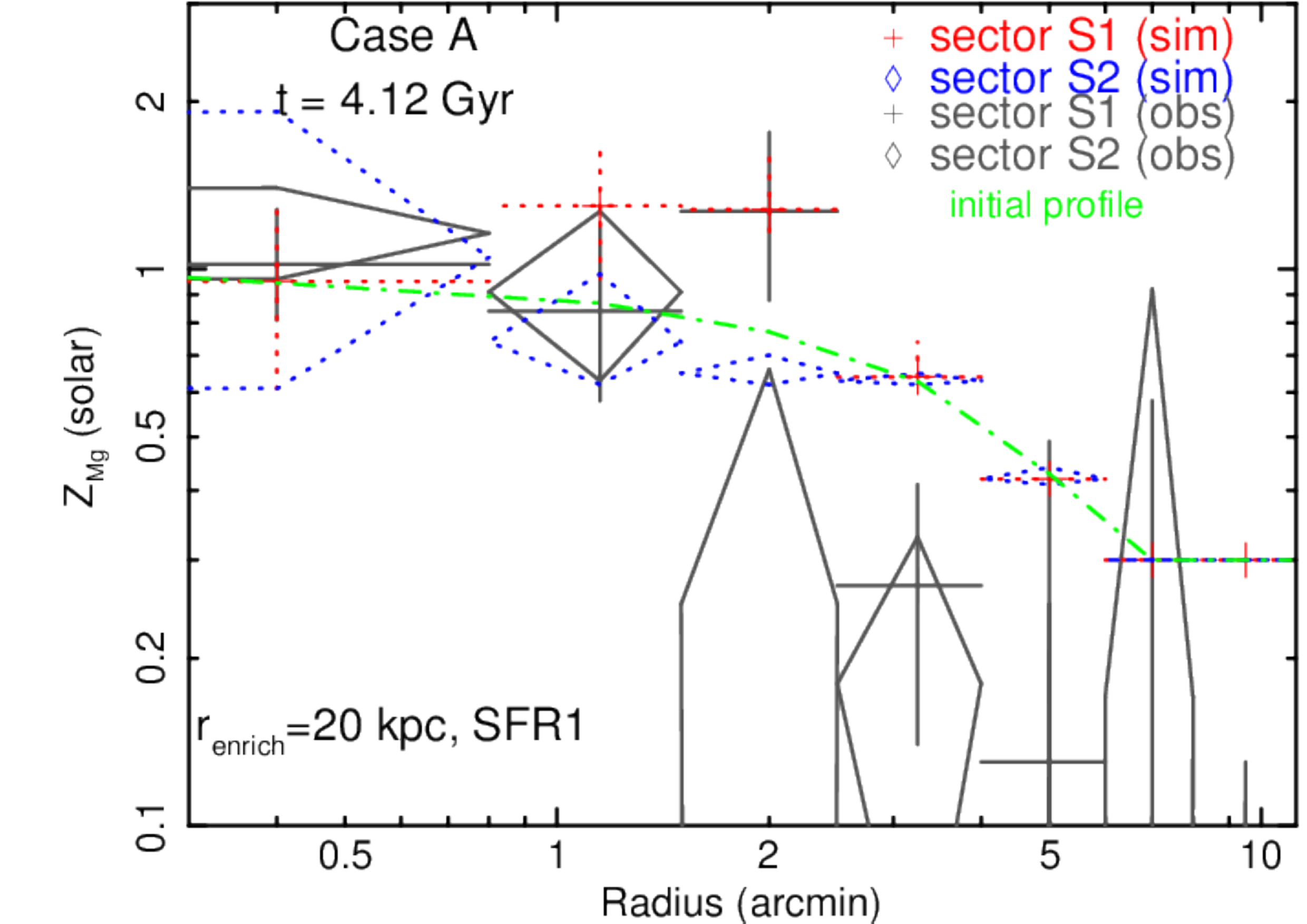}
\caption{Same as Figure~\ref{fig12}, but only for Mg abundance profiles simulated by assuming a constant high star formation rate (SFR2; left) or assuming a centrally peaked initial Mg abundance distribution (right). \label{fig13}}
\end{figure}

%--------------------------------------------------------------------------------------------------

% Figure 14
\begin{figure}
\centering
\graphicspath{{figures/}}
\includegraphics[scale=.28]{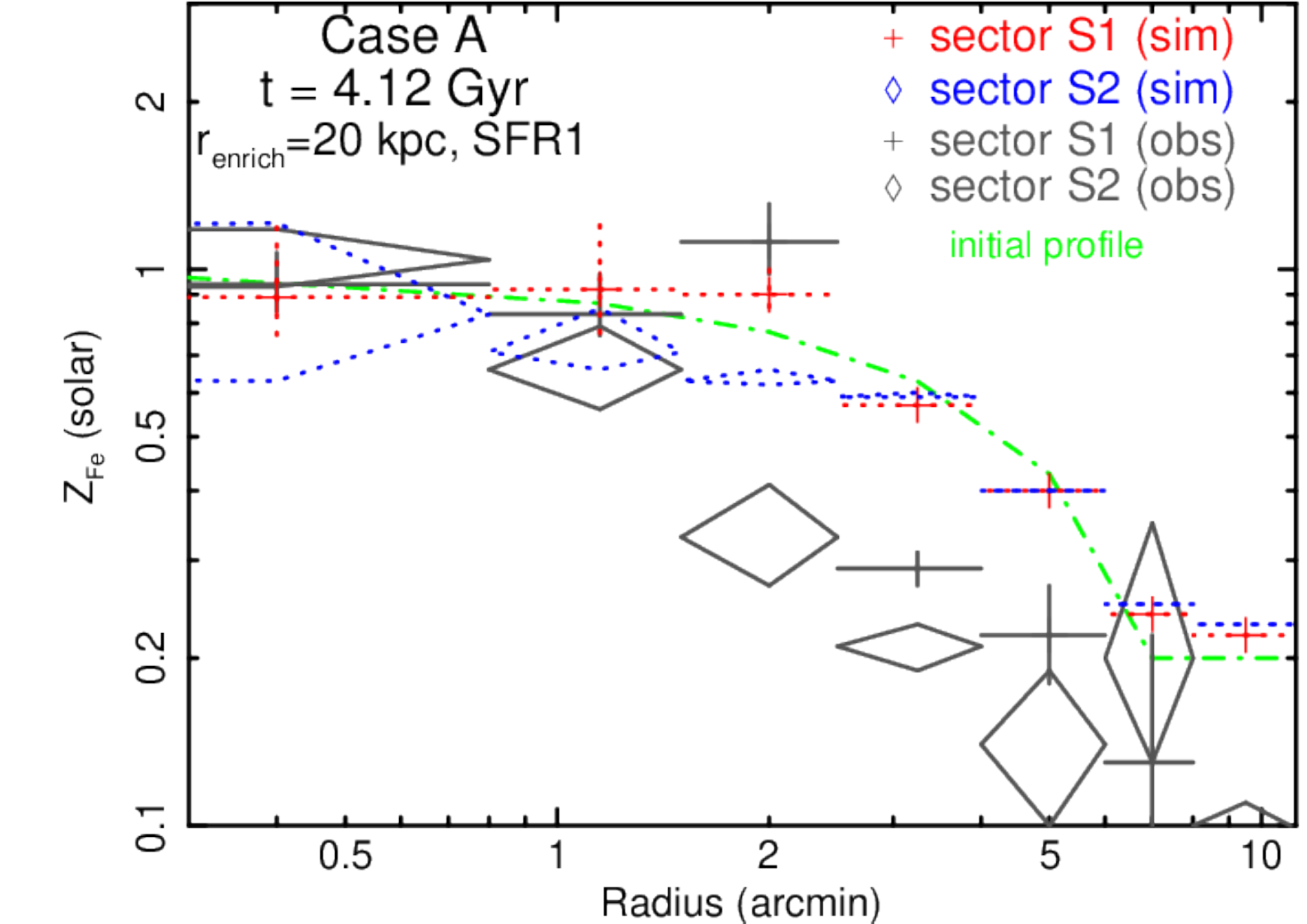}
\includegraphics[scale=.28]{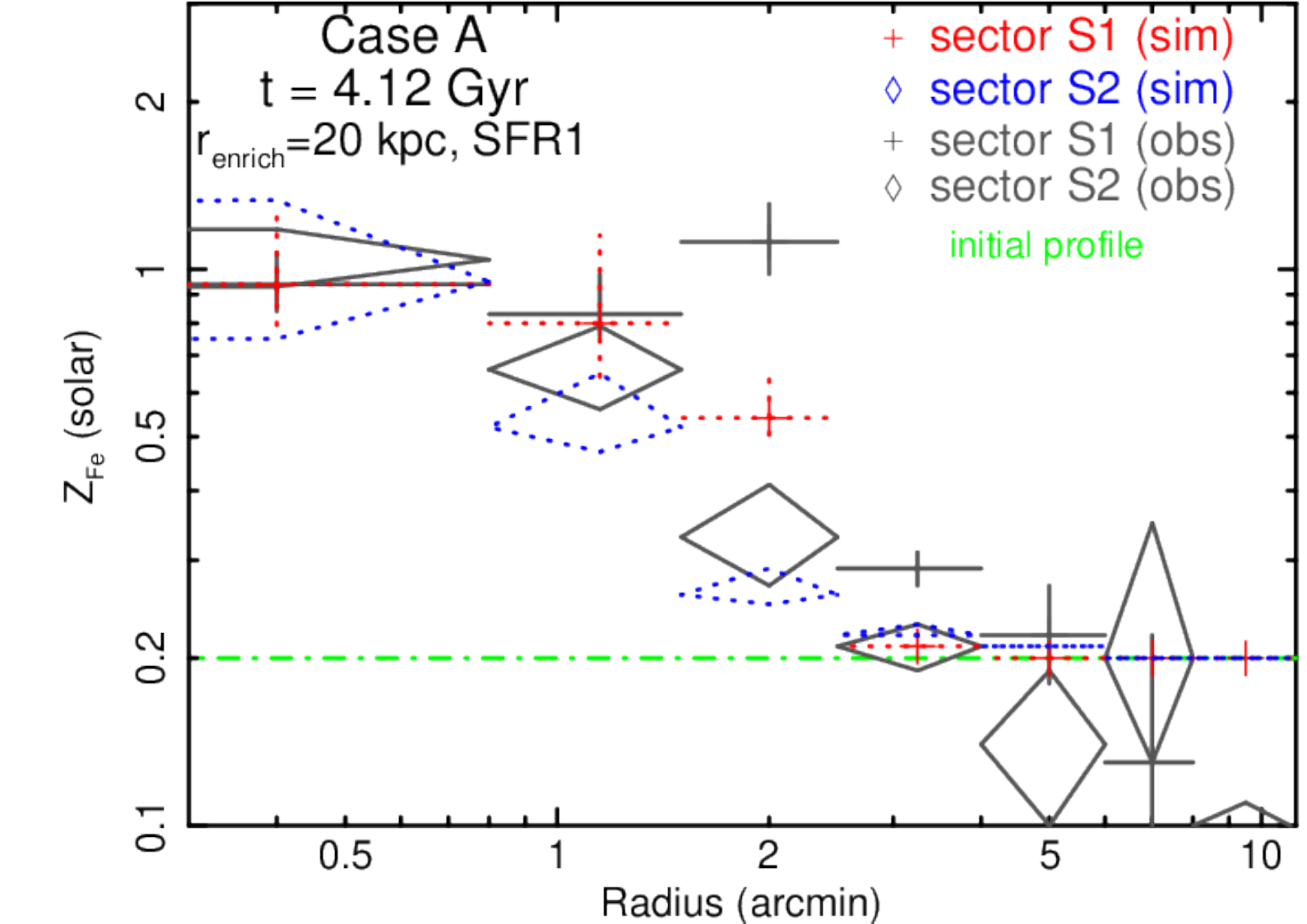}
\includegraphics[scale=.28]{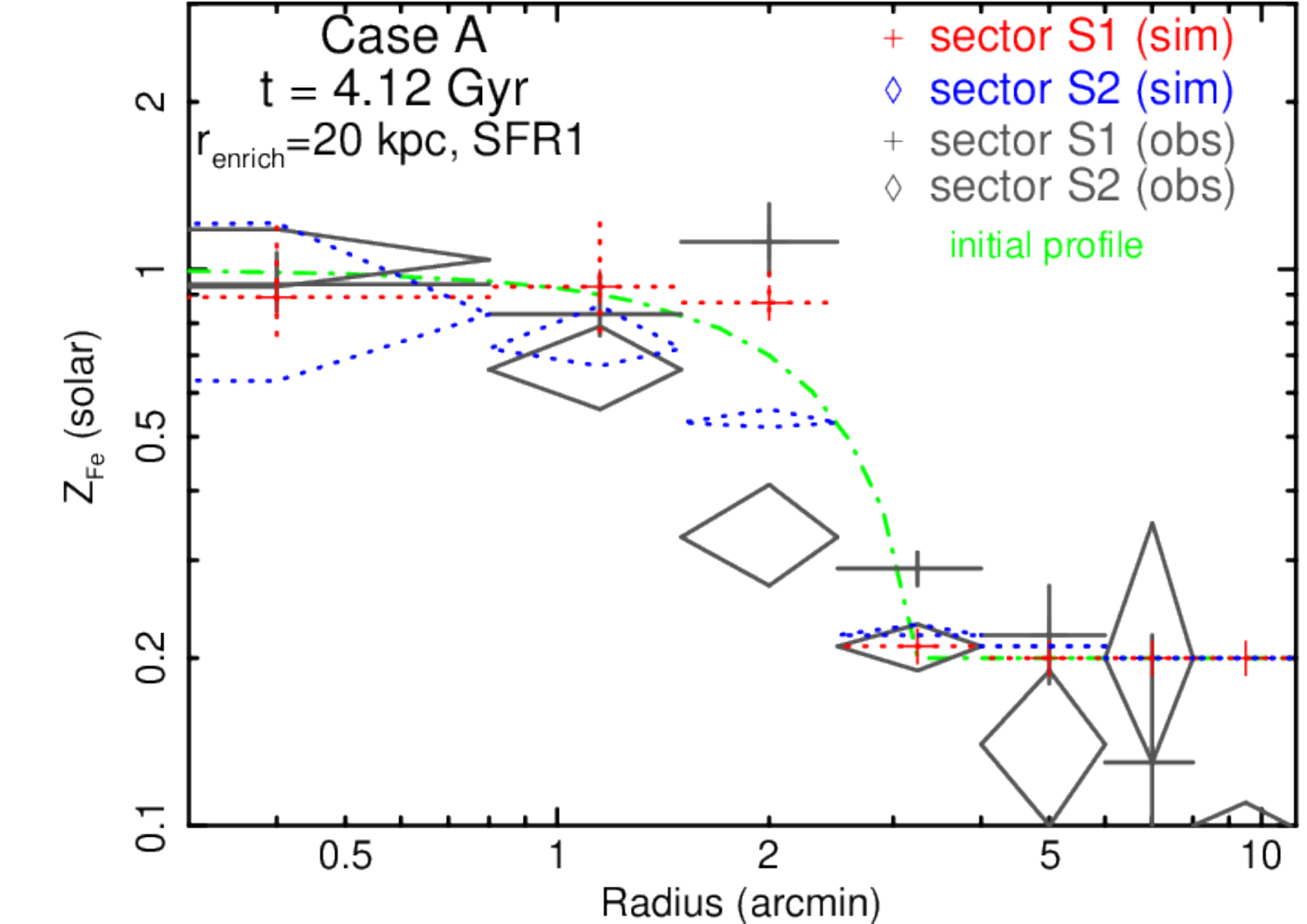}
\caption{ Fe abundance profiles simulated at the `best-match' snapshot with the `best-match' model (Case A, $i_{\rm in}$ = 0) by assuming a centrally peaked (left), a uniform (middle), and a flat-topped (right) initial Fe abundance profile for both gas halos immediately before the second pericentric passage (\S4.2.2). Corresponding Fe abundance profiles of observation are also plotted by dark-grey line. \label{fig14}}
\end{figure}

%--------------------------------------------------------------------------------------------------

% Figure 15
\begin{figure}
\centering
\graphicspath{{figures/}}
\plottwo{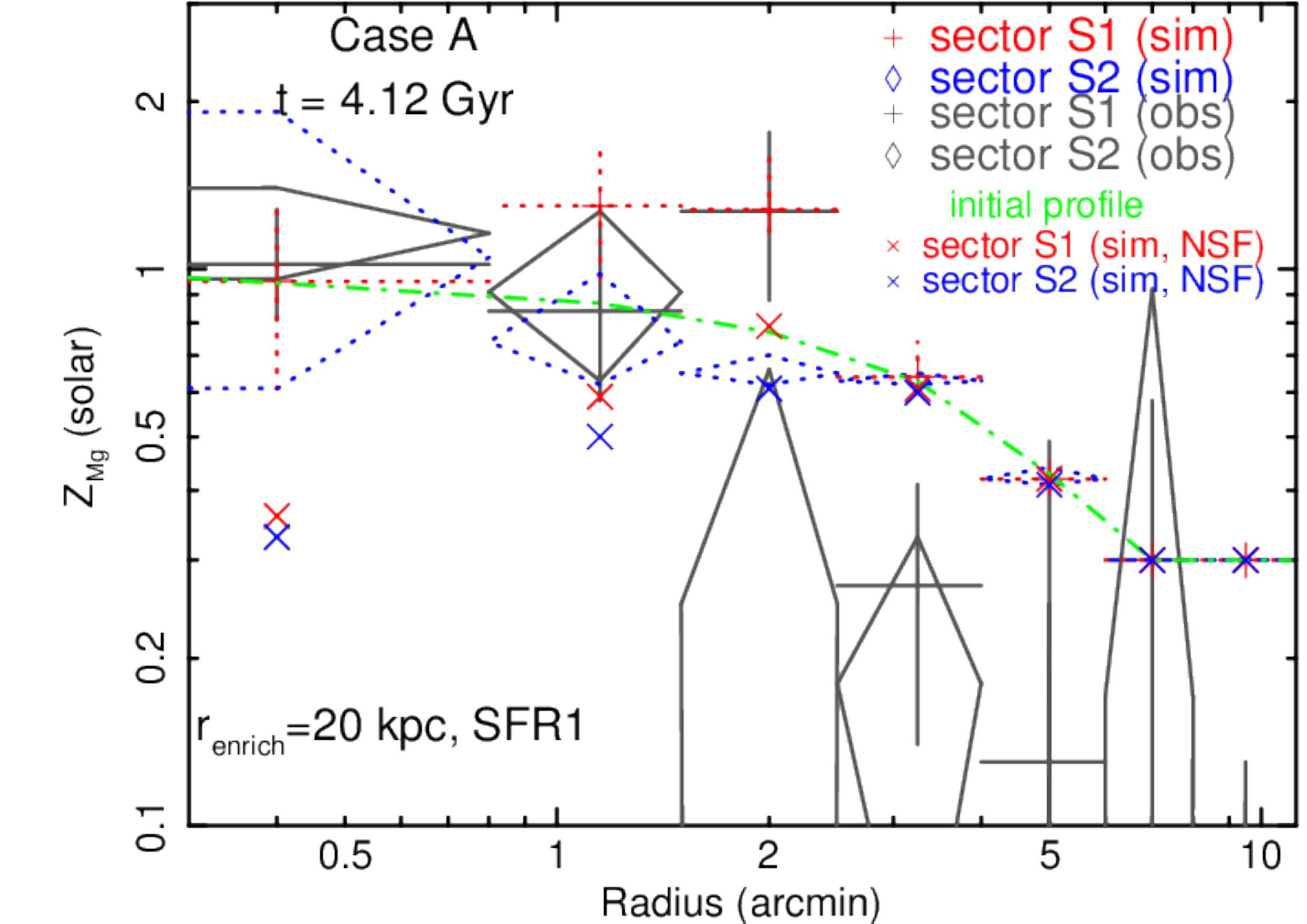}{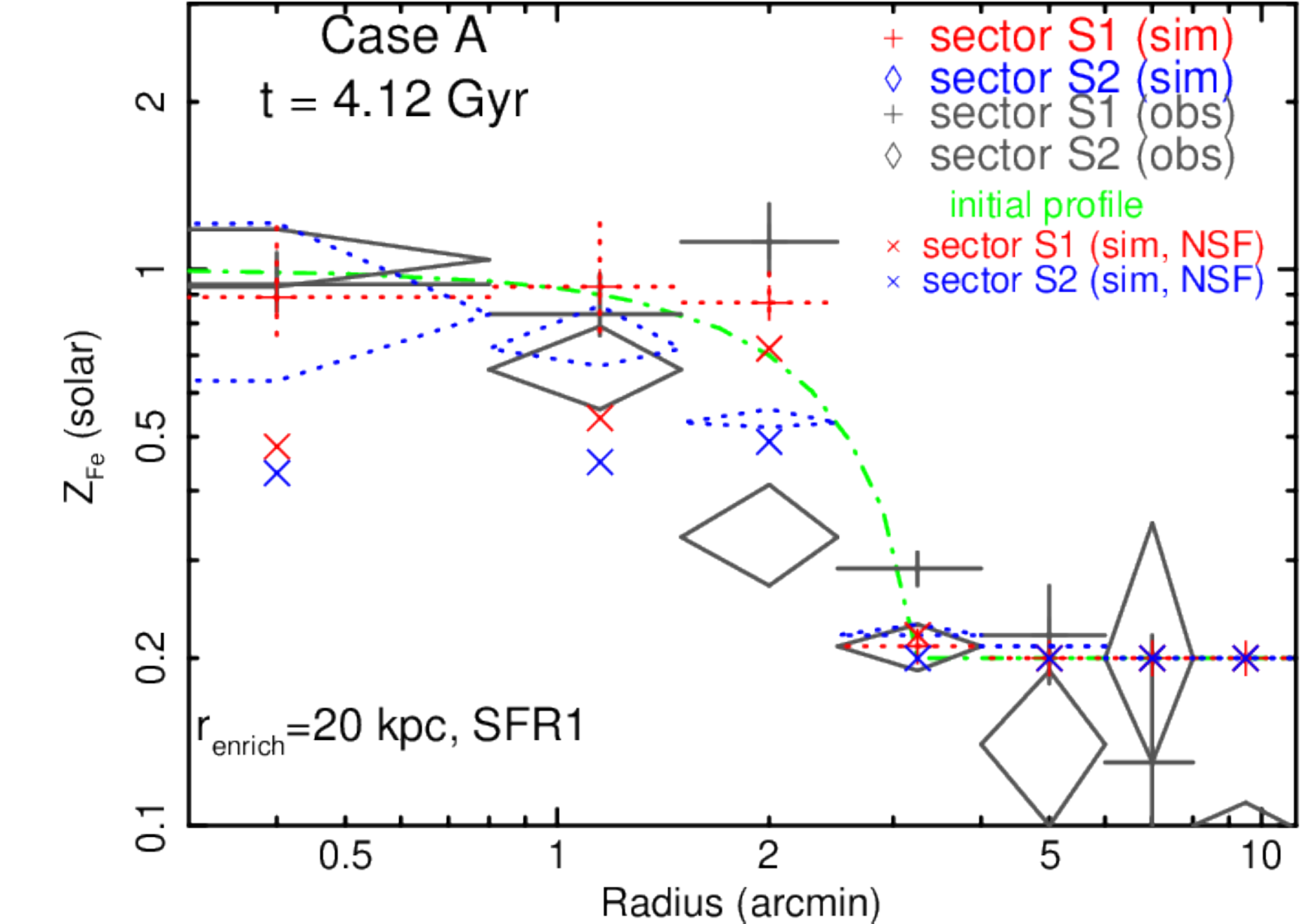}
\caption{ Simulated Mg (left) and Fe (right) abundance profiles at the `best-match' snapshot with the `best-match' model (Case A, $i_{\rm in}$ = 0) when the merger-induced starburst is switched off. The corresponding abundance profiles with the starburst enabled are also plotted for comparison. The marks of `X' with red and blue are represented the simulated abundance without star formation (NSF) process for sector S1 and sector S2, respectively. \label{fig15}}
\end{figure}

\clearpage
%------------------------------------------------------------------------------

\clearpage

%-----------------------------------------------------------------------------------------------

\begin{appendices}

\section{ Uncertainties in Modeling of the \XMM\ Instrumental Background }

Both \XMM\ EPIC-MOS and EPIC-PN detectors produce strong $\rm Al~K\alpha$ lines (at $\simeq 1.5$~keV), which close to the Mg K-lines. In order to investigate whether or not the measurement of Mg abundance is biased if the Al lines are not correctly modeled in the modeling of the instrumental backgrounds, we use Monte Carlo technique to evaluate the impacts of the corresponding uncertainties on our spectral analysis.

For each of the best-fit background models determined in \S 2.3.2, we randomly vary the normalization of each instrumental line (two lines for MOS and six for PN) according to the uncertainty that is obtained by spectral fittings presented in \S 2.3.2, and use them to generate a set of new background templates. After the cross-talk correction and deprojection operation are performed, we jointly fit the observed \XMM\ and \Chandra\ spectra using the new backgrounds. Such a procedure is repeated for 1,000 times, and the derived fitting results are analyzed to determine the best-fit values and corresponding uncertainty ranges for each of the temperature and abundance parameters. Note that the uncertainties estimated in this way (Table~\ref{tbl-4}) contain both the statistical errors and systematic uncertainties caused by the instrumental lines.

By comparing the results to the previous results listed in Table~\ref{tbl-2}, we find that the Mg abundance errors for the outer regions ($>4'$) show apparent changes, which, however, are still much smaller than the corresponding statistical errors. On the other hand, the Mg abundance for the inner regions, the abundances of other elements, and the gas temperatures remain unchanged. Therefore, the uncertainties remaining in the modeling of the instrumental lines have a very limited impact on the measurement of Mg abundance, especially in the inner regions ($< 4'$) where the emission of the galaxy group dominates the spectra \citep[see also][]{mernier16,deplaa17}.

%---------------------------------------------------------------------------------------------------

% Table 4
\begin{table}
\centering
\caption{Deprojected gas temperatures and metal abundances with the single-phase model measured in sectors S1 and S2, after concerning the uncertainties in modeling of the \XMM\ instrumental background (Appendix A). \label{tbl-4} }
\scriptsize
\tablecolumns{7}
\tablewidth{1pc}
\setlength{\tabcolsep}{13pt}
\renewcommand{\arraystretch}{1}
\begin{tabular}{lcccccc}
\tableline\tableline\
 Radius &  kT  &   O   &  Mg  &  Si &  S  &   Fe  \\
(arcmin) &  (keV)  &   (solar)   &   (solar)   &  (solar) &  (solar)  &   (solar)  \\
\tableline\tableline 
\multicolumn{7}{c}{sector S1} \\
\tableline
\multicolumn{4}{l}{Combined} \\
\tableline
$0.0 - 0.8$ & $0.92 \pm 0.01$ & $0.52_{-0.16}^{+0.20}$ & $0.92_{-0.17}^{+0.20}$ & $0.83_{-0.12}^{+0.14}$ & $0.65_{-0.18}^{+0.21}$ & $0.85_{-0.08}^{+0.11}$  \\

$0.8 - 1.5$ & $1.19 \pm 0.02$ & $0.01_{-0.01}^{+0.31}$ & $0.84_{-0.26}^{+0.37}$ & $0.71_{-0.15}^{+0.22}$ & $0.53_{-0.25}^{+0.29}$ & $0.83_{-0.07}^{+0.15}$  \\

$1.5 - 2.5$ & $1.33 \pm 0.01$ & $0.22_{-0.22}^{+0.41}$ & $1.25_{-0.40}^{+0.50}$ & $1.13_{-0.23}^{+0.29}$ & $0.72_{-0.28}^{+0.32}$ & $1.12_{-0.14}^{+0.19}$ \\

$2.5 - 4.0$ & $1.33 \pm 0.01$ & $0.19_{-0.14}^{+0.15}$ & $0.30_{-0.15}^{+0.16}$ & $0.26_{-0.07}^{+0.07}$ & $0.18_{-0.11}^{+0.12}$ & $0.29_{-0.02}^{+0.02}$ \\

\tableline

\multicolumn{4}{l}{\XMM\ EPIC} \\
\tableline

$4.0 - 6.0$ & $1.28 \pm 0.06$ & $0.17_{-0.17}^{+0.42}$ & $0.23_{-0.17}^{+0.42}$ & $0.36_{-0.20}^{+0.22}$ & $0.17_{-0.17}^{+0.35}$ & $0.20_{-0.05}^{+0.07}$  \\

$6.0 - 8.0$ & $1.46_{-0.19}^{+0.20} $ & $0.28_{-0.28}^{+0.80}$ & $0.00_{-0.00}^{+0.61}$ & $0.14_{-0.14}^{+0.31}$ & $0.00_{-0.00}^{+0.36}$ & $0.10_{-0.06}^{+0.10}$  \\

$8.0 - 11.0$ & $0.94 \pm 0.03$ & $0.26_{-0.11}^{+0.13}$ & $0.04_{-0.04}^{+0.12}$ & $0.05_{-0.05}^{+0.07}$ & $0.04_{-0.05}^{+0.20}$ & $0.07_{-0.01}^{+0.01}$  \\

\tableline\tableline

\multicolumn{7}{c}{sector S2} \\
\tableline

\multicolumn{4}{l}{Combined} \\
\tableline
$0.0 - 0.8$ & $0.83 \pm 0.01$ & $0.66_{-0.16}^{+0.19}$ & $1.19_{-0.18}^{+0.23}$ & $0.89_{-0.12}^{+0.15}$ & $0.63_{-0.18}^{+0.21}$ & $0.99_{-0.10}^{+0.13}$  \\

$0.8 - 1.5$ & $1.02 \pm 0.01$ & $0.21_{-0.21}^{+0.30}$ & $0.91_{-0.28}^{+0.36}$ & $0.67_{-0.16}^{+0.21}$ & $0.34_{-0.25}^{+0.29}$ & $0.66_{-0.10}^{+0.13}$  \\

$1.5 - 2.5$ & $1.35_{-0.04}^{+0.11}$ & $0.04_{-0.04}^{+0.45}$ & $0.26_{-0.25}^{+0.41}$ & $0.29_{-0.19}^{+0.22}$ & $0.32_{-0.30}^{+0.34}$ & $0.33_{-0.06}^{+0.08}$ \\

$2.5 - 4.0$ & $1.34 \pm 0.02$ & $0.29_{-0.16}^{+0.17}$ & $0.20_{-0.15}^{+0.16}$ & $0.25_{-0.08}^{+0.08}$ & $0.15_{-0.12}^{+0.12}$ & $0.21_{-0.02}^{+0.02}$ \\

\tableline

\multicolumn{4}{l}{\XMM\ EPIC} \\
\tableline

$4.0 - 6.0$ & $1.32_{-0.08}^{+0.10}$ & $0.00_{-0.00}^{+0.31}$ & $0.00_{-0.00}^{+0.28}$ & $0.16_{-0.16}^{+0.13}$ & $0.00_{-0.00}^{+0.30}$ & $0.12_{-0.06}^{+0.08}$  \\

$6.0 - 8.0$ & $1.40_{-0.09}^{+0.24} $ & $0.33_{-0.33}^{+0.98}$ & $0.28_{-0.28}^{+0.87}$ & $0.37_{-0.32}^{+0.38}$ & $0.24_{-0.24}^{+0.61}$ & $0.16_{-0.09}^{+0.17}$  \\

$8.0 - 11.0$ & $1.02 \pm 0.02$ & $0.27_{-0.10}^{+0.13}$ & $0.00_{-0.00}^{+0.11}$ & $0.12_{-0.07}^{+0.08}$ & $0.14_{-0.18}^{+0.20}$ & $0.10_{-0.01}^{+0.01}$  \\

\tableline\tableline\
\end{tabular}
\end{table}

%\end{appendices}

%-----------------------------------------------------------------------------------------------
%\begin{appendices}
\section{ PSF scattering }

% Table 
\begin{table}[!h]
%\begin{center}
\centering
\tablewidth{1.5pt}
\tablecolumns{10}
\caption{ Percentage contribution of the flux on each pie region from adjacent regions for the PN data in sector S1. Columns refer to the percentage fluxes providing the flux, and the rows refer to the percentage flux receiving flux. For example, 24.08\% (first row, second column) is the percentage of photons which leaked from pie 1 contaminating pie 2 to total photons in pie 2. Note that, `Right' and `Left' in column represent the corresponding pie region is sector S3 and sector S4, respectively. \label{a1}}
\setlength{\tabcolsep}{0.5pt}
\renewcommand{\arraystretch}{0.9}
\begin{tabular}{l@{\hspace{2em}}r@{\hspace{1.5em}}r@{\hspace{1.5em}}r@{\hspace{1.5em}}r@{\hspace{1.5em}}r@{\hspace{1.5em}}r@{\hspace{1.5em}}r@{\hspace{1.5em}}r@{\hspace{1.5em}}r}
\tableline\tableline\
  & pie 1 &  pie 2  &  pie 3  &  pie 4  &  pie 5  &  pie 6  & pie 7 &   Right &  Left \\
\tableline  

 pie 1  &  $-$     &  6.53    &    0.15  &  0.01 & $< 0.01$ & $< 0.01$ & $< 0.01$ & 5.22 &  5.43  \\
 pie 2  & 24.08    &  $-$     &    7.55  & 0.30  & $< 0.01$ & $< 0.01$ & $< 0.01$ & 2.46  & 2.77  \\
 pie 3  & 13.01    & 12.43    &     $-$  & 5.28  &   0.64   & $< 0.01$ & $< 0.01$ & 1.87 &  1.76  \\
 pie 4  & 2.97     & 2.81     &    14.40 &  $-$  &  4.44    & 0.11     & $< 0.01$ & 1.21  & 1.09  \\
 pie 5  & 1.61     & 1.41     &    2.47  & 12.34 &    $-$   & 3.36     &   0.13   & 1.03  & 0.98  \\
 pie 6  & $< 0.01$ & $< 0.01$ &   0.12   & 1.84  &  12.20   &    $-$   &   4.81   & 1.00  & 0.84  \\
 pie 7  & $< 0.01$ & $< 0.01$ & $< 0.01$ & 0.07  &   0.94   & 7.20     &    $-$   & 0.96  & 0.0  \\

\tableline\tableline 
\end{tabular}
%\end{center}
\end{table}

\end{appendices}
%-----------------------------------------------------------------------------------------------

\end{document}